\documentclass[Royal,times]{sagej}

\usepackage{moreverb,url}

\usepackage[colorlinks,bookmarksopen,bookmarksnumbered,citecolor=red,urlcolor=red]{hyperref}

\newcommand\BibTeX{{\rmfamily B\kern-.05em \textsc{i\kern-.025em b}\kern-.08em
T\kern-.1667em\lower.7ex\hbox{E}\kern-.125emX}}

\renewcommand\journalname{Neurosymbolic Artificial Intelligence}
\newcommand\supplement{Supplemental material}

\usepackage{natbib}
\usepackage{graphicx}
\usepackage{hyperref}
\usepackage{color}
\usepackage{epigraph}
\usepackage{csquotes}
\usepackage{amssymb}
\usepackage{amsmath}
\usepackage{amsthm}
\usepackage{bbm}
\usepackage{tikz}
\usetikzlibrary{arrows}
\usetikzlibrary{shapes.geometric}
\usepackage{caption}
\usepackage{subcaption}
\usepackage[algo2e, ruled]{algorithm2e} 
\usepackage{soul}
\usepackage{bbm}
\usepackage{eurosym}
\usepackage{booktabs}
\usepackage{float}
\usepackage{url}
\usepackage{multirow}
\usepackage{multicol}
\usepackage{dsfont}
\usepackage{listings}
\usepackage{pifont}
\usepackage{enumitem}
\usepackage{xcolor}
\usepackage{mdframed}
\usepackage{tablefootnote}
\usepackage[T1]{fontenc}
\usepackage{tikz}
\usetikzlibrary{shapes}
\usetikzlibrary{shapes.geometric}
\usetikzlibrary{arrows,decorations.pathmorphing,backgrounds,positioning,fit}
\usetikzlibrary{arrows,shapes,positioning,shadows,trees}
\usetikzlibrary{shapes,shadows,trees}

\newcommand{\reasonx}[1]{\normalfont {\scshape reasonx}}

\definecolor{codegreen}{rgb}{0,0.6,0}
\definecolor{codegray}{rgb}{0.5,0.5,0.5}
\definecolor{codepurple}{rgb}{0.58,0,0.82}
\definecolor{backcolour}{rgb}{0.95,0.95,0.92}
\definecolor{answertoquery}{rgb}{0.9, 0.9, 0.9}
\lstdefinestyle{mystyle}{
    backgroundcolor=\color{backcolour},   
    commentstyle=\color{codegreen},
    keywordstyle=\color{magenta},
    numberstyle=\tiny\color{codegray},
    stringstyle=\color{codepurple},
    basicstyle=\ttfamily\footnotesize,
    breakatwhitespace=false,         
    breaklines=true,                 
    captionpos=b,                    
    keepspaces=true,                 
    numbers=left,                    
    numbersep=5pt,                  
    showspaces=false,                
    showstringspaces=false,
    showtabs=false,                  
    tabsize=2
}
\def\volumeyear{20XX}

\begin{document}

\runninghead{State, Ruggieri and Turini}

\title{{\reasonx{}}: Declarative Reasoning on Explanations}

\author{Laura State\affilnum{1,}\affilnum{2,}\affilnum{3}, Salvatore Ruggieri\affilnum{1} and Franco Turini\affilnum{1}}

\affiliation{\affilnum{1} University of Pisa, IT\\
\affilnum{2} Scuola Normale Superiore, IT\\
\affilnum{3} Alexander von Humboldt Institute for Internet and Society, GER}

\corrauth{Laura State, Alexander von Humboldt Institute for Internet and Society, Franz\"osische Str. 9, 10117 Berlin, GER.}

\email{laura.state@hiig.de}

\begin{abstract}
Explaining opaque Machine Learning (ML) models has become an increasingly important challenge.
However, current eXplanation in AI (XAI) methods suffer several shortcomings, including insufficient abstraction, limited user interactivity, and inadequate integration of symbolic knowledge.
We propose {\reasonx{}}, an explanation tool based on  expressions (or, queries) in a closed algebra of operators over theories of linear constraints.
{\reasonx{}} provides declarative and interactive explanations for decision trees, which may represent the ML models under analysis or serve as global or local surrogate models for {tabular-data} black-box classifiers.
Users can express background or common sense knowledge as linear constraints. This allows for reasoning at multiple levels of abstraction, ranging from fully specified examples to under-specified or partially constrained ones. {\reasonx{}} leverages Mixed-Integer Linear Programming (MILP) to reason over the features of factual and contrastive instances. We present here the architecture of {\reasonx{}}, which consists of a Python layer, closer to the user, and a Constraint Logic Programming (CLP) layer, which implements a meta-interpreter of the query algebra. The capabilities of {\reasonx{}} are demonstrated through qualitative examples, and compared to other XAI tools through quantitative experiments.
\end{abstract}


\maketitle

\section{Introduction}
\label{sec:reasonx_introduction}

The acceptance and trust of  Artificial Intelligence (AI) applications is hampered by the opacity and complexity of Machine Learning (ML) models, possibly resulting in biased or even socially discriminatory decision-making \citep{DBLP:journals/widm/NtoutsiFGINVRTP20}.
{As a possible solution,} the field of eXplainable Artificial Intelligence (XAI) provides methods to understand how a ML model reaches a decision~\citep{DBLP:journals/csur/GuidottiMRTGP19,DBLP:journals/air/MinhWLN22,DBLP:journals/ijon/MershaLWAK24}.
However, most existing approaches focus on descriptive explanations and rarely support declarative reasoning over the decision rationale.
By reasoning, we mean the possibility for the user to define any number of conditions over explanations (both factual and contrastive ones), which would codify both background knowledge and what-if analyses~\citep{DBLP:journals/corr/abs-2105-10172}, and then query for explanations at the symbolic and intensional level. Answers could be expressed in the same language as the user queries, thus making the query language closed, i.e., (part of) the answer to a query can be used in following queries. This would support the interactivity of the explanation process -- a requirement of XAI~\citep{DBLP:journals/ai/Miller19}.

%


In this paper, we define a closed algebra of operators over sets of linear constraints. Expressions in this algebra represent user queries for computing factual and contrastive explanations {of decision-tree (DT) classifiers}. DTs may be the ML models directly under analysis, {in which case our approach is model-specific}. Alternatively, {provided that they achieve sufficient fidelity}, DTs may serve as global or local surrogates of other {black-box classifiers for tabular data}, with local surrogates approximating the black box in the neighborhood of the instance to be explained. {In this latter setting, our approach is model-agnostic only conditionally on the fidelity of the surrogate to the original model.}

{The motivation for introducing an algebra, rather than a collection of separate explanation procedures, is to provide a uniform, declarative, and compositional representation of explanation problems. Decision-tree paths, instance descriptions, background knowledge, and relationships among factual and contrastive instances can all be expressed in symbolic form as linear constraints. The algebra then provides operators for combining these conditions, checking their consistency, projecting explanations onto selected features, and optimizing user-defined objectives.
Declarativeness allows a query to specify which properties its answers must satisfy while abstracting from the procedure used to compute them. Moreover, the algebra is closed: every operator returns another theory of linear constraints. Therefore, expressions can be freely composed (the meaning of a compound query to be defined from the meanings of its sub-expressions), and the result of a query can be refined or embedded in a subsequent query. This supports interactive symbolic reasoning and enables explanation problems involving under-specified instances, multiple contrastive instances, and multiple models to be expressed within the same high-level framework.
%
}

On top of the algebra, we design {\reasonx{}} (\textit{reason to explain}), an explanation tool that consists of two layers. The first layer is in Python, closer to the data analyst user, where DTs and user queries are parsed and translated. The second layer is in Constraint Logic Programming (CLP) over the reals~\citep{DBLP:journals/jlp/JaffarM94}, where embedding of DTs, and user constraints coding background knowledge are reasoned about in symbolic form.
%
The expressions of the algebra used at the Python layer are evaluated through a CLP meta-interpreter -- a powerful reasoning capability of logic programming~\citep{DBLP:conf/iclp/BrogiMPT91}.
The evaluation of expressions produces linear constraints, which are passed back to the Python layer. 
{The meta-interpreter provides a declarative implementation whose clauses closely mirror the formal semantics of the algebra, thereby establishing a direct correspondence between the specification and execution of a query. Such a correspondence is less immediate and typically requires separate correctness proofs in procedural implementations commonly used in XAI.}

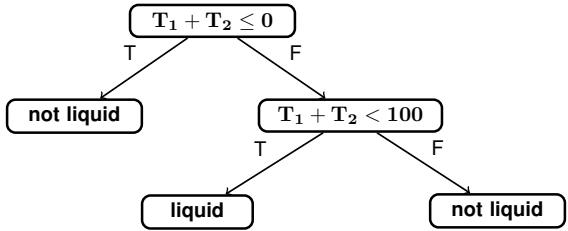
\begin{figure}[t!]
    \centering
		\scalebox{0.8}{
			\begin{tikzpicture}  [every text node part/.style={ align=center},
			every two node part/.style={align=left},
			part/.style  =  every text node part/.style={align=center},
			font={\sf}, rectangle split part fill={white}]
			
			\node  (N1) [rectangle, rounded corners, very thick,draw, text
			width=2.5cm] { $\mathbf{T_1+T_2 \leq 0}$ };
			
			\node  (N2) [rectangle, rounded corners, very thick,draw, text
			width=2cm, below=1cm of N1.south west, xshift=-0.9cm] {
				\textbf{not liquid}
			};			
			\node  (N3) [rectangle, rounded corners, very thick,draw, text
			width=2.8cm, below=1cm of N1.south east,xshift=0.9cm] {
				$\mathbf{T_1+T_2 < 100}$
			};
			
			\node  (N5) [rectangle, rounded corners, very thick,draw, text
			width=1.6cm, below=1cm of N2.south east,xshift=0.9cm] {
				\textbf{liquid}
			};

			\node  (N6) [rectangle, rounded corners, very thick,draw, text width=2cm, below=1cm of N3.south east,xshift=0.9cm]
			{  
				\textbf{not liquid}
			};

			\draw[->,draw,thick] (N1)  -- (N2) node[midway,above left] {T};
			\draw[->,draw,thick] (N1)  -- (N3) node[midway,above right] {F};
			
			\draw[->,draw,thick] (N3)  -- (N5) node[midway,above left] {T};
			\draw[->,draw,thick] (N3)  -- (N6) node[midway,above right] {F};
			
			\end{tikzpicture} }
    \caption{A simple decision tree illustrating the state of a cup of water in a room at temperature $T_1$, which is warmed by a heater contributing an additional temperature $T_2$.}
    \label{fig:cup}
\end{figure}

Consider a simple example illustrating the need for high-level abstraction mechanisms. Figure~\ref{fig:cup} shows a simple DT describing the state of a cup of water (\enquote{liquid} or \enquote{not liquid}), based on room temperature $T_1$ and an additional temperature $T_2$ contributed by a heater (assuming temperatures are measured in Celsius). Traditional XAI tools typically explain predictions only for fully specified instances. E.g., for $T_1 = -10, T_2=20$, the prediction of the decision tree is \enquote{liquid}, and the corresponding factual explanation is the rule $0 < T_1 + T_2 < 100 \rightarrow$ \enquote{liquid}. At a more abstract (intensional) level, we would be interested in explaining a partially specified instance such as $T_1 = -10$. In such a case, the prediction \enquote{liquid} holds for  values of $T_2$ such that $10 < T_2 < 110$, based on the same factual rule, and the prediction \enquote{not liquid} holds for $T_2 \leq 10$ or $T_2 \geq 100$. \reasonx{} enables reasoning at precisely this intensional level.

The contributions of this paper are the following:

\begin{itemize}
    \item We introduce a closed algebra of operators over sets of linear constraints. The algebra is expressive enough to model complex explanation problems over DTs as expressions (or \textit{queries}\footnote{This naming is inspired by relational algebra in database theory \citep{dbscb2008}, which likewise provides a declarative and compositional language whose operators are closed over their domain.}). 

    \item We develop the {\reasonx{}} tool, consisting of two layers. The CLP layer implements an interpreter of the above algebra through meta-programming. The Python layer offers a user-friendly interface specifically designed for ML developers, who can easily integrate models, instances, background knowledge to (silently) generate queries over the algebra and decode back the answers.
    \item We demonstrate the capabilities of {\reasonx{}} through qualitative demonstrations, and a quantitative comparison against other XAI tools.
\end{itemize}

The code of {\reasonx{}} together with the data used in the demonstrations is publicly available at:
\begin{center}
\href{https://github.com/lstate/reasonx}{https://github.com/lstate/reasonx}
\end{center}

This paper is a substantial extension of two previous works: an introductory paper~\citep{DBLP:conf/xai/StateRT23}, targeting an interdisciplinary audience, and a technical paper~\citep{DBLP:conf/jelia/StateRT23}, providing the preliminary architecture of {\reasonx{}}. This extension {adds} the formal definition of the query algebra and its operators, and the definition of explanation problems as queries; 
an in-depth presentation of {an extended} architecture of {\reasonx{}}, and the demonstrators of its capabilities; as well as the comparison with other XAI tools.

The paper is structured as follows: we discuss the background on XAI topics and related work in Section~\nameref{sec:reasonx_background}. To keep the paper self-contained, a brief introduction on CLP and meta-reasoning is provided  in the \supplement. 
In Section~\nameref{sec:reasonx_algebra_operators}, we introduce the algebra of operators, and its capabilities to model explanation problems. The implementation details on \reasonx{} are presented in Section~\nameref{sec:reasonx_details}.
We demonstrate \reasonx{} via some examples in Section~\nameref{sec:reasonx_demonstrations}, present its evaluation in Section~\nameref{sec:reasonx_evaluation}, and close with limitations and future work in Section~\nameref{sec:reasonx_limitations}. Finally, we close with Section~\nameref{sec:reasonx_conclusion}. The \supplement\ reports {a brief introduction to CLP over the reals, further related work,  full experimental settings, and additional experimental results}.


\section{XAI and Related Work}
\label{sec:reasonx_background}

\label{sec:reasonx_xai}

XAI aims at coupling effective ML models with explanations of how they work~\citep{DBLP:journals/csur/GuidottiMRTGP19,DBLP:journals/air/MinhWLN22,molnar2019,DBLP:journals/ijon/MershaLWAK24}.
This can be achieved through novel \textit{explainable-by-design} ML models, or through \textit{post-hoc} explanations of non-interpretable ML models, commonly referred to as \textit{black-box} models. 
{The lack of interpretability of black-box models largely stems from their high complexity and the presence of non-linear relationships.}
Post-hoc explanation methods can be categorized with respect to two aspects~\citep{DBLP:journals/csur/GuidottiMRTGP19}.
One contrasts model-specific \textit{vs.} model-agnostic approaches, depending on whether the explanation method exploits knowledge about the internals of the black-box or not.
The other aspect contrasts local \textit{vs.} global approaches, depending on whether an explanation refers to a specific instance or to the black-box as a whole.

In this work, we focus on post-hoc explanations of classification models {over tabular data}. Our approach is {model-specific to DT models, which can be often too large to be directly understood. As common in the literature in XAI~\citep{DBLP:conf/kdd/Ribeiro0G16,DBLP:journals/expert/GuidottiMGPRT19}, the approach can be conditionally considered model-agnostic, under the assumption that the DT is a surrogate model of the black-box with sufficiently high fidelity, namely that is able to approximate the prediction of the black-box either globally or locally to a specific instance to explain.}

\begin{table}[t!]
    \centering
    \scriptsize
    \setlength{\tabcolsep}{5pt}
    \begin{tabular}{rcccccccccc}
    \toprule
        \multirow{2}{*}{\textbf{Method($^\star$)}} & \multicolumn{2}{c}{\multirow{2}{*}{\textbf{Type}}} & \multicolumn{3}{c}{\textbf{Output}} & {\multirow{2}{*}{\textbf{CON}}} & {\multirow{2}{*}{\textbf{DIV}}}  & \multirow{2}{*}{\textbf{USI}} & \multirow{2}{*}{\textbf{MM}} & \multirow{2}{*}{\textbf{INT}}\\
        & \multicolumn{2}{c}{} & \textbf{FR} & \textbf{CR} & \textbf{CE} & & & \\
        \midrule
        {{\reasonx{}}} & {MA/MS} & {g/l} & {\checkmark} & {\checkmark} & {\checkmark} & {linear} & {\checkmark} & {\checkmark} & {\checkmark} & {\checkmark}\\
        LORE & MA & l & \checkmark & \checkmark & \ding{55} & \ding{55} & n.a. & n.a. & \ding{55} & \ding{55} \\
        GLocalX & MA & g & \multicolumn{2}{c}{rule sets} & \ding{55} & \ding{55} & n.a. & n.a. & \ding{55} & \ding{55} \\
        ANCHORS & MA & l & \checkmark & \ding{55} & \ding{55} & n.a. & n.a. & n.a. & \ding{55} & \ding{55} \\
        Wachter & MA & l & \ding{55} & \ding{55} & \checkmark & \ding{55} & \checkmark & \ding{55} & \ding{55} & \ding{55} \\
        DiCE & MA & l & \ding{55} & \ding{55} & \checkmark & \checkmark & \checkmark & \ding{55} & \ding{55} & \ding{55} \\
        DACE & {MS} & l & \ding{55} & \ding{55} & \checkmark & \checkmark & \ding{55} & \ding{55} & \ding{55} & \ding{55} \\ 
        Cui & MS & l & \ding{55} & \ding{55} & \checkmark & \ding{55} & \ding{55} & \ding{55} & \ding{55} & \ding{55} \\
        Russell & MS & l & \ding{55} & \ding{55} & \checkmark & \checkmark & \checkmark & \ding{55} & \ding{55} & \ding{55} \\ 
        Ustun & MS & l & \ding{55} & \ding{55} & \checkmark & \checkmark & \ding{55} & \ding{55} & \ding{55} & \ding{55} \\ 
        MACE & MA & l & \ding{55} & \ding{55} & \checkmark & \checkmark & \checkmark & \ding{55} & \ding{55} & \ding{55} \\ 
        Karimi & MS & l & \ding{55} & \ding{55} & \checkmark & \checkmark & \ding{55} & \ding{55}& \ding{55} & \ding{55} \\
        Bertossi & MA & l & \ding{55} & \ding{55} & \checkmark & \checkmark &  \ding{55} & \ding{55} & \ding{55} & \ding{55} \\
        Takemura & MS & g & \multicolumn{2}{c}{rule sets} & \ding{55} & \checkmark & n.a. & n.a. & \ding{55} & \ding{55} \\
        LIMEtree & MA & l & \checkmark & \checkmark & \ding{55} & \checkmark & n.a. & n.a. & \ding{55} & \checkmark \\ 
        {T-CREx} & {MA} & {g/l} & {\ding{55}} & {\checkmark} & {\ding{55}} & {\ding{55}} & {n.a.} & {n.a.} & {\ding{55}} & {\ding{55}} \\
        {UGCE} & {MA} & {l} & {\ding{55}} & {\ding{55}} & {\checkmark} & {\checkmark} & {\ding{55}} & {\ding{55}} & {\ding{55}} & {\checkmark}\\
        \bottomrule
    \end{tabular}
    \caption{Comparing {\reasonx{}} against related explainability methods {for tabular data}. Explanation type according to the introduced taxonomy (MA = model-agnostic, MS = model-specific, g = global, l = local). The output can be factual decision rules (FR), contrastive decision rules (CR) or contrastive examples (CE). CON = possibility to enforce constraints on CR/CE, DIV = possibility of optimizing for diversity on a set of CE. USI = under-specified information for a CE. n.a. = not applicable, tab. = tabular, im. = image. MM = explanations over time and multiple models, INT = interactive explanations. \\
    ($^\star$) For space reasons, corresponding papers are listed next:
    LORE~\citep{DBLP:journals/expert/GuidottiMGPRT19},
    GlocalX~\citep{DBLP:journals/ai/SetzuGMTPG21}, ANCHORS~\citep{DBLP:conf/aaai/Ribeiro0G18}, Wachter~\citep{DBLP:journals/corr/abs-1711-00399}, DiCE~\citep{DBLP:conf/fat/MothilalST20}, DACE~\citep{DBLP:conf/ijcai/KanamoriTKA20}, Cui~\citep{DBLP:conf/kdd/CuiCHC15}, Russell~\citep{DBLP:conf/fat/Russell19}, Ustun~\citep{DBLP:conf/fat/UstunSL19}, MACE~\citep{DBLP:conf/aistats/KarimiBBV20}, Karimi~\citep{DBLP:conf/fat/KarimiSV21}, Bertossi~\citep{DBLP:journals/tplp/Bertossi23}, Takemura~\citep{DBLP:journals/tplp/TakemuraI24}, LIMEtree~\citep{DBLP:journals/corr/abs-2005-01427}, {T-CREx~\citep{DBLP:conf/icml/BewleyAMMV24}}, {UGCE~\citep{DBLP:conf/ijcnn/FragkathoulasP25}}.}
\label{tab:reasonx_vs_other_methods}
\end{table}

Table~\ref{tab:reasonx_vs_other_methods} compares {\reasonx{}} with related explainability methods. It shows that our tool uniquely combines reasoning over constraints, rules and contrastive examples, over multiple instances and models, as well as with under-specified instances. We review related work in  this section.

\subsection{Neuro-symbolic XAI}

\cite{DBLP:journals/ia/CalegariCO20} survey how symbolic methods based on some formal logics can be integrated with sub-symbolic models for (neuro-symbolic) explanation purposes. 
\cite{Sabbatini2025} provides an overview of symbolic knowledge extraction methods from black-boxes.
Existing approaches either aim at blending symbolic and sub-symbolic methods (\textit{integration}), or treat them as distinct parts that remain separated but work together (\textit{combination}). The central building blocks of these approaches are decision rules, decision trees, and knowledge graphs. Decision rules have been shown to be effective for both developers \citep{DBLP:journals/pacmhci/PiorkowskiVCDA23} and domain users \citep{DBLP:journals/tvcg/MingQB19,DBLP:journals/ai/WaaNCN21}.

{\reasonx{}} relies on a novel usage of constraint logic programming -- a form of rule-based symbolic logic reasoning -- serving as an explanation tool that adopts the \textit{combination} approach.
Our post-hoc explanation method can be categorized as a \enquote{symbolic [neuro]} approach (type 2 in \citet{DBLP:journals/nca/BhuyanRTS24}), where the black-box model is the neuro subroutine called by the symbolic explanation method locally to an instance to explain or global to the model. The approach is complementary to other neuro-symbolic ones, such as  \enquote{neuro[symbolic]} (type 6) that embeds a symbolic thinking engine directly into a neural engine. Such forms of integration aim at making AI models explainable-by-design.

Other explanation methods exist that borrow concepts from logics but do not allow users to reason over those concepts. 
The closest ones to our approach are those using (propositional) logic rules as forms of model-agnostic explanations both locally~\citep{DBLP:journals/tvcg/MingQB19,DBLP:journals/expert/GuidottiMGPRT19,DBLP:conf/aaai/Ribeiro0G18} and globally~\citep{DBLP:journals/ai/SetzuGMTPG21}.
Another prominent example is TREPAN~\citep{DBLP:conf/nips/CravenS95}, which computes a decision tree with m-of-n split binary conditions.
Further related work in this category is presented by a series of papers by Sokol et al.~\citep{sokol2021intelligible,DBLP:journals/corr/abs-2005-01427,DBLP:conf/ijcai/SokolF18a}, based on local surrogate regression  trees.
Furthermore, Answer Set Programming (ASP), another powerful logic programming extension, has been adopted for local and global explanations of tree ensembles~\citep{DBLP:journals/tplp/TakemuraI24}, and for contrastive explanations from the angle of actual causality~\citep{DBLP:journals/tplp/Bertossi23}. 

Model-specific logic approaches to XAI offer formal guarantees of rigor, giving rise to the sub-field of formal XAI~\citep{DBLP:conf/rweb/Silva22}. 
In the \supplement, we compare sufficient reasons from the formal XAI literature to the notion of factual explanations supported by \reasonx{}.
%
Recently, epistemic logics have been used to model the justification (i.e.,~a \enquote{proof}) for a belief embedded in an AI model, e.g., for the statements made by Large Language Models (LLMs)~\citep{Kristina2025}. The justification logic~\citep{Artemov2019}, for instance, has been used to provide personalized explanations~\citep{DBLP:conf/clar/LuoSD23} in a multi-agent conversation framework.

\subsection{Decision Trees and Factual Explanations}
\label{sec:reasonx_embeddings_dt}

Decision trees (DTs)~\citep{DBLP:books/sp/datamining2005/RokachM05,DBLP:journals/air/CostaP23} are classification models{, trained over tabular data,} that predict or describe a nominal feature (\textit{the class}) on the basis of predictive features, which can be nominal, ordinal or continuous.
A decision tree is a tree data structure comprised of internal nodes and leaves.
Each internal node points to a number of child nodes, the degree of the node, based on the possible outcomes of a split condition at the node. The split conditions are defined over the predictive features. Here, we will consider binary split conditions in the form of linear inequality constraints.
Leaf nodes terminate the tree and are associated with an estimated probability distribution over class labels. The class label with the highest estimated probability is predicted by the DT (\textit{Bayes' rule}) for instances satisfying all split conditions from the root node to the leaf. The highest estimated probability is called the \textit{confidence} of the prediction. {For surrogate DTs, such a confidence is the \textit{fidelity} at the leaf, namely the estimated probability that the black-box outputs the same prediction.}

DTs are inherently transparent and intelligible~\citep{Rudin2019_StopExplainingML}, unless their size is large. A path from the root note to a leaf corresponds to a conjunction of split conditions that describe the decision rule for class prediction at the leaf. The decision rule of the path followed by an input instance is an explanation for the prediction of the DT over such an instance. This rule-based definition of \linespread{} \textit{(factual) explanations} is shared with most rule-based explainers~\citep{DBLP:journals/tvcg/MingQB19,DBLP:journals/expert/GuidottiMGPRT19,DBLP:conf/aaai/Ribeiro0G18,DBLP:journals/ai/SetzuGMTPG21,DBLP:conf/nips/CravenS95}.
%
\textit{Sufficient reasons}~\citep{DBLP:journals/jair/IzzaIM22} (or subset-minimal abductive explanations) from the sub-field of formal XAI consist of a minimal subset of feature values of the instance to explain, such that the prediction of the DT is the same as for the instance regardless of the values of the remaining features. See~\citet{DBLP:conf/ijcai/AmgoudMT23} for a discussion of the relation to the rule-based definition. 

A trending research topic in XAI consists of the design of DT learning algorithms that can reproduce the behavior of complex and opaque models, e.g., of a tree ensemble~\citep{DBLP:journals/jbd/WeinbergL19,DBLP:conf/icml/VidalS20,DBLP:conf/dis/BonsignoriGM21,DBLP:conf/ijcai/DudyrevK21,DBLP:conf/ifip12/FerreiraCB22}. Such algorithms can be used for training surrogate models, either globally or locally to an instance to explain. Alternative approaches rely on the synthetic generation of a plausible dataset (such as in the neighborhood of the instance to explain) for training surrogate DTs~\citep{DBLP:journals/expert/GuidottiMGPRT19,LOREsa,DBLP:conf/gecco/BarbosaSF24}. Our research is orthogonal to these topics. We assume a given DT to reason about, independently of whether it is the black-box itself or a global or local surrogate model of a black-box.
Nevertheless, high \textit{fidelity} of a surrogate DT w.r.t.~the black-box is a key property that we need to assume in order to support the reasoning over DTs as an approximation of the reasoning over the black-box.

In our approach, we reason over a DT by encoding it as a set of linear constraints. This problem, known as \textit{embedding}~\citep{DBLP:conf/cpaior/BonfiettiLM15}, requires to satisfy $c(x, y) \Leftrightarrow f(x) = y$,
where $f(x)$ is the predicted class by the DT, $x$ is the vector of predictive features, and $c$ is a logic expression. In our approach, $c$ is a disjunction of linear constraints, one for each path from the root to a leaf.
This rule-based encoding takes space in $O(n \log{n})$ where $n$ is the number of nodes in the DT.
%
Other DT encodings, such as Table and MDD from~\citet{DBLP:conf/cpaior/BonfiettiLM15}, require a discretization of continuous features, thus losing the expressive power of reasoning on linear constraints.

\subsection{Contrastive Explanations}
\label{sec:background_contrastive_explanations}

Factual explanations, or simply explanations, provide answers to \enquote{Why did the input data lead to the provided prediction?}.
Contrastive explanations (CEs)\footnote{
Contrastive explanations are closely related to the concept of counterfactuals as understood in the statistical causality literature. However, it is not the same, and to avoid confusion, we use -- in contrast to other literature in XAI -- the term contrastive explanation (later also \enquote{contrastive example/rule/instance}) instead of counterfactuals. For further discussion on this, see e.g.,~\citet{DBLP:journals/ai/Miller19}.
} refer to data instances similar to the one being explained but with a different predicted outcome.
They answer \enquote{What should the input data look like, in order to obtain a different output?} which can be easily translated to \enquote{How should the current input change to obtain a different output?}, or to answer \textit{what-if} questions, such as \enquote{\textit{What} happens to the output \textit{if} the input changes that way?}.
In recent years, contrastive explanations have received considerable attention, as demonstrated by an increasing number of surveys, e.g.,~\citep{DBLP:journals/csur/KarimiBSV23,Guidotti2022,DBLP:conf/ijcai/KeaneKDS21,DBLP:journals/access/StepinACP21}. 
%
%
CEs can be represented not only as data instances but also as decision rules~\citep{DBLP:journals/expert/GuidottiMGPRT19};
they are considered \textit{local}, \textit{model-agnostic} explanations.
Contrastive explanations have been introduced to XAI by~\cite{DBLP:journals/corr/abs-1711-00399}. The authors propose to generate them through the following optimization problem:

\begin{equation} \label{equ:wachter}
    \arg \min_{x_{ce}} \max_{\lambda} \lambda(f(x_{ce})-y_{ce})^2 + d(x_f, x_{ce})
\end{equation}
where $f()$ is the ML model, $x_{ce}$ refers to the contrastive instance, $x_f$ the original instance, $y_{ce}$ to the  prediction to change to, $\lambda$ denotes a tuning parameter, and $d(.,.)$ a distance measure. 
Intuitively, a contrastive example is the more \enquote{realistic} the closer it is to the original instance, understood as the change of as few feature values as possible
while $f(x_{ce})$ must align with $y_{ce}$.
The optimization function (Equation~\ref{equ:wachter}) can be augmented by constraints, e.g.,~to make the changes actionable~\citep{DBLP:journals/csur/KarimiBSV23} or for computing a diverse set of contrastive examples~\citep{DBLP:conf/fat/MothilalST20,DBLP:conf/fat/LaugelJLMD23}.
Such an extension can be understood as adding \textit{background knowledge} to the explanation.
%
Solutions to the optimization problem can be found by reducing it to satisfiability (SAT) or mixed
integer linear programming (MILP) problems~\citep{DBLP:conf/fat/UstunSL19,DBLP:conf/fat/Russell19,DBLP:conf/ijcai/KanamoriTKA20}. 
{In particular, \cite{DBLP:conf/fat/Russell19} operates model-specifically on linear classifiers and minimizes a weighted $L_1$ distance, whereas \reasonx{} reasons over decision-tree path constraints, supports both $L_1$ and $L_\infty$ optimization, and can return contrastive rules and under-specified regions in addition to point-valued examples. Moreover, Russell generates diversity sequentially through exclusion constraints, while \reasonx{} jointly optimizes proximity and pairwise diversity over a set of contrastive instances.
In the sub-field of formal XAI,} contrastive explanations (or CXp's) in the sense of~\cite{DBLP:journals/jair/IzzaIM22,DBLP:conf/ecai/AudemardLMS23} consist of a minimal subset 
of features of the instance to explain, such that the prediction of the DT is changed for some value of such features, all the other feature values of the instance being unchanged.

{\reasonx{}} is the first approach adopting CLP for generating contrastive explanations. We consider a rule-based definition guided by a DT and user-defined constraints linking the instance to explain to its contrastive instance.

\subsection{Group Explanations and Under-specified Instances}\label{sec:usi}

Local explanations focus on predictions for single instances. Global explanations focus on the decision logic of a black-box for the whole population. \textit{Group explanations} research considers sub-populations of similar instances with the same predictions, sometimes called cohorts or profiles, that share similar (contrastive) explanations. {Most of the} existing approaches consist of clustering factual explanations~\citep{DBLP:journals/ai/SetzuGMTPG21} and contrastive examples~\citep{DBLP:conf/iccbr/WarrenDGK24}. Other works extend the optimization problem of Equation \ref{equ:wachter}~\citep{DBLP:journals/eswa/CarrizosaRM24}; focus on explanations that refer to an ensemble of decisions~\citep{DBLP:journals/corr/abs-2205-08974}; or connect group contrastive explanations to actionable recourse~\citep{DBLP:conf/nips/RawalL20}. {The T-CREx approach \citep{DBLP:conf/icml/BewleyAMMV24} provides model-agnostic local and global contrastive explanations by learning generalized recourse rules from axis-parallel tree-based surrogate models. While T-CREx summarizes recourse options for groups through learned rules, {\reasonx{}} allows groups to be specified intensionally through linear constraints.}

The literature for local and group explanations deals with an \textit{extensional} format of instances, i.e.,~instances are points in a dimensional space, and sub-populations are simply sets of instances. 
Instead, the symbolic approach of {\reasonx{}} deals with instances in an \textit{intensional} format as linear constraints \citep{Cook2009}. Fully specified instances are defined by setting equality constraints for all features, e.g.,~\texttt{age = 20}. In addition, \textit{under-specified instances} can be defined, e.g.,~bounding the feature values, e.g.,~\texttt{age >= 20}, or even setting no constraint on that feature. 
An under-specified instance allows for computing (contrastive) explanations of a specific sub-population. 
This aspects distinguishes our approach from existing methods.

\subsection{Background Knowledge}\label{sec:bk}

Exploiting background knowledge in computing an explanation has the potential to significantly improve its quality~\citep{DBLP:journals/corr/abs-2105-10172,DBLP:conf/iclp/State21}. 
Background (or prior) knowledge is any information that is relevant in the decision context, but that does not (explicitly) emerge through the data.
Not only do simple facts count as such, but we would also consider a natural law as such knowledge, or a specific restriction or requirement a user has, related to her living reality (e.g.,~a minimum credit amount in a credit application scenario).
%
%
Background knowledge integration also draws a connection between our work and the field of \textit{actionable recourse}~\citep{DBLP:journals/csur/KarimiBSV23}:
this research field proposes explanations not only to explain the model outcome to the affected individual, but also to recommend specific actions so that an undesirable outcome can be changed. Integrating background knowledge then helps to better tailor the recommendation to the needs of the user. 

{\reasonx{}} incorporates knowledge in the form of linear constraints over the reals. 
Some examples of knowledge integration from the literature include:
a local explanation tool for medical data that incorporates an ontology to generate a meaningful local neighborhood for explanation generation~\citep{DBLP:conf/fat/PaniguttiPP20};  
a discrimination discovery approach~\citep{DBLP:journals/tkdd/RuggieriPT10}, where knowledge in the form of association rules is used to detect cases of indirect discrimination;
and, in the sub-field of formal XAI, computational complexity investigations of (subset-minimal) abductive and contrastive explanations under domain theories~\citep{DBLP:conf/ijcai/AudemardLMS24a} and of constrained classifiers~\citep{DBLP:journals/ai/CooperS23}.


\subsection{Interactivity}

Although being acknowledged as a key property of explanation tools~\citep{DBLP:journals/ai/Miller19,DBLP:journals/cacm/WeldB19,DBLP:journals/corr/abs-2202-01875},
few XAI methods incorporate interactivity.
\cite{DBLP:journals/ki/SokolF20} outline the usefulness of interactivity prominently in one of their articles: \enquote{Truly interactive explanations allow the user to tweak, tune and personalise them (i.e.,~their content) via an interaction, hence the explainee is given an opportunity to guide them in a direction that helps to answer selected questions}.
We also point to~\cite{DBLP:journals/corr/abs-2202-01875}, presenting an interview study with practitioners and revealing that interactivity for explanations is preferred over static explanations.
%
A notable tool that provides interactive explanations is the glass-box tool~\citep{DBLP:conf/ijcai/SokolF18a}, which can be queried by voice or chat and provides explanations in natural language. 
Recent advances of LLMs are promising to transform explanations into natural, human-readable narratives~\citep{DBLP:journals/corr/abs-2405-06064,DBLP:journals/corr/abs-2401-13110}. {With regard to computational performance, \cite{DBLP:conf/ijcnn/FragkathoulasP25} propose UGCE, which incrementally updates contrastive explanations when users revise feasibility constraints.}

{\reasonx{}} natively offers an interactive interface, where explanations can be refined by asserting and retracting constraints, by adding and removing factual and contrastive instances, by projecting to features of interest, and by optimizing tailored distance functions. Since the output of a query to {\reasonx{}} consists of linear constraints,
part of an answer can be readily embedded in subsequent queries by the user. This enables a two-way flow of information, commonly referred to as \textit{interaction}, between the system and the user. 
{With regard to computational performance, {\reasonx{}} does not currently reuse intermediate results across successive interactions, and each revised query is evaluated from scratch.}

\subsection{Reasoning over Time and Multiple Models}

In operational deployment of ML models, it is common that the model changes over time. A change can be induced, for example, by new incoming data that lead to a retraining.
Explanation methods that contrast different models are rare. \cite{DBLP:journals/dss/MalandriMMS24} present (global) model-contrastive factual explanations of symbolic models, such as decision rules or trees, by computing conditions satisfied by one model and not satisfied by another one.
A specific issue is that of contrastive examples that are invalidated by model changes~\citep{DBLP:conf/fat/BarocasSR20}. An invalidation can be critical if the contrastive explanation is intended for the user to seek recourse, i.e.,~to change the unfavorable model decision by actions proposed by the contrastive explanation.
%
\cite{DBLP:journals/access/FerrarioL22} call such cases \enquote{unfortunate counterfactual events}, and suggest a data augmentation strategy that relies on contrastive explanation generated previously.

A related case is to compare explanations of the same instance classified by different ML models \textit{at the same time}. For example, model developers are interested in reasoning on explanations of the same instance with respect to two different but similarly performing models. 
%
\cite{DBLP:conf/uai/PawelczykBK20, DBLP:conf/ijcai/AlfanoGL0T25} discuss the generation of contrastive explanations 
under predictive multiplicity. This phenomenon refers to different models that exhibit very similar performances for a specific prediction problem.
%
%

{ \cite{DBLP:conf/ijcai/JiangL0T24} survey robustness of contrastive explanations to model changes, model multiplicity, noisy execution, and input changes. {\reasonx{}} addresses only a restricted form of the first two by reasoning over explicitly supplied models, without providing robustness guarantees against unspecified future changes or perturbations. Specifically,
} {\reasonx{}} allows to declare factual and contrastive instances over different models, and to relate them via constraints. This is a natural by-product of the query language {\reasonx{}} is built on, which is able to cover the case of the same instance and two (or more) different models, either referring to different training times or to different black-box models.


\section{A Query Language over Linear Constraints for XAI}
\label{sec:reasonx_algebra_operators}

We define an algebra of operators over sets of linear constraints. Expressions over the operators define a language of queries. 
{The purpose of the query language is to express different explanation problems within a single declarative and compositional framework. Conventional explanation methods commonly define a dedicated procedure for each task, such as computing a factual rule, generating a contrastive example, or minimizing the distance between factual and contrastive instances. In our approach, these tasks are instead specified by composing a small set of operators over a common symbolic representation.
A decision-tree path identifies a region of the feature space through a conjunction of linear constraints. Descriptions of factual and contrastive instances, user-provided background knowledge, and relationships among instances can be represented in the same form. A query may then combine these conditions, eliminate inconsistent alternatives, retain only selected variables, or optimize an objective. Thus, a query specifies the properties that the requested explanations must satisfy rather than the computational steps required to obtain them.
The language is designed according to four main principles. First, it provides a uniform representation: model paths, instance descriptions, and background knowledge belong to the same constraint domain. Second, it supports intensional reasoning: an instance may denote a region of the feature space rather than a fully specified point. Third, it is compositional: complex explanation problems are constructed from simpler expressions. Fourth, it is closed: every expression denotes a theory of linear constraints and may therefore appear as a subexpression of another query.
}

\subsection{Linear Constraints and Decision Trees}

A primitive linear constraint is an expression $a_1 \cdot x_1 + \ldots a_n \cdot x_n \simeq b$,
where $\simeq$ is in $\{<, \leq, =, \geq, > \}$, $a_1, \ldots, a_n$ are constants in ${\mathbb{R}}$ and
$x_1, \ldots, x_n$ are variables. We will use the inner product form by rewriting it as ${\bf
a}^T{\bf x} \simeq b$, where ${\bf x} = x_1, \ldots, x_n$ is the vector of variables and ${\bf a} = a_1, \ldots, a_n$ the vector of constants. A linear constraint, denoted by $c$, is a sequence of primitive linear constraints,
whose interpretation is their conjunction. A path from the root node to a leaf in a decision tree with linear split conditions trivially maps to a linear constraint.

We write 
$\models \psi$, to denote the validity of the
first order formula $\psi$ in the domain of the reals. A constraint $c$ is satisfiable if $\models \exists\ c$, namely if there exists  an assignment of all of its variables to real values that evaluates to true in the domain of reals -- also called a solution of $c$. The solutions of a linear constraint $c$ are a polyhedron in the space of the variables~$\mathbf{x}$. {If the polyhedron is bounded along every dimension, it is called a polytope.} The projection of $c$ over a subset $\mathbf{w}$ of variables is the linear constraint equivalent to existentially quantifying over all variables except $\mathbf{w}$, or in formula $\exists_{-\mathbf{w}}\ c$. The linear projection can be computed by the Fourier-Motzkin projection method. {We refer to \cite{Sch:theory} for a standard introduction to linear constraints and polyhedral theory.} 

We assume any number of decision tree identifiers $DT_1$, $DT_2$, $\ldots$, each decision tree is defined over the features $\mathbf{x} = x_1, \ldots, x_n$. 
Also, we assume any number of data instance identifiers (or, simply, \textit{instances}) $I_1, I_2, \ldots$. 
For each instance $I_i$, and feature~$x_j$, the variable identifier $I_i.x_j$ is given and called an \textit{instance feature}. 
In other words, instance features are variables whose identifiers are prefixed by the instance identifier and suffixed by the feature names. 
We consider sets of linear constraints over variables including at least the instance features set $\mathcal{F} = \{ I_i.x_j \ |\ i \geq 1, j \in [1, n] \}$.

We call a set of linear constraints a \textit{theory}. Each constraint in a theory is interpreted as a conjunction of linear inequalities over instance features. The theory as a whole is interpreted as the disjunction of the constraints in the theory. {Theories constitute the semantic domain of the algebra: every operator introduced below denotes a theory. This common result type is what makes the algebra closed and its expressions compositional.}

\subsection{An Algebra of Operators}\label{sec:algebra}

{We now introduce the syntax and semantics of the operators. The formal definitions specify the logical meaning of each expression independently of its current implementation by the CLP meta-interpreter. The next sub-section shows how factual, contrastive, and minimal contrastive explanation problems are obtained by composing these operators.}
%
The syntax of expressions $E$ in the algebra is  shown next:
\begin{small}
\begin{eqnarray*}\label{def:syn}
E & ::= & \mathit{inst}(I_i, DT_k, l, pr) \ |\ \{U\}\ |\ \mathit{cross}(E+) \ | \ \mathit{sat}(E) \  | \ \pi_{\mathbf{w}}(E)\ | \ \mathit{inf}(E, f(\mathbf{w})) \ | \ \mathit{relax}(E)
\end{eqnarray*}    
\end{small}
where $l$ is a constant that denotes a class label, $pr$ is a constant that denotes a probability in $[0, 1]$, $\mathbf{w} \subseteq \mathcal{F}$ is a set of instance features, and $f()$ a linear function.

The semantics $Th(E)$ of an expression $E$ is a theory, defined {inductively} as follows:
\begin{itemize}
 \item $Th(\mathit{inst}(I_i, DT_k, l, pr))$ is the set $\{ \wedge_{\mathbf{a}^T \mathbf{x} \simeq b \in P} \ \mathbf{a}^T I_i.\mathbf{x} \simeq b \ |\ P \in DT_k(l, pr) \}$ where $DT_k(l, pr)$ are the paths in $DT_k$ leading to a leaf with class label $l$ predicted with confidence at least $pr$. 
 {Intuitively, the $\mathit{inst}$ operator allows to answer the explanation question: \textit{Which constraints lead a decision tree to predict a selected class with a required confidence?}}
 \item $Th(\{U\})$ is a singleton linear constraint $\{ \wedge_{c \in U} c \}$ where $U$ is a sequence of (user-provided) linear constraints.
  {Intuitively, the $\mathit{\{U\}}$ operator answers: \textit{Which constraints are required by the user?}}
 \item $Th(\mathit{cross}(E_1, \ldots, E_k)) = \{ c_1 \wedge \ldots \wedge c_k \ |\ c_1 \in Th(E_1), \ldots, c_k \in Th(E_k) \}$ is the cross-product of constraints in the theories $Th(E_1)$, $\ldots$, $Th(E_k)$. 
  {Intuitively, the $\mathit{cross}$ operator answers: \textit{Which constraints must hold simultaneously?}}
 \item $Th(\mathit{sat}(E)) = \{ c \in Th(E) \ |\ \models \exists\ c \}$ is the set of satisfiable constraints $c$ in $Th(E)$ 
  {Intuitively, the $\mathit{sat}$ operator answers: \textit{Which constraints are feasible?}}
 \item $Th(\pi_{\mathbf{w}}(E)) = \{ \exists_{-\mathbf{w}}\ c \ |\ c \in Th(E) \}$ is the set of projected constraints $c$ in $Th(E)$ 
  {Intuitively, the $\mathit{projection}$ operator answers: \textit{Which constraints remain when restricting to selected features or instances?}}
 \item $Th(\mathit{inf}(E, f(\mathbf{w}))) = \{ c \wedge ( f(\mathbf{w})= (\mathit{inf}{f(\mathbf{w})\ \textrm{s.t.}\,c})) \ |\ c \in Th(E)\}$ is the set of  constraints in $Th(E)$  extended with the minimization of $f()$, 
  {Intuitively, the $\mathit{inf}$ operator answers: \textit{Which constraints minimize a selected (distance) function?}}
 \item $Th(\mathit{relax}(E)) = \{ c^{=} \ |\ c \in Th(E)\}$ where $c^{=}$ is obtained by replacing strict inequalities in $c$ with inequalities: $\mathbf{a}^T{\bf x} < b$ is replaced by $\mathbf{a}^T{\bf x} \leq b-\epsilon$, and $\mathbf{a}^T{\bf x} > b$ is replaced by $\mathbf{a}^T{\bf x} \geq b+\epsilon$, where $\epsilon > 0$ is a machine precision parameter.
  {Intuitively, the $\mathit{relax}$ operator answers: \textit{How can strict inequalities be relaxed to include boundary points when a minimization problem has an infimum but no minimum?}}
\end{itemize}
Closedness of the algebra is immediate for all operators, with the exception of $\mathit{inf}$. 
In such a case, if the problem $\mathit{inf}{f(\mathbf{w})}$ $\textrm{s.t.}\,c$ is unbounded or infeasible, then a linear constraint equivalent to $f(\mathbf{w}) = (\mathit{inf}{f(\mathbf{w})}$ $\textrm{s.t.}\,c)$ is any unsatisfiable one, e.g.,~$0=1$. If the problem has a solution $v$, then $f(\mathbf{w}) = v$ is a linear inequality since $f()$ is linear. 

\subsection{Explanations as Queries}\label{sec:expquery}

\begin{table}[t]
    \centering
    \resizebox{\textwidth}{!}{
    \begin{tabular}{|l|l|}
        \toprule
        \textbf{Factual} & \\
        Base expression & $e_1 = \mathit{sat}(\mathit{cross}(\mathit{inst}(I_1, DT_1, 0, 0.95), \{U_1\}))$ \\
        Factual constraints & $c \in Th(\mathit{inst}(I_1, DT_1, 0, 0.95))$ \\
        Factual rule & $c \rightarrow l=0\ [p]$ \\
        \midrule
        \textbf{Contrastive} & \\
        Base expression & $e_2 = \mathit{sat}(\mathit{cross}(\mathit{inst}(I_1, DT_1, 0, 0.95), \mathit{inst}(I_2, DT_1, 1, 0.95), \{U_2\}))$ \\
        Contrastive constraint & $c \in Th(e_3)$, where $e_3 = \pi_{I_2.\mathbf{x}}(e_2)$ \\
        Contrastive example & any solution of $c$ \\
        Contrastive rule & $c \rightarrow l=1\ [f]$ \\
        \midrule
        \textbf{Minimal contrastive} & \\
        Base expression & $e_4 = \mathit{{inf}}(\mathit{relax}(e_2), f(I_1.\mathbf{x}, I_2.\mathbf{x}))$ \\
        Minimal contrastive constraint & $c \in Th(e_5)$, where $e_5 = \pi_{I_2.\mathbf{x}}(e_4)$ \\
        Minimal contrastive example & any solution of $c$ \\
        Minimal contrastive rule & $c \rightarrow l=1\ [f]$ \\
        \bottomrule
    \end{tabular}
    }
    \caption{Example expressions for factual, contrastive and minimal contrastive explanations.}
    \label{tab:my_label}
\end{table}


{By composing the algebra operators, the same language can express factual, contrastive, and minimal contrastive explanations; queries over under-specified instances; and relationships among explanations produced by multiple models. Thus, the algebra should be understood not merely as notation for the implementation, but as a declarative specification language for explanation problems.}
We substantiate such a claim in this sub-section. See Table~\ref{tab:my_label} for a summary of example expressions.
For ease of presentation, we assume here that the features are continuous. In the next section, we will discuss how \reasonx{} deals with discrete features.

\paragraph*{Factual explanations.} Consider the following expression:
\begin{equation} \label{q:e1}
e_1 = \mathit{sat}(\mathit{cross}(\mathit{inst}(I_1, DT_1, 0, 0.95), \{U_1\}))
\end{equation}
where $U_1$ is $I_1.x_1 = v_1, \ldots, I_1.x_n=v_n$, i.e.,~it fixes the values of the features of the instance $I_1$ to constants $v_1, \ldots, v_n$. 
Then $c \in Th(e_1)$ iff $c$ consists of the constraints appearing in a path of $DT_1$ predicting the class label $0$ with confidence $p \geq 0.95$, and such that the constraints in the path are consistent with the values $v_i$'s for the features $x_i$'s. In other words, $e_1$ tests for a factual explanation for the fully-specified instance $I_1.\mathbf{x} = \mathbf{v}$, where $\mathbf{v} = v_1, \ldots, v_n$.
If $Th(e_1)$ is not empty, such a factual explanation exists. In such a case, the sub-expression $\mathit{inst}(I_1, DT_1, 0, 0.95)$ states why $DT_1$ classifies $I_1$ with label $l=0$ and confidence $p \geq 0.95$. We call a constraint $c \in Th(\mathit{inst}(I_1, DT_1, 0, 0.95))$ that contributes to $Th(e_1)$ a \textit{factual constraint}, and the rule $c \rightarrow l=0\ [p]$ a \textit{factual (explanation) rule}. {$p$ is also called the \textit{precision} of the rule.\footnote{{\reasonx{} also reports the \textit{coverage} of each factual rule, which can be readily computed from summary statistics stored in the DT. Coverage is defined as the estimated probability that an instance satisfies all split conditions along the corresponding root-to-leaf path. We omit further details to keep the notation simple.}}}

\paragraph*{\textit{Example.}}

Consider a decision tree $DT_1$ with a single split condition, $x_1 + x_2 < 5$ that perfectly separates two classes. Instances that match this condition belong to class $l=0$ with confidence $1.0$, while instances that do not match this condition belong to class $l=1$ with confidence $1.0$.
We set $U_1$ to $I_1.x_1 = 2, I_1.x_2 = 2$, and we obtain from the tree $Th(inst(I_1,DT_1,0,0.95)) = \{I_1.x_1 + I_1.x_2 < 5 \}$. 
The expression $e_1$ evaluates to:
\[
    Th(e_1) = \{I_1.x_1 + I_1.x_2 < 5 \wedge I_1.x_1 = 2 \wedge I_1.x_2 = 2\} = \{I_1.x_1 = 2 \wedge I_1.x_2 = 2\}
\]
which is non-empty.
The factual rule $I_1.x_1 + I_1.x_2 < 5 \rightarrow l=0\ [1.0]$ then describes why $DT_1$ classifies the instance $I_1$ with label $0$ and confidence $1.0$. {Recall that confidence denotes the estimated probability that the factual rule prediction agrees with the ground-truth label when the DT is the model under analysis, and the estimated fidelity of the factual rule to the black-box classifier when the DT serves as a surrogate.}

\paragraph*{Contrastive explanations.} Consider the following expression: 
\[ e_2 = \mathit{sat}(\mathit{cross}(\mathit{inst}(I_1, DT_1, 0, 0.95), \mathit{inst}(I_2, DT_1, 1, 0.95), \{U_2\})) \]
where $U_2$ extends $U_1$ by further constraints relating the features of $I_1$ and $I_2$. 
Here, it is assumed that the instance $I_2$ is classified by $DT_1$ with class label $1$, hence it is a contrastive instance w.r.t.~$I_1$. 
Then $Th(e_2)$ is the set of satisfiable constraints specifying: for $I_1$ a path in $DT_1$ leading to class label $0$ (factual constraints for $I_1$); for $I_2$ a path in $DT_1$ leading to class label $1$ (factual constraints for $I_2$); the satisfiability of the constraints between $I_1$ and $I_2$ stated in $U_2$. By projecting over $I_2.\mathbf{x}$, that is:
\[ e_3 = \pi_{I_2.\mathbf{x}}(e_2) \]
we obtain constraints over the variables of the contrastive instance $I_2$. We call a constraint $c \in Th(e_3)$ a \textit{contrastive constraint}, and any solution of $c$ a \textit{contrastive example}. Intuitively, contrastive examples are the extensional counterpart of the (intensional) contrastive constraints{: a contrastive example is a data instance satisfying the user constraints with regard to some factual instance  and for which the decision tree outputs a different prediction.}
Moreover, we call the rule $c \rightarrow l=1\ [f]$, {where $f$ is the (estimated) fidelity of the rule},
%
%
a \textit{contrastive (explanation) rule}. Notice that this refers to $I_2$ contrasted to $I_1$. One can still apply the definition of factual explanation for $I_2$, resulting in the factual constraints in
$Th(\mathit{inst}(I_2, DT_1, 1, 0.95))$, {and a factual rule for $I_2$. If the DT is the model under analysis, $f=1.0$ because $e_3$ keeps the constraints in $\mathit{inst}(I_2, DT_1, 1, 0.95)$, namely those appearing in a path with prediction $1$. If the DT serves as a surrogate, such an $f$ must be estimated -- we will discuss the estimation procedure implemented by \reasonx{} in the next section.}
Also, notice that more than one instance per class label can be defined. In general, an instance is contrastive w.r.t.~any other instance with a different class label.\,This extends the 1-to-1 settings of one factual and one contrastive rule/example to a $m$-of-$n$ setting of $m$ factual and $n$ contrastive rules/examples. E.g.,~one can look for a contrastive instance common to two given instances. Such a $m$-of-$n$ setting in XAI is novel.

\paragraph*{\textit{Example.}}

We revisit the previous example and keep the same decision tree $DT_1$ and the same instance $I_1$. Additionally, we define the instance $I_2$ such that $Th(inst(I_2,DT_1,1,0.95)) = \{I_2.x_1 + I_2.x_2 \geq 5 \}$. Moreover, $U_2$ is defined as $U_1$ plus $I_2.x_1 = I_1.x_1$, namely we add immutability to feature $x_1$.
Then, the previously introduced expressions evaluate to the following:
\begin{eqnarray*}
    Th(e_2) &=& \{I_2.x_1 + I_2.x_2 \geq 5 \wedge I_1.x_1 = 2 \wedge I_1.x_2 = 2 \wedge I_2.x_1 = I_1.x_1 \} \\
    &=& \{I_2.x_2 \geq 3 \wedge I_1.x_1 = 2 \wedge I_1.x_2 = 2 \wedge I_2.x_1 = 2 \} \\
    Th(e_3) &=& \{I_2.x_1 = 2 \wedge I_2.x_2 \geq 3 \}
\end{eqnarray*}
The contrastive rule $I_2.x_1 = 2 \wedge I_2.x_2 \geq 3 \rightarrow l=1\ [1.0]$ describes then a condition under which $DT_1$ would classify the instance $I_2$  with label $1$ (and confidence $1.0$) and for which the immutability constraint holds.  Notice that $I_2.x_1 + I_2.x_2 \geq 5 \rightarrow l=1\ [1.0]$ is a factual rule for the instance $I_2$. {The fidelity of the contrastive rule is $1.0$ when the DT is the model under analysis, as previously observed; in this special case, the fidelity is $f=1.0$ also when the DT serves as a surrogate model. In fact, when the confidence of the factual rule is $1.0$, adding further constraints cannot decrease fidelity (on the sample used to estimate it), since the contrastive constraint identifies a subset of the instances satisfying the corresponding factual constraint. For fidelity values strictly below $1.0$, however, restricting the region may change fidelity, possibly substantially, so fidelity must be re-estimated over the restricted region.
}
%

\paragraph*{Minimal contrastive explanations.} 
Consider the following expression: 
\[ e_4 = \mathit{inf}(e_2, f(I_1.\mathbf{x}, I_2.\mathbf{x})) \]
where $f()$ is a linear (or linearizable) distance function between the instances $I_1$ and $I_2$. Then $Th(e_4)$ includes the constraints that, in addition to $Th(e_2)$, minimize the distance between $I_1$ and the contrastive instance $I_2$. Given:
\[ e_5 = \pi_{I_2.\mathbf{x}}(e_4) \]
we call a constraint $c \in Th(e_5)$ a \textit{minimal contrastive constraint}, and the rule $c \rightarrow l=1\ [f]$, where $f$ is the {fidelity of the rule}, a \textit{minimal contrastive (explanation) rule}.

\paragraph*{\textit{Example}}
We refer to the previous example, and keep the same decision tree $DT_1$, instances $I_1$ and $I_2$ and  constraints $U_2$. Further, we define $f(I_1.\mathbf{x}, I_2.\mathbf{x}) = \sum_i | I_2.x_i - I_1.x_i |$, i.e.,~we use the $L_1$~norm as the distance function. The minimal value of $f(I_1.\mathbf{x}, I_2.\mathbf{x})$ occurs when $f(I_1.\mathbf{x}, I_2.\mathbf{x}) = |I_2.x_2 - 2| + |2 - 2| = I_2.x_2 - 2 = 1$.
The expressions $e_4$ and $e_5$ evaluate to:
\begin{eqnarray*}
    Th(e_4) & = & \{I_1.x_1 = 2 \wedge I_1.x_2 = 2  \wedge I_2.x_1 = 2 \wedge I_2.x_2 = 3 \}\\
    Th(e_5) & = & \{ I_2.x_1 = 2 \wedge I_2.x_2 = 3\}
\end{eqnarray*}
Hence, $I_2.x_1 = 2 \wedge I_2.x_2 = 3 \rightarrow l=1\ [1.0]$ is a minimal contrastive explanation.

\paragraph{Explanations for under-specified instances.} Reconsider the previous examples under the revised assumption that $U_1$ consists of $I_1.x_1 = 2$ only, i.e.,~we want to reason not on a fully specified instance, but on a set of instances. 
It turns out that:
\begin{eqnarray*}
    Th(e_1) &=& \{ I_1.x_1 = 2 \wedge I_1.x_2 < 3\}\\
    Th(e_2) &=& \{I_2.x_1 + I_2.x_2 \geq 5 \wedge I_1.x_1 = 2 \wedge I_1.x_2 < 3 \wedge I_2.x_1 = I_1.x_1 \} \\
    &=& \{I_2.x_2 \geq 3 \wedge I_1.x_1 = 2 \wedge I_1.x_2 < 3 \wedge I_2.x_1 = 2 \} \\
    Th(e_3) &=& \{I_2.x_1 = 2 \wedge I_2.x_2 \geq 3 \}
\end{eqnarray*}
with $Th(e_3)$ unchanged. Hence, {we have the same  contrastive rule $I_2.x_1 = 2 \wedge I_2.x_2 \geq 3 \rightarrow l=1\ [1.0]$. Intuitively, the contrastive constraint $I_2.x_1 = 2 \wedge I_2.x_2 \geq 3$ characterizes contrastive examples for the under-specified condition $I_1.x_1 = 2$ -- or, in other words, for the set of solutions of the constraint $I_1.x_1 = 2$.}
Regarding the minimal contrastive rule, now $|I_2.x_2 - I_1.x_2| = I_2.x_2 - I_1.x_2 > 0$ since $I_1.x_2 < 3$ (see $Th(e_1)$) and $I_2.x_2 \geq 3$ (see $Th(e_3)$). The $\mathit{inf}$ operator then imposes $I_2.x_2 = I_1.x_2$ {because $\mathit{inf} |I_2.x_2 - I_1.x_2| = 0$}. 
This leads to an unsatisfiable theory, as it includes both $I_2.x_2 = I_1.x_2$, $I_1.x_2 < 3$, and  $I_2.x_2 \geq 3$. The issue is that the minimum of $I_2.x_2 - I_1.x_2$, under the constraint that it is greater than $0$, does not exist. In practice, $\mathit{inf}$ implements an infimum, but its solution is no longer consistent with the other constraints. We could stop here, accepting no answer to the query. Considering the uncertainty behind decision node splits, we propose instead to relax strict inequalities to inequalities:
\[ e'_4 = \mathit{inf}( \mathit{relax}(e_2), f(I_1.\mathbf{x}, I_2.\mathbf{x})) \quad\quad\quad
e'_5 = \pi_{I_2.\mathbf{x}}(e'_4)\]
We then have:
\begin{eqnarray*}
    Th(\mathit{relax}(e_2)) & = & \{I_2.x_2 \geq 3 \wedge I_1.x_1 = 2 \wedge I_1.x_2 \leq 3-\epsilon \wedge I_2.x_1 = 2 \} \\
    Th(e'_4) & = & \{I_1.x_1 = 2 \wedge I_1.x_2 = 3  \wedge I_2.x_1 = 2 \wedge I_2.x_2 = 3-\epsilon \}\\
    Th(e'_5) & = & \{ I_2.x_1 = 2 \wedge I_2.x_2 = 3-\epsilon\}
\end{eqnarray*}
The theory of $e'_4$ contains and therefore constraints both instances $I_1$ and $I_2$ in such a way as to minimize their distance. Since $I_1$ is now under-specified, i.e.,~a set of instances, this means the minimization will look for the subset of such instances for which the distance is minimal. In this example, such a subset turns out to be a singleton. {Overall, $e'_4$ characterize the pairs of data points, one in the under-specified instace and one in contrastive constraints, such that their distance is minimized.}

\paragraph{On the $\mathit{relax}$ operator.}

The need to relax strict inequalities occurs in general only in the case of expressions using the $\mathit{inf}$ operator, since it is actually implemented as an infimum. 
{This is a standard treatment of strict inequalities in polyhedral and numerical optimization~\citep{DBLP:conf/issac/NalbachAK21}, where closed feasible regions defined by non-strict inequalities are typically required. The $\mathit{relax}$ operator transforms an open feasible region into a closed inner approximation and ensures that candidate solutions remain at least $\epsilon$ away from the strict boundary. For a sufficiently small $\epsilon$, e.g., on the order of machine precision, the transformation has a negligible effect on the region covered by the analysis while avoiding the numerical and optimization issues.}
%

\paragraph{Explanations over different models: contrastive explanations.}

We consider contrastive explanations of a same instance over two different decision trees. These trees can represent either: two models built on different training sets (e.g.,~trained at different points of time); or, two different models trained on the same training set at the same point of time. 
The expressions:
\begin{eqnarray*}
    & e_6 = \mathit{sat}(\mathit{cross}(\mathit{inst}(I_1, DT_1, 0, 0.95), \mathit{inst}(I_1, DT_2, 0, 0.95),\\
    & \hspace{2cm} \mathit{inst}(I_2, DT_1, 1, 0.95), \mathit{inst}(I_2, DT_2, 1, 0.95), \{U_3\}))\\
    & e_7 = \pi_{I_2.\mathbf{x}}(e_6) 
\end{eqnarray*}
look for conditions for which $I_1$ is an instance with the same class label in $DT_1$ and $DT_2$, and $I_2$ is a contrastive instance for both $DT_1$ and $DT_2$. Such conditions regard contrastive explanations that remain unchanged between $DT_1$ and $DT_2$.

\paragraph*{Example}
Let us consider $DT_1$ fixed as in the previous examples, and $DT_2$ be defined with a single split condition $x_1 + x_2 < 6$. {All instances matching this condition belong to class $l=0$.} Let $U_3$ include $I_1.x_1 = 2, I_1.x_2 = 2$ plus $I_2.x_1 = I_1.x_1$. 
The expressions $e_6$ and $e_7$ evaluate to:
\begin{eqnarray*}
Th(e_6) & = & \{ I_2.x_1 + I_2.x_2 \geq 6 \wedge I_1.x_1 = 2 \wedge I_1.x_2 = 2 \wedge I_2.x_1 = I_1.x_1 \}\\
Th(e_7) & = & \{I_2.x_1 = 2 \wedge I_2.x_2 \geq 4\}
\end{eqnarray*}
Comparing $Th(e_7)$ to $Th(e_3)$, we notice that the minimal contrastive examples for $I_1$  that are also minimal contrastive examples for $DT_2$ are those for which $I_2.x_1 = 2 \wedge I_2.x_2 \geq 4$. The ones that are not anymore contrastive in $DT_2$ are those for which $I_2.x_1 = 2 \wedge 4 > I_2.x_2 \wedge I_2.x_2 \geq 3$.

\section{{{\reasonx{}}}: Reason to explain}
\label{sec:reasonx_details}

{\reasonx{}} is a tool for reasoning over explanations and is based on the query language over constraints introduced in the previous section. 
Here, we present the architecture of {\reasonx{}}, the translation of user inputs into queries and of answer constraints into explanations, and the execution of the above queries through a CLP($\mathbb{R})$ meta-interpreter.



\begin{figure}[t]
    \begin{center}
    \includegraphics[width = 1.0\linewidth]{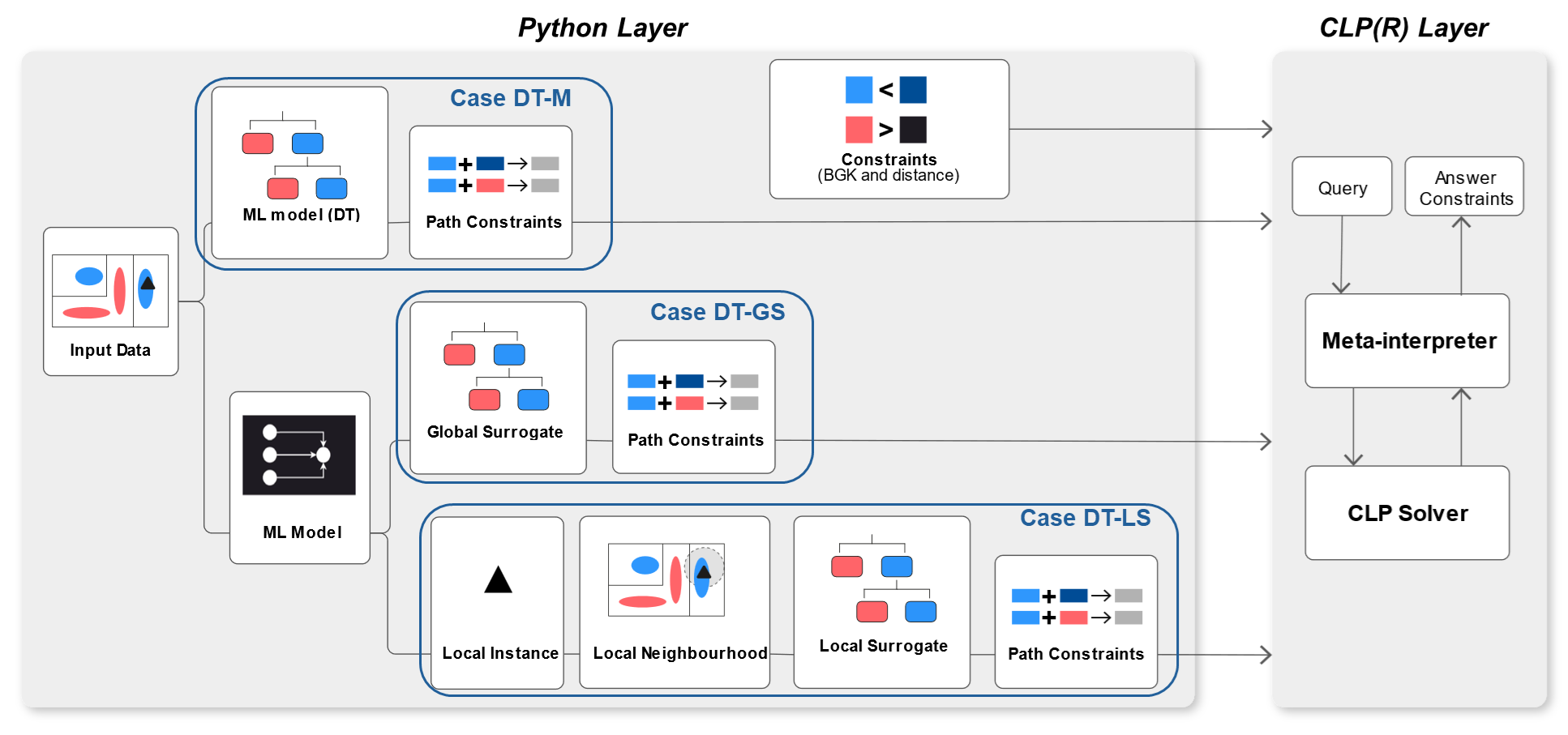} %
    \end{center}
    \caption{Workflow of data, models and explanations for {\reasonx{}}. To be read from left to right. 
    The three paths show the cases \textbf{(DT-M)}, \textbf{(DT-GS)} and \textbf{(DT-LS)} for the base model: base model is a DT model, or a global, or a local surrogate. 
    The explanations are created from the answer constraints produced by a meta-interpreter of the query language over queries generated from user inputs (background and distance), and embeddings of the base model (path constraints).
    }
    \label{fig:workflow}
\end{figure}

\subsection{{{\reasonx{}}} Architecture}
\label{sec:reasonx_architecture}

Figure~\ref{fig:workflow} shows the workflow from data, models, and user constraints to queries, and back from answer constraints to explanations. 
There are two main layers in {\reasonx{}}.
The core layer is written in CLP($\mathbb{R}$) as provided by the SWI Prolog system~\citep{DBLP:journals/tplp/WielemakerSTL12}. 
It implements a meta-interpreter for computing the theory $Th(e)$ of an expression $e$ over the language of queries.
On top of it, there is a Python layer, which is constructed such that a user -- here, a developer -- can easily name instances, link them to decision trees, 
and state user constraints.
The Python layer takes care of the meta-data (name and type of features) and of the ML model. It is designed for an easy integration with the \texttt{pandas} and \texttt{scikit-learn} standard libraries for data storage and ML model construction. {The interface between Python and SWI-Prolog is managed by the Janus library, a bi-directional bridge that provides high-throughput inter-language calls~\citep{DBLP:journals/corr/abs-2308-15893}.}

{\reasonx{}} is strongly guided by a decision tree, called the \textit{base model}. Such a decision tree can be: \textbf{(DT-M)} the model to be explained and reasoned about; \textbf{(DT-GS)} a global surrogate model of a {tabular-data black-box classifier}, such as a neural network or an ensemble; or \textbf{(DT-LS)} a local surrogate model trained to mimic a {tabular-data black-box classifier} in the neighborhood instances of the instance to explain. 
Decision trees are directly interpretable only if their size/depth is limited. Large decision trees are hard to reason about, especially in a contrastive explanation scenario. Case \textbf{(DT-M)} can be used in such contexts.
In cases \textbf{(DT-GS)} and \textbf{(DT-LS)}, the surrogate model is assumed to have high fidelity in reproducing the decisions of the original ML model. This is reasonable for local models, i.e.,~in case \textbf{(DT-LS)}, since learning an interpretable model over a neighborhood has been a common strategy in perturbation-based explainability methods such as LIME~\citep{DBLP:conf/kdd/Ribeiro0G16} and LORE~\citep{DBLP:journals/expert/GuidottiMGPRT19}. {However, local explanations also depend on the neighborhood-generation procedure and should therefore be interpreted under the premise that they are valid ``in the neighborhood of the instance''. In our experiments with \reasonx{}, for example, the neighborhood is generated by keeping nominal and ordinal features fixed while allowing continuous features to vary. Consequently, the resulting local explanations implicitly rely on the assumption that all nominal and ordinal features are immutable. Strictly speaking, however, local-neighborhood generation is not part of \reasonx{}, but rather an orthogonal preprocessing step.} 

In summary, explanations produced by {\reasonx{}} are  global in cases~\textbf{(DT-M)} and~\textbf{(DT-GS)}, local in case~\textbf{(DT-LS)}, model-specific in case~\textbf{(DT-M)}, and model-agnostic in cases~\textbf{(DT-GS)} and~\textbf{(DT-LS)}.

\subsection{Python Layer}
\label{sec:python_layer}

Consider the Python layer. Data is stored in tabular form as \texttt{pandas} data frames. One or more decision trees are obtained by one of the cases \textbf{(DT-M)}, \textbf{(DT-GS)}, and \textbf{(DT-LS)} discussed above.
The user can define one or more instances, with the expected class label, and associate each instance to a specific decision tree.
Moreover, the user can directly state linear constraints over the features of one or more instances. In addition, the user can fully or partially specify the value of some of the features of an instance.
The Python layer transforms decision trees, user constraints, and additional (called, \textit{implicit}) constraints into CLP($\mathbb{R}$) facts through an embedding procedure described later. The implicit constraints are used to encode the nominal features, which are not directly accounted for in the query language, in a way that is transparent to the user.
Then, the Python layer submits one or more expressions over the query language to the CLP($\mathbb{R}$) layer, parses back the answer constraints, and returns the results to the user.

\paragraph*{Query generation.}\label{sec:query}

Let $I_1, \ldots, I_m$ be the instances declared, and $DT_1, \ldots, DT_m$ the linked decision trees. Moreover, let $\Phi$ be the user constraints and $\Psi$ the implicit constraints. 
The following query expressions are general versions of the ones presented in Section~\nameref{sec:expquery}:

\begin{itemize}
    \item $S = \mathit{sat}(\mathit{cross}(inst(I_1, DT_1, l_1, pr_1), \ldots, inst(I_m, DT_m, l_m, pr_m), \{\Phi, \Psi\}))$ tests satisfiability of the constraints generated from decision trees, stated by the users, and imposed by the data types;

    \item {a sub-expression} $F_i = inst(I_i, DT_i, l_i, pr_i)$ {of $S$} is used to generate factual rules of the form $c \rightarrow l=l_i\ [p]$, {where $c \in Th(F_i)$. Such a factual rule is associated with any answer constraint of $S$ that includes $c$, i.e., an answer constraint of the form $c \wedge c' \in Th(S)$}; 

    \item $CE_i = \pi_{I_i.\mathbf{x}}(S)$ generates contrastive constraints $c \in Th(CE_i)$, and contrastive rules of the form $c \rightarrow l=l_i\ [p]$. Here, contrastive refers to the other instances $I_j$ with the class label $l_j \neq l_i$. Also, $p$ is the confidence of the prediction of $DT_i$ for the path that contributes to $c$;

    \item $M = \mathit{inf}(\mathit{relax}(S), f(I_h.\mathbf{x}, I_k.\mathbf{x}))$ and $\mathit{MCE}_i = \pi_{I_i.\mathbf{x}}(M)$ generalize $S$ and $CE_i$ to the case of minimal solutions. In particular, minimal contrastive rules are of the form $c \rightarrow l=l_i\ [f]$, where $c \in Th(\mathit{MCE}_i)$ is a minimal contrastive constraint;

    \item $P = \pi_{\mathbf{w}}(M)$ or $P = \pi_{\mathbf{w}}(S)$ are the most general expressions generated\footnote{Notice that $\pi_{\mathbf{w}}(S)$ is not a special case of $\pi_{\mathbf{w}}(M)$, e.g., by setting $f()$ to a constant. First, $\mathit{relax}$ is not applied in $\pi_{\mathbf{w}}(S)$, while it is in $\pi_{\mathbf{w}}(M)$. Second, the actual implementation of $\mathit{inf}$ grounds integer variables, as it will be discussed later on, 
    which is not necessarily the case for $\pi_{\mathbf{w}}(S)$.}, including the projection of the answer constraints over a set of features $\mathbf{w} \subseteq \mathcal{F}$ required by the user.    
\end{itemize}

The Python layer submits one of the two queries $P = \pi_{\mathbf{w}}(M)$ or $P = \pi_{\mathbf{w}}(S)$ to the CLP($\mathbb{R}$) layer, depending on whether the user requires minimization or not. If the user does not require projection, then $\mathbf{w} = \mathcal{F}$ and $P$ boils down to $M$ or to $S$ respectively.
The result of the submitted query, and of the factual rules from $F_i$ are returned by the CLP($\mathbb{R}$) layer. For efficiency reasons, the actual implementation does not submit separate queries for the $F_i$'s and $CE_i$'s, but keeps track of factual and contrastive rules while executing the submitted query.  
Finally, the user can obtain a contrastive rule $CE_i$ by querying $P = \pi_{\mathbf{w}}(S)$ with $\mathbf{w} = I_i.\mathbf{x}$. 

\paragraph*{Estimating fidelity of contrastive explanations.}\label{sec:fidelity}

{
Let us now consider how to estimate the fidelity $f$ of a contrastive rule
$c \rightarrow l=1 [f]$. As discussed in Section~\nameref{sec:expquery}, if the DT is
the model under analysis, i.e., case \textbf{(DT-M)}, then $f=1.0$. We therefore focus
on the cases in which the DT serves as a surrogate model, namely \textbf{(DT-GS)}
and \textbf{(DT-LS)}. At the Python layer, the user can associate each base model
with the black-box model it approximates, provided as a Python object.
For a constraint $c \in CE_i$ or $c \in \mathit{MCE}_i$, we estimate $f$ by 
sampling points from the region defined by $c$, using Markov Chain Monte Carlo
(MCMC) methods for uniform sampling from polytopes, for which several algorithms have been
proposed in the literature~\citep{DBLP:journals/jossw/SunC25}. \reasonx{} uses the Hit-and-Run algorithm as provided by the \texttt{hopsy} package~\citep{DBLP:journals/bioinformatics/PaulJSWN24}. We sample
$N=10{,}000$ data points, or another user-specified number, from the solutions of
$c$, unless $c$ fixes all features to constant values, in which case $N=1$.
For each sampled data point, we compute the prediction of the black-box model. The
fidelity $f$ is then estimated as the fraction of the $N$ sampled points for which
the black-box predicts $l=1$. For $N=1$, this boils down to checking the only contrastive example is actually predicted as $l=1$.
In the general case, the sampling algorithms from the literature assume continuous and bounded domain for the features. In the
presence of nominal or ordinal features, the corresponding sampling problem becomes
computationally intractable (\#P-hard)~\citep{DBLP:journals/tcs/JerrumVV86}. We
therefore impose the following \textit{boundedness condition}: $c$ implies that
nominal and ordinal features are constant, while continuous features are bounded.
Notice that boundedness of continuous features can always be enforced by the user
through the \textit{consistency constraints} (see later on).
Moreover, for $c \in \mathit{MCE}_i$, nominal and ordinal features are always
constant. This follows from a separate computational hardness issue arising in the
computation of the infimum of MILP problems, which will be discussed later when
presenting the implementation of the $\mathit{inf}$ operator. If the boundedness
condition is not satisfied, $f$ is set to NA (Not Available).
Finally, observe that uniform sampling does not account for dependencies or correlations among features. Consequently, the resulting estimate of $f$ may differ from the fidelity measured with respect to the actual data distribution. An alternative would be to estimate a multivariate data distribution from an auxiliary dataset and sample from the resulting density restricted to the region defined by the constraint $c$. Our approach avoids requiring such an additional dataset. Moreover, even when such a dataset is available, it may contain too few observations in the region identified by $c$ to support a reliable local density estimate.
}

\subsection{From Python to CLP: Embeddings} 
\label{sec:reasonx_embeddings}

Let us discuss here how base models, instances, implicit and user constraints are mapped from Python to CLP($\mathbb{R}$) terms and facts.

First, we point out that {\reasonx{}} is agnostic about the learning algorithm used to produce the base models. 
We make the following assumptions. 
Predictive features can be nominal, ordinal, or continuous. 
Ordinal features are coded as consecutive integer values. 
Nominal features can be one-hot encoded or not. If not, we force one-hot encoding before submitting the queries to CLP($\mathbb{R}$), and silently decode back from the answer constraints to nominal features before returning the results to the user. In particular, a nominal feature
$x_i$ is one-hot encoded into features $x_i^{v_1}, \ldots, x_i^{v_k}$ with $v_1, \ldots, v_k$ being the distinct values in the domain of $x_i$. It is important to emphasize that the choice of encoding methods for ordinal and nominal features is non-trivial, as it can significantly impact the performance and fairness of machine learning models~\citep{DBLP:conf/aies/MouganCRS23}, as well as the quality of the explanations generated.

We assume that binary split conditions from a parent node to a child node in a decision tree are of the form $\mathbf{a}^T \mathbf{x} \simeq b$, where $\mathbf{x}$ is the vector of all features $x_i$'s. The following common split conditions are covered by such an assumption:
\begin{itemize}
    \item axis-parallel splits based on a single continuous or ordinal feature, i.e.,~$x_i \leq b$ or $x_i > b$;
    \item linear splits over features,i.e.,~${\bf a}^T{\bf x} \leq b$ or ${\bf a}^T{\bf x} > b$;
    \item (in)equality splits for nominal features, i.e.,~$x_i = v$ or $x_i \neq v$; in terms of one-hot encoding, they respectively translate into  $x_i^v = 1$ or $x_i^v = 0$.
\end{itemize}
Axis parallel and equality splits are used in CART~\citep{DBLP:books/wa/BreimanFOS84} and C4.5~\citep{DBLP:books/mk/Quinlan93}. Linear splits are used in oblique~\citep{DBLP:journals/jair/MurthyKS94,DBLP:conf/iclr/LeeJ20} and optimal decision trees~\citep{DBLP:journals/ml/BertsimasD17}. Linear model trees combine axis parallel splits at nodes and linear splits at leaves~\citep{DBLP:journals/ml/FrankWIHW98}. {The current implementation of \reasonx{} accepts CART (as implemented by \texttt{scikit-learn}) and linear decision trees (as implemented by \texttt{linear-tree}\footnote{\url{https://github.com/cerlymarco/linear-tree}}). In experiments, we fix CART as base models.}

\paragraph*{Base models.}
\label{sec:translation_model_to_facts}
The base models are encoded into a set of Prolog facts, one for each path in the decision tree from the root node to a leaf node: 
\begin{center}
\texttt{path($d$,[$\mathbf{x}$], [$\mathbf{a}_1^T \mathbf{x} \simeq b_1$, $\ldots$, $\mathbf{a}_k^T \mathbf{x} \simeq b_k$], $l$, $p$).}
\end{center}
where $d$ is an id of the decision tree,  \texttt{[$\mathbf{x}$]} is a list of CLP($\mathbb{R}$) variables, one for each feature, $l$ the class predicted at the leaf, $p$ the confidence of the prediction,
and \texttt{[$\mathbf{a}_1^T \mathbf{x} \simeq b_1$, $\ldots$, $\mathbf{a}_k^T \mathbf{x} \simeq b_k$]} is the list of the 
split conditions from the root to the leaf. {Notice that this encoding also applies to rule-set classifiers~\citep{DBLP:series/cogtech/FurnkranzGL12}, provided that the rules are non-overlapping and exhaustive, i.e., every instance is classified by exactly one rule. The non-overlapping requirement excludes rule-voting mechanisms, which are not accounted for by the algebra.}

\paragraph*{Instances.}
Each declared instance is encoded by a list of CLP($\mathbb{R}$) variables, one for each feature. The mapping between the feature name and the variable is positional. 
All the instances are collectively represented by a list of lists of variables. 


\paragraph*{Implicit constraints ($\Psi$).}
A number of constraints on the features $\mathbf{x}$ of each instance naturally derive from the data type of the features. We call them implicit constraints, because the system can generate them from meta-data about features:
\begin{itemize}
    \item for continuous features: $x_i \in {\mathbb{R}}$;
    \item for ordinal features: $x_i \in \mathbb{Z}$ and $m_i \leq x_i \leq M_i$ where $dom(x_i) = \
    \{m_i, \ldots, M_i\}$ is the domain of $x_i$;
    \item for one-hot encoded nominal features: $x^{v_1}_i, \ldots, x^{v_k}_i \in \mathbb{Z}$ and $\wedge_{j=1}^k (0 \leq x^{v_j}_i \leq 1)$ and $\sum_{j=1}^k x^{v_j}_i = 1$;
\end{itemize}

We denote by $\Psi$ the conjunction of all implicit constraints.
%

\paragraph*{User constraints ($\Phi$).}
A number of user constraints can be readily expressed in {\reasonx{}}. We provide examples and use $F$ to refer to a factual instance and $CE$ to refer to a corresponding contrastive instance. {The most basic user constraints are those binding features to values, ranges or bounds, such as $F.\mathit{age} = 20, F.\mathit{income} \leq 5000.0$. A special case is when $F$ is fully-specified to the values of a data point. Moreover, forms of} \textit{background knowledge}, loosely categorized as in~\citet{DBLP:journals/csur/KarimiBSV23}, 
and extended based on work by~\citet{DBLP:conf/fat/KarimiSV21,DBLP:conf/fat/UstunSL19,DBLP:conf/fat/MothilalST20,mahajan2019preserving}
are covered. 
\textit{Feasibility} constraints concern the possibility of feature changes between the factual and contrastive instance, and how changes depend on previous values or 
other features.
A feature that is unconstrained in feasibility is \textit{actionable without any condition}. Among the feasibility constraints, we can~distinguish:
\begin{itemize}
\item \textit{Immutable features:}
    a feature cannot or must not change between the factual and contrastive instances, e.g., the birthplace: $CE.{\mathit{birthplace}} = F.{\mathit{birthplace}}$.
\item \textit{Mutable but not actionable features:}
    the change of a feature from the factual to the contrastive instance is only a result of changes in the features it depends upon. 
    An example is the change of unit scale, e.g.,~from Euro to US Dollars, which can be encoded as~$CE.{\text{\officialeuro}} = 1.16 \cdot CE.{\$}$ -- assuming a conversion rate of $1.16$. 
\item \textit{Actionable but constrained features:}
    a feature can be changed only under some conditions, e.g., age can only increase~:~$CE.{\mathit{age}} \geq F.{\mathit{age}}$.
\end{itemize}

Another class regards \textit{consistency constraints}, which aim at bounding the domain values of a feature, e.g.,~$0 \leq CE.{\mathit{age}} \leq 120$ limits the age feature in the range $[0, 120]$.

\paragraph*{Encoding distance functions.}\label{sec:encdist}

The assumption that instances have a specific predicted class label (see operator $\mathit{inst}$), allows us to remove the first term from Equation~\ref{equ:wachter}, hence reducing the problem of computing minimal contrastive explanations to minimize the distance function between the factual and contrastive instances.
{\reasonx{}} offers two distance functions that can be used with the $\mathit{inf}$ operator: $L_1$ combined with a matching norm for the nominal variables (referred to as as $L_1$ norm for the remainder of this paper) and $L_{\infty}$, both over normalized features.
The $L_1$ norm penalizes the average change over all features, while the $L_{\infty}$ norm penalizes the maximum changes over all features. 
See also~\citet{DBLP:journals/corr/abs-1711-00399,DBLP:conf/aistats/KarimiBBV20} for a discussion of how the different norms affect the generated contrastive explanations. 
In the \supplement, 
we show how to linearize such norms, i.e.,~how to express them using linear constraints only, possibly introducing  slack variables and adding further implicit constraints to $\Psi$.

Regarding the diversity of the contrastive explanations, we combine the approach of~\citet{DBLP:journals/ml/LampridisSGR23} and the diversity evaluation of~\citet{DBLP:conf/fat/MothilalST20}. Assume that we have a (factual) instance $I_f$ and a large pool of contrastive instances. From this pool, our objective is to select a diverse subset of $|S|$ instances.
{\reasonx{}} encodes the following distance:

\begin{eqnarray} \label{equ:diversity}
    f(x_f, S) = \frac{\lambda}{|S|} \sum_{i \in S} dist(I_f.\mathbf{x}, I_i.\mathbf{x}) - \frac{1}{|S|^2} \sum_{i \in S} \sum_{j \in S} dist(I_i.\mathbf{x}, I_j.\mathbf{x})
\end{eqnarray} 
where $S$ is a set of indices of contrastive instances w.r.t.~$I_f$, and $\lambda$ is a tuning parameter.
%
%
The first term measures the distance between the factual and the contrastive instances (\textit{proximity}), the second term the distance within the set of contrastive instances (\textit{diversity}). Thus, minimization of the function optimizes the trade-off between proximity and diversity.
%

\subsection{CLP Layer and the Meta-interpreter}
\label{sec:reasonx_meta_interpreter}

The core engine of {\reasonx{}} is a CLP($\mathbb{R})$ meta-interpreter of expressions over the query language presented in Section~\nameref{sec:reasonx_algebra_operators}. {This provides a declarative implementation because the clauses of the interpreter closely follow the semantic definitions of the operators, while constraint solving and simplification (such as removing redundant inequalities) is delegated to the underlying CLP system. The correspondence between the algebra and the interpreter can therefore be reasoned about compositionally over the structure of expressions. A procedural implementation would instead require an explicit correctness argument showing that its control flow, intermediate representations, and solver operations preserve the semantics of every operator. A brief introduction to the syntax and semantics of CLP($\mathbb{R})$ is reported in the \supplement.}

Meta-reasoning is a powerful technique that allows a logic program to manipulate programs encoded as terms. 
In CLP, meta-reasoning is extended by encoding also constraints as terms~\citep{DBLP:journals/tplp/WielemakerSTL12}.
For example, consider Listing~\ref{lst:satisfiable_}. 

\begin{lstlisting}[caption = {Definition of predicates \texttt{satisfiable} and \texttt{tell\_cs}.}, label = {lst:satisfiable_}, basicstyle=\ttfamily\scriptsize]
satisfiable(P) :-
    copy_term(P, CopyP),
    tell_cs(CopyP).	   
tell_cs([]).
tell_cs([C|Cs]) :-  
    {C}, 
    tell_cs(Cs).
\end{lstlisting}

The query \texttt{tell\_cs([X >= 0, Y = 1 - X])} asserts the linear constraints in the list, adding them to the constraint store. Linear constraints can be the result of some manipulation beforehand. 
For example, the query \texttt{satisfiable([X = 0, Y >= 1 - X])} first makes a copy of the list by renaming all variables into fresh ones, 
and then it asserts the copy to the constraint store. 
In this way, the assertion succeeds if and only if the constraints are satisfiable, but without altering the original constraints, e.g.,~without fixing \texttt{X} to $0$.

Here, we discuss a simplified version of the {\reasonx{}} interpreter using the predicate \texttt{solve($E$, $Vars$, $C$)}, where $E$ is the expression to interpret, $Vars$ is the list of lists of variables (one list per instance), and $C$ is the returned answer constraint consisting of $Th(E)$. We proceed with the interpretation of each operator.


%


\paragraph*{($\{U\}$)}
The interpretation of the $\{U\}$ operator is split into two base operators: \texttt{userc} for the  user constraints and \texttt{typec} for the implicit constraints that cover the data types.
Listing~\ref{lst:userc} shows that the former simply retrieves the list of constraints $\Psi$ embedded in the \texttt{user\_constraints} predicate by the Python layer. The latter appends the lists of constraints for nominal variables and ordinal variables. Details on the called predicates are omitted.

\begin{lstlisting}[caption = {Interpretation of the $\{U\}$ operator.}, label = {lst:userc}, basicstyle=\ttfamily\scriptsize]
solve(userc, Vars, Cons) :-
    user_constraints(Vars, Cons).

solve(typec, Vars, Cons) :-
    cat_constraints(Vars, CCat),
    ord_constraints(Vars, COrd),
    append(CCat, COrd, Cons).
\end{lstlisting}

\paragraph*{($\mathit{inst}$)}
The operator $\mathit{inst}(I, DT, l, pr)$ is interpreted in Listing~\ref{lst:inst}.
The position $N$ of $I$ in the list of all instances is decoded through the \texttt{data\_instance} predicate asserted by the Python layer. The interpreter  accesses the variables $V$ of $I$ (predicate \texttt{nth0}), and matches them with the embedding of a path in the $DT$ decision tree that ends in a leaf with class label $l$ and with predicted accuracy $P$. Finally, only paths for which $P$ is at least the required probability $pr$ are considered.

\begin{lstlisting}[caption = {Interpretation of the $\mathit{inst}$ operator.}, label = {lst:inst}, basicstyle=\ttfamily\scriptsize]
solve(inst(I, DT, L, Pr), Vars, Cons) :-
    data_instance(N, I),
    nth0(N, Vars, V),
    path(DT, V, Cons, L, P),
    P >= Pr.
\end{lstlisting}

\paragraph*{($\mathit{cross}$)}
The interpretation of the $\mathit{cross}(L)$ operator is stated by recursively solving the tail list of $L$ and the head of $L$, and then calculating the cross-product of the constraints in the returned results.

\begin{lstlisting} [caption = {Interpretation of the $\mathit{cross}$ operator.}, label = {lst:cross}, basicstyle=\ttfamily\scriptsize]
solve(cross([]), _, []).
solve(cross([T|Ts]), Vars, Cons) :-
    solve(cross(Ts), Vars, TsCons),
    solve(T, Vars, TCons),
    append(TCons, TsCons, Cons).
\end{lstlisting}

\paragraph*{($\mathit{sat}$)} The $\mathit{sat}(T)$ operator is interpreted as shown in Listing~\ref{lst:satisfiable_mi}. After solving the sub-expression $T$, the resulting constraints are checked for satisfiability through the \texttt{satisfiable} predicate. Such a predicate takes also as input the list of variables in the domain of the integers. By using the meta-programming capabilities of CLP($\mathbb{R}$), after making a fresh copy of its inputs through the built-in predicate \texttt{copy\_term}, the satisiability is checked by: (1) asserting the copy of the constraints using \texttt{tell\_cs} (see Listing~\ref{lst:satisfiable_}); and (2) checking that among the solutions there is one assigning integer values to integer variables using the \texttt{bb\_inf} predicate. In this last call, the argument \texttt{0} regards the (constant) function to minimize, and the unnamed variable \enquote{\texttt{\_}} concerns the returned minimum value (not used).

\begin{lstlisting}[caption = {Interpretation of the $\mathit{sat}$ operator.}, label = {lst:satisfiable_mi}, basicstyle=\ttfamily\scriptsize]
solve(sat(T), Vars, Cons :-
    solve(T, Vars, Cons),
    int_vars(Vars, IntVars),
    satisfiable(Cons, IntVars).

satisfiable(Cons, Ints) :-
	copy_term(Cons-Ints, CopyCons-CopyInts),
	tell_cs(CopyCons),
	bb_inf(CopyInts, 0, _).
\end{lstlisting}

\paragraph*{($\pi$)} The interpretation of the $\pi_{\mathbf{w}}(T)$ operator is provided in Listing~\ref{lst:project}. After solving $T$, the projection of the solution over the features $\mathbf{w}$ is implemented by the meta-predicate \texttt{project} designed by~\citet{DBLP:journals/tplp/BenoyKM05}.

\begin{lstlisting} [caption = {Interpretation of the $\pi$ operator.}, label = {lst:project}, basicstyle=\ttfamily\scriptsize]
solve(project(T, PVars), Vars, PCons) :-
    solve(T, Vars, Cons),
    project(PVars, Cons, PCons).    
\end{lstlisting}

\paragraph*{($\mathit{inf}$)} Finally, let us consider $\textit{inf(T, F)}$ in Listing~\ref{lst:minimize}. After solving $T$ into constraints $\mathit{TCons}$, the interpreter linearizes the expression $F$ regarding the distance function into a linear expression $\mathit{LinF}$, possibly generating additional constraints $\mathit{ConF}$. Satisfiability of the union of $\mathit{TCons}$ and $\mathit{ConF}$ is checked with an extended version of \texttt{satisfiable}. Such a variant also computes the infinum value of $\mathit{LinF}$ taking into account the integer variables $\mathit{IntVars}$. The infinum value $\mathit{Min}$ is returned by \texttt{satisfiable} together with integer assignments for the integer variables. Such assignments are turned into equality constraints, and together with the key equality $\mathit{LinF}=\mathit{Min}$ from the definition of the $\mathit{min}$ operator (see~Section \nameref{sec:algebra}) appended to the other constraints.
  
\begin{lstlisting}[caption = {Interpretation of the $\mathit{min}$ operator.}, label = {lst:minimize}, basicstyle=\ttfamily\scriptsize]
solve(inf(T, F), Vars, Cons)  :-
    solve(T, Vars, TCons),
    exp_eval(F, Vars, LinF, ConF),
    append(ConF, Cons, ConsMin),
    int_vars(Vars, IntVars),
    satisfiable(ConsMin, IntVars, LinF, Min, IntValues),
    eq_con(IntVars, IntValues, ConsEq),
    append([LinF=Min|ConF], Cons, Cons1),
    append(CEq, Cons1, Cons).
\end{lstlisting}

\paragraph*{Integer-linear constraints.}
The variant of \texttt{satisfiable} used in the interpretation of $\mathit{inf}$ is based on the MILP built-in predicate \texttt{bb\_inf($C$, $\mathit{IntVars}$, $\mathit{LinF}$, $\mathit{Min}$, $\mathit{IntValues}$)}. The extension to MILP problems is required because (one-hot encoded) nominal and ordinal features lie in the domain of the integers. We discuss here the implications of such an extension on the query language of Section~\nameref{sec:reasonx_algebra_operators}.

For the given integer variables $\mathit{IntVars}$, a call to the \texttt{bb\_inf} predicate returns only one answer, i.e.,~without backtracking (for efficiency reasons, since there may be an exponential number of solutions), with ground values $\mathit{IntValues}$. Such values have the property that the linear function $\mathit{LinF}$ reaches its minimum over $C$ at the minimum of $C \wedge (\mathit{IntVars}=\mathit{IntValues})$. The latter turns out to be a linear constraint. Instead, the minimization of $C$ in presence of integer variables is an NP-hard problem that cannot always be expressed as a linear constraint~\citep{DBLP:conf/coco/Karp72}. The implication for {\reasonx{}} is that 
the returned answer constraint {sets nominal and ordinal features to fixed constant values. Hence, it provides} a correct but not complete characterization of the space of solutions to the query. For instance, consider the query:
\[ \mathit{min}(\{ 0 \leq x \leq 1, 0 \leq y \leq 1, x-y \leq z, y-x \leq z \}, z) \]
with the assumption that $x, y \in \mathbb{Z}$. This query is intended to constrain $z$ to $|x-y|$, the absolute value of the difference between $x$ and $y$. Such constraints naturally arise in the modeling of $L_1$ and $L_{\infty}$ distances. The variables $x$ and $y$ are set to the ground values returned by \texttt{bb\_inf}, namely $x=0, y=0$, and the minimum is reached for $z=0$. Hence, the interpreter returns $x=0, y=0, z=0$. This is a correct answer constraint, but it does not cover all possible solutions. For example, $x=1, y=1, z=1$ is another solution which is never explored.




\section{Demonstrations}
\label{sec:reasonx_demonstrations}

Here, we present several use-cases to illustrate the functionalities of {\reasonx{}}.
%
%
The majority of these demonstrations is based on case \textbf{(DT-M)}, i.e.,~under the assumption that the base model is also the ML model. {The first subsection considers a synthetic dataset. All other subsections consider the Adult Income dataset.} For an overview of the corresponding experimental settings, see Section~\nameref{app:expsettings} in the \supplement, which also reports additional demonstrations in Section~\nameref{app:moredemo}.


\subsection{Synthetic Dataset}
\label{sec:reasonx_synthetic_dataset}


We define a small synthetic dataset, consisting of two features (\texttt{feature1} and \texttt{feature2}) and a binary class with values \texttt{0} and \texttt{1}, each class having $1,000$ data instances. These are sampled from different bivariate normal distributions that are almost separable by an axis-parallel decision tree.
We choose a data instance from class \texttt{0}, called the \textit{factual instance} (\enquote{factual} in the figures), to illustrate some of the functionalities of {\reasonx{}} at the Python layer.
To initialize {\reasonx{}}, the following code is needed. 

\begin{lstlisting}[basicstyle=\ttfamily\scriptsize]
USER:    r = reasonx.ReasonX(pred_atts, target, df_code)
         r.model(clf1)
\end{lstlisting}

The constructor takes as input the list of features and the class name, and an object (\texttt{df\_code}) that decodes categorical features.
The base DT \texttt{clf1} is set via the \texttt{model} method. {The same method accepts as an optional argument the black-box model \texttt{clf2} whose \texttt{clf1} is a surrogate (cases \textbf{DT-GS} and \textbf{DT-LS}): \texttt{r.model(clf1, bb=clf2)}. The black-box model is used in the estimation of the fidelity of contrastive explanations.}

As a first operation, we set a factual instance of interest, named \texttt{F}, using the method \texttt{instance}, and query {\reasonx{}}  through the \texttt{solveopt} method. 

\begin{lstlisting}[basicstyle=\ttfamily\scriptsize]
USER:    r.instance('F', features=[1100, 661], label=0)
         r.solveopt()
\end{lstlisting}

\texttt{F} is fully specified as all feature values are set.
The query generated and solved by the CLP layer is of the form $e_1$ as in (\ref{q:e1}). There is a single constraint in $Th(e_1)$.
The output of {\reasonx{}} shows such a constraint (called \textit{answer constraint}) and the respective factual rule $c \rightarrow l=0$ with confidence $1.0$.

\begin{lstlisting}[basicstyle=\ttfamily\scriptsize,backgroundcolor=\color{answertoquery}]
REASONX: Answer constraint: 
           F.feature1=1100.0,F.feature2=661.0
         Rule satisfied by F: 
           IF F.feature1>1004.0,F.feature2> -55.5 THEN class 0 [1.0000, 0.2493]
\end{lstlisting}

\begin{figure}[t!]
    \centering
    \includegraphics[width=0.48\linewidth]{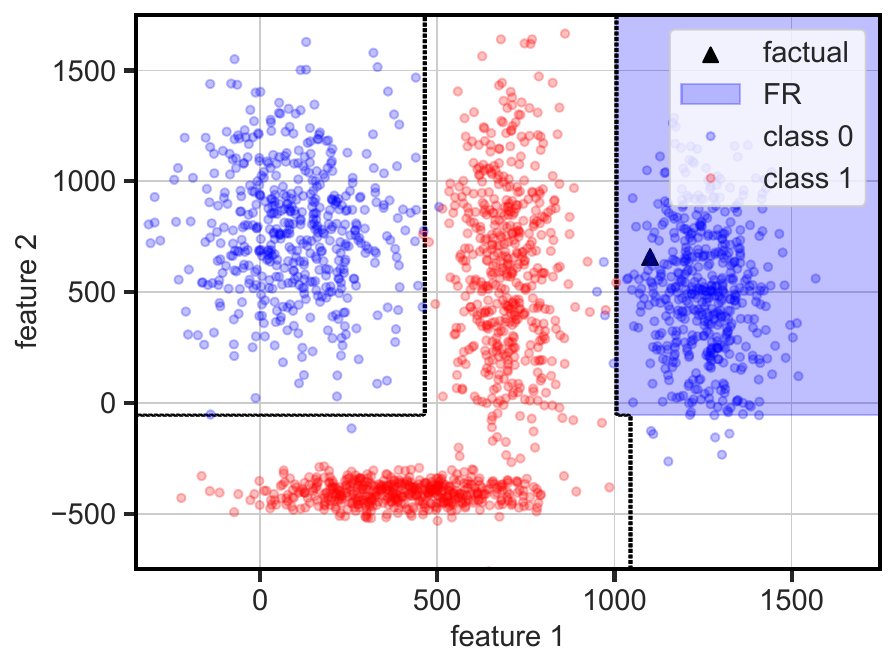}\hfill
    \includegraphics[width=0.48\linewidth]{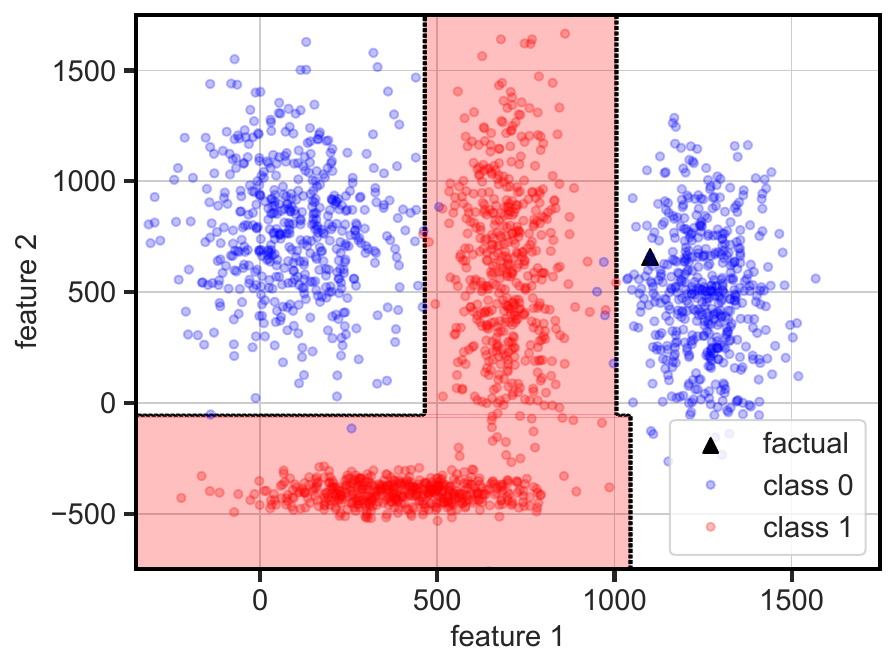}
    \caption{Left: factual region (FR) as provided by {\reasonx{}}. Right: contrastive regions (CR) as provided by {\reasonx{}}. Grey lines refer to the decision boundary of the base DT.}
    \label{fig:factual_ce_simple}
\end{figure}

{The second value in square brackets denotes the rule coverage. In this example, the probability that an instance drawn from the ground-truth distribution satisfies the premise of the rule is estimated to be $24.93\%$. Coverage can be readily computed from summary statistics stored in the DT and is optionally reported by \reasonx{}. For simplicity, we omit it from the subsequent examples.} To help intuition, we also plot the results. The left-hand side $c$ of the factual rule is shown as factual region in Figure~\ref{fig:factual_ce_simple} (left).

Next, we ask for the contrastive constraints, i.e.,~feasible regions of contrastive instances.
We code the problem by declaring an additional instance with the name \texttt{CE}, with class label \texttt{1}, and now requiring a minimum confidence of $0.8$ for its factual rule.

\begin{lstlisting}[basicstyle=\ttfamily\scriptsize]
USER:    r.instance('CE', label=1, minconf=0.8)
         r.solveopt()
\end{lstlisting}
\begin{lstlisting}[basicstyle=\ttfamily\scriptsize,backgroundcolor=\color{answertoquery}]
REASONX: Answer constraint:
           F.feature1=1100.0,F.feature2=661.0,CE.feature2<= -55.5,CE.feature1<=1043.5
         Rule satisfied by F: 
           IF F.feature2> -55.5,F.feature1>1004.0 THEN class 0 [1.0]
         Rule satisfied by CE: 
           IF CE.feature2<= -55.5,CE.feature1<=1043.5 THEN class 1 [0.9974]
--
         Answer constraint: 
           F.feature1=1100.0,F.feature2=661.0,CE.feature2> -55.5,CE.feature1>466.0,
         CE.feature1<=1004.0
         Rule satisfied by F: 
           IF F.feature2> -55.5,F.feature1>1004.0 THEN class 0 [1.0]
         Rule satisfied by CE: 
           IF CE.feature2> -55.5,CE.feature1>466.0,CE.feature1<=1004.0 
           THEN class 1 [0.9939]
\end{lstlisting}

\begin{figure}[t!]
    \centering
    \includegraphics[width=0.48\linewidth]{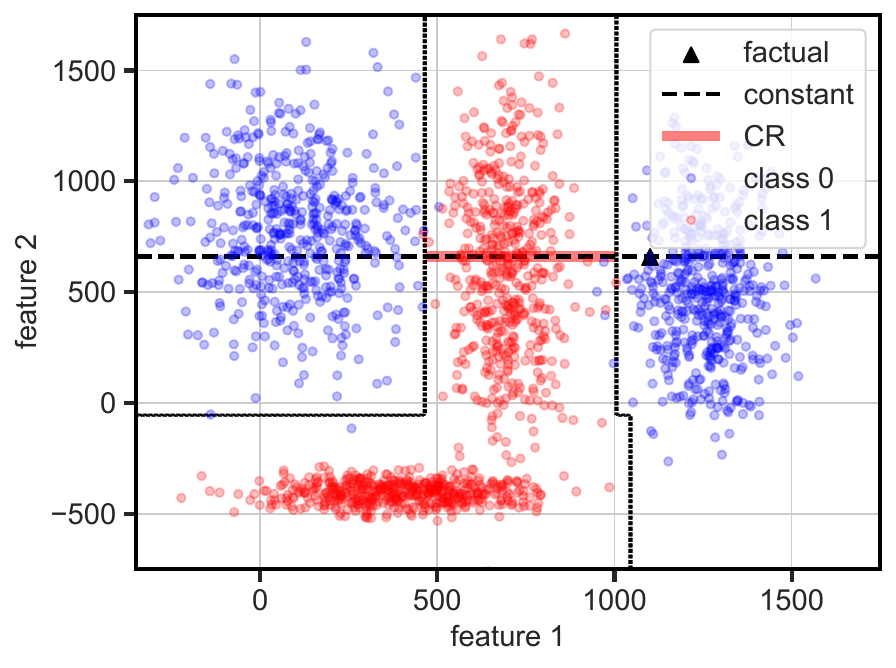}\hfill
    \includegraphics[width=0.48\linewidth]{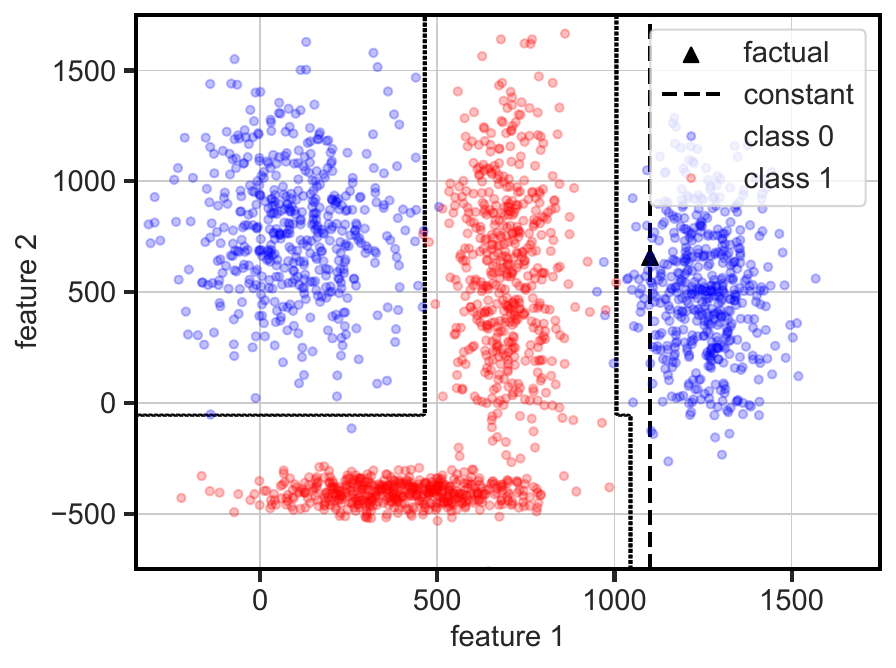}
    \caption{Left: contrastive region (CR) as provided by {\reasonx{}}, and given a constraint on \texttt{feature2}. Right: given the constraint on \texttt{feature1}, no solution exists. The dashed line refers to the enforced constraint. Grey lines refer to the decision boundary of the DT.}
    \label{fig:ce_constant}
\end{figure}

Figure~\ref{fig:factual_ce_simple}~(right) shows the two contrastive regions denoted by the left-hand sides of the two rules for \texttt{CE}. 
%
Notice that the instances \texttt{F} and \texttt{CE} are not related, apart from being predicted by the DT with different classes.
This means that while we interpret instance \texttt{CE} as contrastive to instance \texttt{F}, it could be also interpreted as a different factual instance.
We then use a constraint to ensure that {\small \texttt{feature2}} stays constant between the factual instance and the contrastive one (see Figure~\ref{fig:ce_constant} (left)), or to ensure that instead \texttt{feature1} stays constant (Figure~\ref{fig:ce_constant} (right)). While the first constraint leads to an admissible contrastive region as a line ({\small \texttt{CE.feature2 = 661.0, CE.feature1 > 466.0, CE.feature1 <= 1004.0}}), the second constraint leads to no solution.
The following code show the first case. 

\begin{lstlisting}[basicstyle=\ttfamily\scriptsize]
USER:    r.constraint('CE.feature2 = F.feature2') 
         r.solveopt()
\end{lstlisting}
\begin{lstlisting}[basicstyle=\ttfamily\scriptsize,backgroundcolor=\color{answertoquery}]
REASONX: Answer constraint:
           F.feature1=1100.0,F.feature2=661.0,CE.feature2=661.0,CE.feature1>466.0,
         CE.feature1<=1004.0
         Rule satisfied by F: 
           IF F.feature2> -55.5,F.feature1>1004.0 THEN class 0 [1.0]
         Rule satisfied by CE: 
           IF CE.feature2> -55.5,CE.feature1>466.0,CE.feature1<=1004.0 
           THEN class 1 [0.9939]
\end{lstlisting}

{By projecting the answer constraint onto the instance \texttt{CE}, we obtain the corresponding contrastive rule together with its fidelity, as summarized in Table~\ref{tab:my_label}. Since we are considering case \textbf{(DT-M)}, the fidelity is $1.0$. Notice that the output includes also the factual rule for \texttt{CE}, which will omit in next examples for ease of presentation.}

\begin{lstlisting}[basicstyle=\ttfamily\scriptsize]
USER:    r.solveopt(project=['CE'])
\end{lstlisting}
\begin{lstlisting}[basicstyle=\ttfamily\scriptsize,backgroundcolor=\color{answertoquery}]
REASONX: Answer constraint:
           CE.feature2=661.0,CE.feature1>466.0,CE.feature1<=1004.0
         Contrastive rule satisfied by CE: 
           IF Answer constraint THEN class 1 [1.0000]
         Rule satisfied by CE: 
           IF CE.feature2> -55.5,CE.feature1>466.0,CE.feature1<=1004.0 
           THEN class 1 [0.9939]
\end{lstlisting}

{Notice that the boundedness conditions from Section~\nameref{sec:fidelity} are trivially satisfied by the answer constraint in this example, so there is no need to impose additional bounds on the features.}

Before the constraint on fixing \texttt{feature1} to be constant can be asserted, the first constraint needs to be retracted.
Retraction facilitates interactivity: asserted constrains accumulate one after the other, until they are retracted\footnote{\texttt{r.retract(last=True)} retracts the last asserted constraint, and \texttt{r.reset(keep\_model=True)} retracts all instances and constraints, but it keeps the DTs.}.

\begin{lstlisting}[basicstyle=\ttfamily\scriptsize]
USER:    r.retract('CE.feature2 = F.feature2') 
         r.constraint('CE.feature1 = F.feature1')
         r.solveopt()
\end{lstlisting}
\begin{lstlisting}[basicstyle=\ttfamily\scriptsize,backgroundcolor=\color{answertoquery}]
REASONX: No answer.
\end{lstlisting}

Now, we extend the example by asking for the \textit{closest} contrastive points lying over the identity line ({\small \texttt{CE.feature2 = CE.feature1}}). 
The minimization of the $L_1$ norm between \texttt{F} and \texttt{CE} is passed as an argument to {\small \texttt{solveopt}}. 

\begin{lstlisting}[basicstyle=\ttfamily\scriptsize]
USER:    r.retract(last=True)
         r.constraint('CE.feature2 = CE.feature1')
         r.solveopt(minimize='l1norm(F, CE)', project=['CE'])
\end{lstlisting}
\begin{lstlisting}[basicstyle=\ttfamily\scriptsize,backgroundcolor=\color{answertoquery}]
REASONX: Answer constraint:
           CE.feature1= -55.5,CE.feature2= -55.5
         Contrastive rule satisfied by CE: 
           IF Answer constraint THEN class 1 [1.0000]
         Min value: 0.8937
--
         Answer constraint:
           CE.feature1=1004.0,CE.feature2=1004.0
         Contrastive rule satisfied by CE: 
           IF Answer constraint THEN class 1 [1.0000]
         Min value:0.1849
\end{lstlisting}

\begin{figure}[t!]
    \centering
    \includegraphics[width=0.48\linewidth]{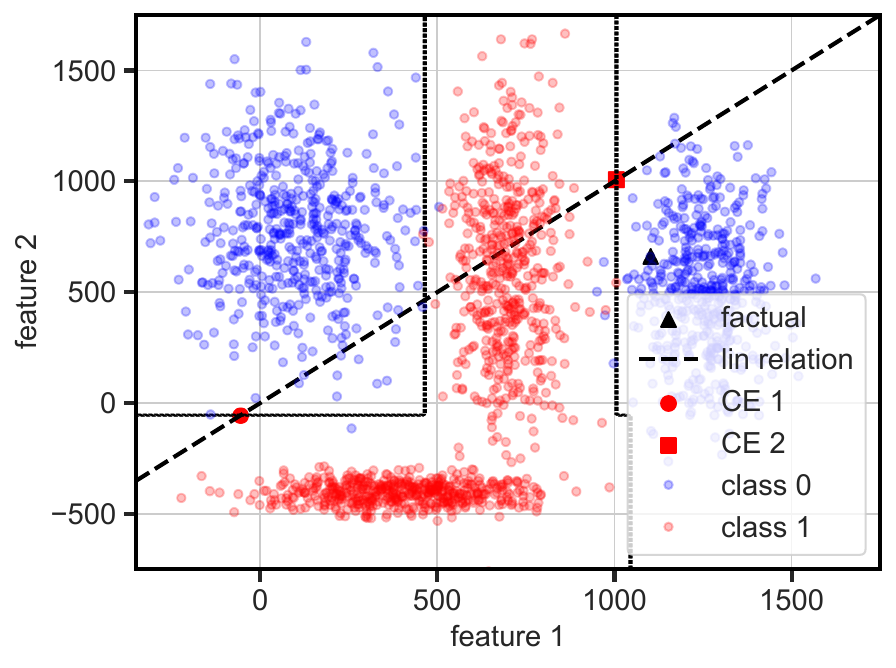}\hfill
    \includegraphics[width=0.48\linewidth]{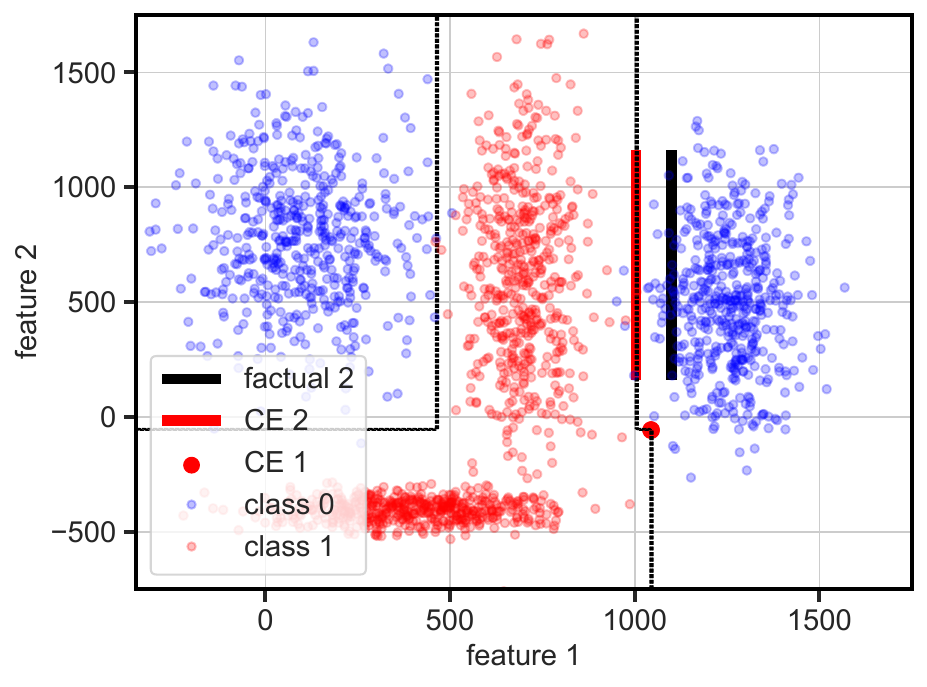}
    \caption{Left: minimal contrastive examples provided by {\reasonx{}}, under the constraint denoted by the identity line. Right: minimal contrastive examples provided by {\reasonx{}} for under-specified instances. Grey lines refer to the decision boundary of the DT.}
    \label{fig:ce_closest_region}
\end{figure}

Figure~\ref{fig:ce_closest_region} (left) reports the two solutions returned by {\reasonx{}} -- shown as red dots. Both of them stay in the identity line. They are the closest instances for each of the contrastive regions characterized by leaves of the DT. In fact, {\reasonx{}} solves the optimization problem for each of those independently and provides the user therefore not with the global optimum, but with two local optima. 
This 
leaves more flexibility to the user. Further processing of the results, e.g.,~filtering for the global optimum {based on the returned \texttt{Min value}}, can be implemented. 
The data point {\small \texttt{CE.feature1 = 1004.0, CE.feature2 = 1004.0}} denoted by \textit{CE 2} in Figure~\ref{fig:ce_closest_region} (left) is the global optimum.
The distance between such pairs is the one reported in the output of {\reasonx{}}.

Let us now allow {\small \texttt{feature2}} of the factual instance to be under-specified, by setting it to a range instead of a fixed value. We reset all constraints, and assert the value for {\small \texttt{feature1}} and bounds for {\small \texttt{feature2}}. In the {\small \texttt{solveopt}} we also add another parameter to project the answer constraint over the \texttt{CE} instance only.
Figure~\ref{fig:ce_closest_region} (right) shows the region of the two answer constraints: a single point and a line. Each point in the line represent a closest point to some of the possible values for \texttt{F}.

\begin{lstlisting}[basicstyle=\ttfamily\scriptsize]
USER:    r.reset(keep_model=True)
         r.instance('F', label=0)
         r.constraint('F.feature1 = 1100, F.feature2 >= 161, 
         F.feature2 <= 1161')
         r.instance('CE', label=1, minconf=0.8)
         r.solveopt(minimize='l1norm(F, CE)', project=['CE'])
\end{lstlisting}
\begin{lstlisting}[basicstyle=\ttfamily\scriptsize,backgroundcolor=\color{answertoquery}]
REASONX: Answer constraint: 
           CE.feature1=1043.5,CE.feature2= -55.5
         Contrastive rule satisfied by CE: 
           IF Answer constraint THEN class 1 [1.0000]
         Min value: 0.1146
--
         Answer constraint: 
           CE.feature1=1003.99999,CE.feature2>=161.0,CE.feature2<=1161.0
         Contrastive rule satisfied by CE: 
           IF Answer constraint THEN class 1 [1.0000]
         Min value: 0.0510
\end{lstlisting}

{Notice that the infimum value of the $L_1$ function is reported as \texttt{CE.feature1=1003.99999}. Although strict inequalities are present, their relaxation by the $\mathit{relax}$ operator does not affect the minimization problem in this case. The reported value is instead due to numerical rounding in the MILP optimization.}


\subsection{Reasoning over Time}
\label{sec:reasonx_time}

{\reasonx{}} can be run on several models and instances. Here, we assume the following case: a model is used over a long period of time, and it undergoes re-training due to new, incoming data instances. The re-training induces small model changes.

We simulate this case by splitting the Adult Income dataset into two datasets of the same size, and training DTs under the same parameter settings on these two splits{, in particular with maximum tree depth $D=3$}. Let $DT_0$ and $DT_1$ be the resulting DTs.

\paragraph*{Independent explanations.}
First, we define two {fully-specified} instances \texttt{F0} and \texttt{F1}, assigning the same data values to them, but a different model -- $DT_0$ and $DT_1$ respectively. 
We independently query for factual, contrastive rules, and minimal contrastive rules under the $L_1$~norm. {Since we start from a fully-specified instance, the minimal contrastive rules turn out to be single data points, namely constrastive examples. The contrastive instances are constrained to keep the features \texttt{race}, \texttt{sex}, and \texttt{workclass} unchanged, and not to decrease \texttt{education} or \texttt{age}.}
Results are displayed in Table~\ref{tab:reasonx_different_models_results_1} for the factual rules and in Table~\ref{tab:reasonx_different_models_results_2} (1)-(8) for the (minimal) contrastive rules.
For both DTs, two contrastive rules and two minimal contrastive rules are generated. 
{Factual rules differ slightly in the upper bound for \texttt{capitalgain}. Overcoming such bounds turns out to lead both DTs to change the prediction.}

\paragraph*{Intersecting explanations.}
Now, we use the reasoning capabilities of {\reasonx{}} to better understand how these two sets of explanations relate to each other.
In order to do so, we use the output produced for the contrastive instance {\texttt{CF0} relative to $DT_0$ to further constrain the contrastive instance \texttt{CF1} relative to $DT_1$}. 
%
As a result, we obtain {two contrastive rules that are not only contrastive w.r.t.~$DT_0$ but also to w.r.t.~$DT_1$ and thus in their intersection. They are shown as rules (9) and (10). It turns out they are equivalent to (5) and (6) respectively. This is expected, since (5) and (6) for $DT_1$ are stricter rules than (1) and (2) for $DT_0$, respectively. As a consequence, data points that satisfy (1) but not (5) are contrastive examples w.r.t.~$DT_0$ but not w.r.t.~$DT_1$. Such data points are characterized by the condition \texttt{5095.5 < capitalgain <= 5119.0}.}
If we remind ourselves of the time dimension, it means that the change in the model over time became indeed visible in its explanation and that some of the contrastive rules and examples derived at the first time point (model $DT_0$) are no longer valid {in full} at the second time point (model $DT_1$). 

\begin{small}
\begin{table}[t]
    \centering
    \scriptsize
    \begin{tabular}{cl}
        \toprule
        \textbf{ID} & \textbf{Factual explanation rule} \\
        \midrule
        (1) & \texttt{IF F1.capitalgain<=5095.5,F1.education<=12.5,F1.age>30.5} \\
        & \texttt{THEN <=50K [0.8071]} \\
        (2) & \texttt{IF F0.capitalgain<=5119.0,F0.education<=12.5,F0.age>30.5} \\
        & \texttt{THEN <=50K [0.8012]} \\
        \bottomrule
    \end{tabular}
    \caption{{Factual rules for a same instance and different DTs, both under case \textbf{(DT-M)}. (1) for $DT_0$. (2) for $DT_1$.}}   \label{tab:reasonx_different_models_results_1}
\end{table}    
\end{small}

\begin{small}
\begin{table}[t]
    \centering
    \scriptsize
    \begin{tabular}{ccl}
        \toprule
        \textbf{ID} & $L_1$ & \textbf{(Minimal) contrastive rule} \\
        \midrule        
        (1) &  & \texttt{IF $c_0$,CE0.capitalgain>5095.5,CE0.capitalgain<=5316.5} \\
        && \texttt{THEN >50K [1.0000]} \\
        (2) &  & \texttt{IF $c_0$,CE0.capitalgain>7055.5 THEN >50K [1.0000]} \\
        (3) & $0.0510$ & \texttt{IF CE0.race=White,CE0.sex=Male,CE0.workclass=Private,} \\
        &  & \texttt{\ \ CE0.age=61.0,CE0.education=9.0,CE0.capitalgain=5095.51,} \\
        && \texttt{\ \ CE0.capitalloss=0.0,CE0.hoursperweek=50.0} \\
        && \texttt{THEN >50K [1.0000]} \\
        (4) & $0.0706$ & \texttt{IF CE0.race=White,CE0.sex=Male,CE0.workclass=Private,} \\
        &  & \texttt{\ \ CE0.age=61.0,CE0.education=9.0,CE0.capitalgain=7055.51,} \\
        && \texttt{\ \ CE0.capitalloss=0.0,CE0.hoursperweek=50.0} \\
        && \texttt{THEN >50K [1.0000]} \\
        (5) &  & \texttt{IF $c_1$,CE1.capitalgain>5119.0,CE1.capitalgain<=5316.5 } \\
        && \texttt{THEN >50K [1.0000]} \\
        (6) & & \texttt{IF $c_1$,CE1.capitalgain>7073.5 THEN >50K [1.0000]} \\
        (7) & $0.0512$ & \texttt{IF CE1.race=White,CE1.sex=Male,CE1.workclass=Private,} \\
        &  & \texttt{\ \ CE1.age=61.0, CE1.education=9.0,CE1.capitalgain=5119.01,} \\
        && \texttt{\ \ CE1.capitalloss=0.0,CE1.hoursperweek=50.0 } \\
        && \texttt{THEN >50K [1.0000]} \\
        (8) & $0.0707$ &  \texttt{IF CE1.race=White,CE1.sex=Male,CE1.workclass=Private,} \\
        &  & \texttt{\ \ CE1.age=61.0,CE1.education=9.0,} \\
        && \texttt{\ \ CE1.capitalgain=7073.51,CE1.capitalloss=0.0,} \\
        && \texttt{\ \ CE1.hoursperweek=50.0} \\
        && \texttt{THEN >50K [1.0000]} \\
        (9) &  & \texttt{IF $c_1$,CE1.capitalgain>5119.0,CE1.capitalgain<=5316.5} \\
        && \texttt{THEN >50K [1.0000]} \\
        (10) &  & \texttt{IF $c_1$,CE1.capitalgain>7073.5 THEN >50K [1.0000]} \\
        \bottomrule
    \end{tabular}
    \caption{{(Minimal) contrastive rules for a same instance and different DTs, both under case \textbf{(DT-M)}. The contrastive instance \texttt{CE0} is constrained to $c_0 =$  \texttt{CE0.race=White,} \texttt{CE0.sex=Male,} \texttt{CE0.workclass=Private,} \texttt{CE0.age>=61.0,} \texttt{CE0.education>=9.0}.
    We denote by $c_1$ a similar constraint but for \texttt{CE1}. 
    (1) - (4) refer to $DT_0$. (5) - (8) refer to $DT_1$. (9) - (10) refer to contrastive rules that hold in both $DT_0$ and $DT_1$. The $L_1$ column reports the minimized distance for minimal contrastive rules.}}
    \label{tab:reasonx_different_models_results_2}
\end{table}    
\end{small}

\subsection{Reasoning over Different Model Types} 
\label{sec:reasonx_models}

In this section, we show how {\reasonx{}} can be used to reason over different model types. 
As an example, we compare a DT model {with maximum depth of $D=3$} (case \textbf{(DT-M)}) with a ML model that is explained via a global surrogate model, i.e.,~case \textbf{(DT-GS)}. A simple comparison between a random forest (RF), a multi-layer perceptron (MLP), and an XGBoost (XGB) model shows that the latter slightly outperforms the former models (an accuracy of $0.851$ against $0.814$ for the MLP and $0.804$ for the RF, in comparison $0.802$ for the DT). 
We compare in $DT_0$ the DT from case \textbf{(DT-M)} and in $DT_1$ the {surrogate DT with depth 3 of the} XGB model from case \textbf{(DT-GS)}. The fidelity {over an hold-out dataset}, i.e.,~the ratio of correct predictions of $DT_1$ w.r.t. prediction of the XGB model is $\mathit{fidelity} = 0.921$. {Higher fidelity can be achieved by deeper surrogate DTs; however, this would produce more complex explanation rules.}

\begin{table}[t]
    \centering
    \scriptsize
    \begin{tabular}{cl}
        \toprule
        \textbf{ID} & \textbf{Factual explanation rule} \\
        \midrule
        (1) & \texttt{IF F0.capitalgain>7055.5,F0.age>20.0 THEN >50K [0.9882]} \\
        (2) & \texttt{IF F1.capitalgain>7073.5,F1.education<=12.5 THEN >50K [0.9940]} \\        
        \bottomrule
    \end{tabular}
    \caption{{Factual rules for a same instance and different models. (1) for $DT_0$ under case \textbf{(DT-M)}. (2) for $DT_1$ surrogate model of XGB under case \textbf{(DT-GS)}.}}\label{tab:reasonx_different_models_results_3}
\end{table}

\begin{small}
\begin{table}[t]
    \centering
    \scriptsize
    \begin{tabular}{ccl}
    \toprule
    \textbf{ID} & $L_1$ & \textbf{(Minimal) contrastive rule} \\
    \midrule
         (1) & & \texttt{IF $c_0$,CE0.capitalgain<=5119.0  THEN <=50K [1.0000]} \\     
        (2) & $0.0218$ & \texttt{IF $c_0$,CE0.age=41.0,CE0.capitalgain=5119.0} \\
        &  & \texttt{\ \ CE0.capitalloss=0.0,CE0.hoursperweek=50.0}\\
        && \texttt{THEN >50K [1.0000]} \\
         (3) &  & \texttt{IF $c_1$,CE1.age<=90.0,} \\ 
         && \texttt{\ \ CE1.capitalgain>= 0.0,CE1.capitalgain<=4243.5,} \\
         && \texttt{\ \ CE1.capitalloss>= 0.0,CE1.capitalloss<=1805.0,} \\
         && \texttt{\ \ CE1.hoursperweek>=1.0,CE1.hoursperweek<=99.0} \\
         && \texttt{THEN <=50K [0.9982]} \\        
        (4) & $0.0305$ & \texttt{IF $c_1$,CE1.age=41.0,CE1.capitalgain=4243.5,} \\
        &  & \texttt{\ \ CE1.capitalloss=0.0,CE1.hoursperweek=50.0} \\
        && \texttt{THEN >50K [1.0000]} \\
    \bottomrule
    \end{tabular}
    \caption{{(Minimal) contrastive rules for a same instance and different model types. {The contrastive instance \texttt{CE0} is constrained to $c_0 =$  \texttt{CE0.race=AmerIndianEskimo,} \texttt{CE0.sex=Male,} \texttt{CE0.workclass=Private,}\texttt{CE0.age>=41.0,} \texttt{CE0.education=10.0}. We denote by $c_1$ a similar constraint but for \texttt{CE1}.}  (1)-(2) refer to $DT_0$ under case \textbf{(DT-M)}. (3)-(4) refer to $DT_1$ surrogate model of XGB under case \textbf{(DT-GS)}. The $L_1$ column reports the minimized distance for minimal contrastive rules.}}    \label{tab:reasonx_different_models_results_4}
\end{table}    
\end{small}

We compare the factual rules for a same sample data instance. The results are shown in Table~\ref{tab:reasonx_different_models_results_3}. 
The reasons behind the classification of the data instance, i.e.,~the features appearing on the left-hand side of the rules, are different -- even if both models correctly classify the instance as class \texttt{>50K}. Both models use the feature \texttt{capitalgain}, but while the $DT_0$ relies on the feature \texttt{age}, $DT_1$ (alias the XGB model) uses the feature \texttt{education}.
We expect these differences to be reflected also in the (minimal) contrastive rules, which are reported in Table~\ref{tab:reasonx_different_models_results_4}.
{As in the previous sub-section, the contrastive instances are constrained to keep the features \texttt{race}, \texttt{sex}, and \texttt{workclass} unchanged, and not to decrease \texttt{education} or \texttt{age}.}
{The fidelity of the contrastive rule (3) and of the minimal rule (4) for $DT_1$ is computed as described in Subsection~\nameref{sec:fidelity}. To enable this computation, all continuous features had to be additionally bounded and the ordinal feature \texttt{education} fixed to a constant value. Overall, the two contrastive rules (1) and (3) differ in the bounds imposed on \texttt{capitalgain}, while the rule for $DT_1$ additionally includes a bound on \texttt{capitalloss} and on \texttt{hoursperweek}. In both cases, the upper bound on \texttt{capitalgain} determines the corresponding minimal contrastive rules (2) and (4) respectively.}

\subsection{Diversity Optimization}
\label{sec:reasonx_demonstration_diversity}


In this experiment, we select
{three}
contrastive explanations from an admissible set of contrastive explanations, using the optimization function discussed in Section~\nameref{sec:reasonx_embeddings}, paragraph \nameref{sec:encdist}.
%
We run this experiment on the Adult Income dataset, case \textbf{(DT-M)} and for a decision tree with a depth of $depth = 5$. This parameter is changed to obtain a higher number of contrastive examples. Furthermore, we set $\lambda = 0.5$ in (\ref{equ:diversity}) and repeat for $50$ instances.

Results are compared to the classical approach of selecting contrastive instances, i.e., optimizing only by proximity. In both cases, we compute the value of the optimization function $f(x_f, S)$ and of the proximity and diversity term separately.
We plot the results in Figure~\ref*{fig:diversity_optimization} (see the \supplement\ in~\nameref{appendix:diversity}).
The classical approach leads to a distribution of values of the optimization function that has a sharp peak and is centered around zero. When considering the diversity of generated contrastive explanations in the optimization, the distribution of the function changes: values are much more diverse, there is no sharp peak.
To confirm that these values indeed belong to different distributions, we calculated the Kolmogorov-Smirnow test ($p$-value $\approx 0$).


\subsection{Detecting Biases}
\label{sec:reasonx_detecting_bias}

Here, we demonstrate how the generation of contrastive examples and the use of constraints in {\reasonx{}} can support the detection of societal biases. 
This case is close to the definition of \textit{explicit bias} as in \citet{DBLP:journals/ml/GoethalsMC24} and may be useful in the context of \textit{direct discrimination} and \textit{prima facie} discrimination.
We investigated a DT and an XGB model, and possible discrimination based on the sensitive attributes \texttt{age}, \texttt{race}, \texttt{sex}, and pairwise combinations.

In summary, we find that {the DT in case \textbf{DT-M} exhibits a stronger dependence on sensitive attributes for reversing predictions, whereas the surrogate DT of XGB in case \textbf{DT-GS} appears less affected by this form of bias.}
%
Details on the experiment, more information on bias detection through XAI and definitions, as well as details on the experimental results can be found in the \supplement\ in~\nameref{appendix:detecting_biases}.

\section{Quantitative Evaluation}
\label{sec:reasonx_evaluation}

In this section, we present a quantitative evaluation of {\reasonx{}}. We first evaluate {\reasonx{}} in isolation. We then compare its (minimal) contrastive rules with those generated by DiCE~\citep{DBLP:conf/fat/MothilalST20}, and its factual rules with those generated by ANCHORS~\citep{DBLP:conf/aaai/Ribeiro0G18}. Finally, we compare the runtime of the three approaches. {The evaluation targets selected capabilities rather than providing a comprehensive state-of-the-art benchmark. The broader theoretical contributions of {\reasonx{}}, including reasoning over under-specified instances and user-defined constraints, are illustrated qualitatively in Section~\nameref{sec:reasonx_demonstrations}. Accordingly, we do not claim that {\reasonx{}} outperforms methods specialized in individual explanation tasks; instead, the experiments focus on two explanation types for which direct comparisons with established baselines are possible.}

Information on datasets{, metrics, and experimental settings are reported} in Section~\nameref{app:expsettings} in the \supplement.
In particular, the metrics used in this section are standard metrics~\citep{DBLP:journals/frai/ViloneL21,DBLP:conf/nips/PawelczykBHRK21,DBLP:journals/expert/GuidottiMGPRT19} adapted to the characteristics of {\reasonx{}}. 
%
%

\subsection{Results}

\paragraph*{{{\reasonx{}}} only.}
Results for {\reasonx{}} are reported in Table~\ref{tab:reasonx}. They depend on the dataset and experimental case, {as indicated in the first and second columns, respectively. For all cases, we report average values over $100$ fully specified test instances. No user constraint is set. The results for minimal contrastive explanations also depend on the selected distance function, namely $L_1$ or $L_{\infty}$. All DTs are trained with a maximum depth of $D=3$. Despite their limited depth, they achieve relatively high fidelity, particularly in the \textbf{DT-LS} case, where the decision boundary to be approximated is local and therefore generally simpler.
The fraction $S/N$ of instances for which at least one contrastive rule can be generated is close to the accuracy $acc$ in case \textbf{DT-M}, and to the fidelity in case \textbf{DT-GS}. This behavior is expected, since unsuccessful cases are associated with instances for which the base DT does not reproduce the ground-truth value or the target prediction. For \textbf{DT-LS}, however, $S/N$ can be substantially lower than the corresponding fidelity. This can be attributed to the synthetic neighborhood-generation procedure: if the generated neighborhood is too narrow and does not cross the decision boundary, no contrastive rule can be obtained even when the local surrogate has high fidelity.
The average length $l_F$ of factual rules is close to the maximum depth $D=3$ for \textbf{DT-M} and \textbf{DT-GS}, whereas it is smaller for \textbf{DT-LS}. This provides further evidence that the decision boundary in the neighborhood of a given instance can often be represented by a simpler DT than the global boundary. The number $N_{\mathit{CE}}$ of contrastive rules per factual instance and their average length $l_{\mathit{CE}}$ vary across datasets and experimental cases, reflecting differences in both class distribution and tree structure. The minimized $L_1$ and $L_{\infty}$ distances are generally smaller for \textbf{DT-LS} than for the other two cases, indicating that local surrogate models tend to identify closer contrastive examples. Finally, for fully specified factual instances, minimization with respect to $L_1$ always yields a single data point, whereas minimization with respect to $L_{\infty}$ may yield higher-dimensional regions of equally optimal contrastive explanations.}

{Section~\nameref{sec:appendix_parameter_testing} in the {\supplement} contains additional experimental results at the variation of $D$ and of the number of user constraints.}

\begin{table}[t]
    \centering
    \scriptsize
    \setlength{\tabcolsep}{4pt}
    \begin{tabular}{ccccccccccccc}
    \toprule
    & \textbf{Case} & $acc$/$f$ & $S/N$ & $l_{F}$ & $N_{\mathit{CE}}$ &  $l_{\mathit{CE}}$  & $d_{\mathit{MCE}, L_1}$ &
    $d_{\mathit{MCE}, L_{\infty}}$ & $\mathit{dim}_{L_1}$ &
    $\mathit{dim}_{L_{\infty}}$ \\
    \midrule
    \multirow{3}{*}{\rotatebox[origin=c]{90}{\texttt{ADULT}}}
& DT-M & 0.802 & 0.82 & 2.99 & 1.68 & 2.02 & 0.069 & 0.066 & 0 & 4\\
& DT-GS & 0.921 & 0.93 & 2.98 & 4.34 & 2.82 & 0.361 & 0.304 & 0 & 4\\
& DT-LS & 0.946 & 1.00 & 1.27 & 2.46 & 2.41 & 0.084 & 0.067 & 0 & 3\\
    \midrule
    \multirow{3}{*}{\rotatebox[origin=c]{90}{\texttt{SGC}}}
& DT-M & 0.733 & 0.76 & 3.00 & 2.42 & 3.00 & 0.824 & 0.441 & 0 & 3\\
& DT-GS & 0.767 & 0.77 & 2.93 & 2.30 & 2.54 & 0.655 & 0.417 & 0 & 3\\
& DT-LS & 0.952 & 0.65 & 2.03 & 1.26 & 2.76 & 0.121 & 0.085 & 0 & 2\\
    \midrule
    \multirow{3}{*}{\rotatebox[origin=c]{90}{\texttt{GMSC}}} 
& DT-M & 0.936 & 0.95 & 3.00 & 0.95 & 2.00 & 0.020 & 0.015 & 0 & 9\\
& DT-GS & 0.983 & 0.99 & 2.97 & 2.99 & 2.67 & 0.018 & 0.015 & 0 & 9\\
& DT-LS & 0.874 & 1.00 & 2.93 & 3.28 & 2.83 & 0.018 & 0.014 & 0 & 9\\
    \midrule
    \multirow{3}{*}{\rotatebox[origin=c]{90}{\texttt{DCCC}}}
& DT-M & 0.780 & 0.74 & 3.00 & 0.80 & 3.00 & 0.113 & 0.140 & 0 & 7\\
& DT-GS & 0.921 & 0.92 & 3.00 & 1.04 & 3.00 & 0.118 & 0.129 & 0 & 7\\
& DT-LS & 0.994 & 0.96 & 2.56 & 1.04 & 2.82 & 0.035 & 0.043 & 0 & 7\\
    \midrule
    \multirow{3}{*}{\rotatebox[origin=c]{90}{\texttt{ACA}}}
& DT-M & 0.845 & 0.83 & 3.00 & 3.74 & 3.00 & 1.174 & 0.875 & 0 & 6\\
& DT-GS & 0.905 & 0.90 & 2.28 & 2.71 & 2.67 & 0.733 & 0.659 & 0 & 6\\
& DT-LS & 0.991 & 0.80 & 1.64 & 1.48 & 2.65 & 0.085 & 0.066 & 0 & 5\\

    \bottomrule
    \end{tabular}
    \caption{Evaluation of {\reasonx{}}. 
    \texttt{ADULT} = Adult Income, 
    \texttt{SGC} = South German Credit, \texttt{GMSC} = Give Me Some Credit, \texttt{DCCC} = Default Credit Card Clients, \texttt{ACA} = Australian Credit Approval. 
    {$acc$ indicates the accuracy of the DT model (case \textbf{DT-M}), $f$ the fidelity of the surrogate DT (cases \textbf{DT-GS} and \textbf{DT-LS}).  $S/N$ indicates the ratio of instances for which an output could be generated. $l_F$ and $l_{\mathit{CE}}$ are the lengths of the factual and contrastive rules, respectively. $N_{\mathit{CE}}$ is the number of contrastive rules. $d_{\mathit{MCE}, L_1}$ and $d_{\mathit{MCE}, L_{\infty}}$ are the minimized distance in minimal contrastive rules, while $d_{\mathit{MCE}, L_1}$ and $d_{\mathit{MCE}, L_{\infty}}$ their maximum dimensionality ($0$ means a data point). In all cases, the maximum depth of DTs is set to $D=3$ and no user constraint is set.}}
    \label{tab:reasonx}
\end{table}

\begin{table}[t]
    \centering
    \scriptsize
    \setlength{\tabcolsep}{5pt}
    \begin{tabular}{lcccccccc}
    \toprule
    &&&&&&&\multicolumn{2}{c}{\textbf{time (s)}}\\
    \textbf{Approach} & $D$ & $f$ & $S/N$ & $N_{CE} (^{\star})$ &  $\min d_{\mathit{MCE}, L_1}$ & $\min d_{\mathit{MCE}, L_{\infty}}$ & wrt $L_1$ & wrt $L_{\infty}$\\
    \midrule
    \multicolumn{8}{l}{\textbf{No constraint}} \\
DiCE & & & 1.00 & 2 & 1.036 & 0.687 & \multicolumn{2}{c}{0.389} \\
DiCE & & & 1.00 & 3 & 0.896 & 0.640 & \multicolumn{2}{c}{0.584} \\
DiCE & & & 1.00 & 4 & 0.855 & 0.621 & \multicolumn{2}{c}{0.689} \\
DiCE & & & 1.00 & 5 & 0.821 & 0.602 & \multicolumn{2}{c}{0.597} \\
REASONX & 3 & 0.921 & 0.96 & 4.34 & 0.065 & 0.060 & 0.078 & 0.168 \\
REASONX & 4 & 0.940 & 0.96 & 7.44 & 0.047 & 0.048 & 0.133 & 0.321 \\
REASONX & 5 & 0.948 & 0.96 & 14.40 & 0.047 & 0.044 & 0.286 & 0.761 \\
\midrule
    \multicolumn{8}{l}{\textbf{Immutability on \texttt{hoursperweek}}} \\
DiCE & & & 1.00 & 2 & 0.998 & 0.681 & \multicolumn{2}{c}{0.662} \\
DiCE & & & 1.00 & 3 & 0.896 & 0.626 & \multicolumn{2}{c}{0.984} \\
DiCE & & & 1.00 & 4 & 0.832 & 0.606 & \multicolumn{2}{c}{1.041} \\
DiCE & & & 1.00 & 5 & 0.846 & 0.636 & \multicolumn{2}{c}{1.979} \\
REASONX & 3 & 0.921 & 0.96 & 4.34 & 0.065 & 0.060 & 0.064 & 0.158 \\
REASONX & 4 & 0.940 & 0.96 & 6.23 & 0.061 & 0.058 & 0.105 & 0.332 \\
REASONX & 5 & 0.948 & 0.96 & 12.29 & 0.049 & 0.047 & 0.262 & 0.920 \\
      \midrule
    \multicolumn{8}{l}{\textbf{Immutability on \texttt{race} and \texttt{sex}}} \\
DiCE & & & 1.00 & 2 & 0.727 & 0.549 & \multicolumn{2}{c}{1.034} \\
DiCE & & & 1.00 & 3 & 0.629 & 0.477 & \multicolumn{2}{c}{1.668} \\
DiCE & & & 1.00 & 4 & 0.593 & 0.454 & \multicolumn{2}{c}{1.189} \\
DiCE & & & 1.00 & 5 & 0.532 & 0.408 & \multicolumn{2}{c}{1.682} \\
REASONX & 3 & 0.921 & 0.96 & 3.86 & 0.067 & 0.060 & 0.051 & 0.188 \\
REASONX & 4 & 0.940 & 0.96 & 6.51 & 0.047 & 0.048 & 0.102 & 0.641 \\
REASONX & 5 & 0.948 & 0.96 & 12.48 & 0.047 & 0.044 & 0.261 & 2.106 \\
    \bottomrule
    \end{tabular}
    \caption{Contrastive examples: comparison between DiCE and {\reasonx{}}{ (case \textbf{DT-GS}) on the \texttt{ADULT} dataset. $D$ indicates the depth of the surrogate DT, $f$ its fidelity, $S/N$ the ratio of instances for which an output could be generated,  $N_{CE}$ the number of generated contrastive examples (DiCE) and rules ({\reasonx{}}). ($^{\star}$) In case of DiCE, the number of contrastive examples is a parameter. For each instance, the minimum distance with any of its contrastive examples (minimal contrastive rule for {\reasonx{}}) is computed. Columns $\min d_{\mathit{MCE}, L_1}$ and $\min d_{\mathit{MCE}, L_{\infty}}$ report the mean values over the 100 test instances and w.r.t.~$L_1$ and $L_{\infty}$ norm respectively. Finally, the last two columns report the elapsed times: for {\reasonx{}}, these are different w.r.t.~$L_1$ and $L_{\infty}$. Time is reported in seconds, and does not account for initializations.}
    }
    \label{tab:dice}
\end{table}

\begin{table}[t]
    \centering
    \scriptsize
    \setlength{\tabcolsep}{5pt}
    \begin{tabular}{lccccccccc}
    \toprule
    \textbf{Approach} & $D$ & $f$ & $\mathit{MC}_F$ & $S/N$ & \textbf{$l_F$} & \textbf{$p$} & $c$ & \textbf{time (s)}\\
    \midrule
ANCHOR & & & 0.80 & 1.00 & 1.62 & 0.887 & 0.234 & 0.293 \\
ANCHOR & & & 0.90 & 1.00 & 1.72 & 0.893 & 0.225 & 0.294 \\
ANCHOR & & & 0.95 & 1.00 & 2.66 & 0.930 & 0.174 & 0.303 \\
\midrule
REASONX & 3 & 0.921 & 0.80 & 0.81 & 2.98 & 0.964 &  0.607 & 0.015 \\
REASONX & 3 & 0.921 & 0.90 & 0.76 & 2.97 & 0.974 &  0.645 & 0.009 \\
REASONX & 3 & 0.921 & 0.95 & 0.70 & 2.97 & 0.979 &  0.694 & 0.009 \\
REASONX & 4 & 0.940 & 0.80 & 0.82 & 3.15 & 0.973 &  0.583 & 0.015 \\
REASONX & 4 & 0.940 & 0.90 & 0.76 & 3.09 & 0.980 &  0.627 & 0.010 \\
REASONX & 4 & 0.940 & 0.95 & 0.71 & 3.03 & 0.982 &  0.667 & 0.009 \\
REASONX & 5 & 0.948 & 0.80 & 0.92 & 4.17 & 0.967 &  0.511 & 0.016 \\
REASONX & 5 & 0.948 & 0.90 & 0.81 & 4.06 & 0.985 &  0.567 & 0.011 \\
REASONX & 5 & 0.948 & 0.95 & 0.76 & 4.00 & 0.988 &  0.602 & 0.010 \\
    \bottomrule
    \end{tabular}
    
    \caption{Factual rules: comparison between  ANCHORS and {\reasonx{}}{ (case \textbf{DT-GS}) on the \texttt{ADULT} dataset. $D$ indicates the depth of the surrogate DT, $f$ its fidelity, $S/N$ the ratio of instances for which an output could be generated, $\mathit{MC}_F$ the minimum confidence required by the user, $p$ the  precision of the rule found, $l_F$ the length of the rule, and $c$ its coverage.  Time is reported in seconds, and it does not account for initializations.}
    %
    }
    \label{tab:anchors}
\end{table}

\paragraph*{Constrative examples.}
A comparison between {\reasonx{}}{, case \textbf{DT-GS},} and DiCE{, as offered by the \texttt{dice-ml} Python library,} is provided in Table~\ref{tab:dice}{, averaged over $100$ test instances. We compare the two approaches with no additional constraint, with an immutability constraint on the feature \texttt{hoursperweek}, and with an immutability constraint on both \texttt{race} and \texttt{sex}.}
Distances of {\reasonx{}} are smaller than those obtained with DiCE, i.e.,~the contrastive examples are closer to the original data instance.  {However, DiCE always find some contrastive examples (metric $S/N$), while \reasonx{} may fail to return an answer (due to insufficient fidelity of the surrogate DT).}
When constraints are enforced, less solutions (metric $N_{CE}$) are found by {\reasonx{}}. Such a behavior is expected as adding constraints cuts down on the admissible paths.
Notice that in DiCE the number of contrastive examples $N_{\mathit{CE}}$ is an input parameter. 
{Finally, increasing the maximum depth $D$ of surrogate models, increase fidelity and decrease the distance with minimal contrastive rules. On the other hand, time increases and also the length of explanation rules increases. See  Section~\nameref{sec:appendix_parameter_testing} in the {\supplement} for additional experiments at the variation of $D$.}
%


\paragraph*{Factual rules.}
The summary results for the factual rules produced by {\reasonx{}} and ANCHORS{, as implemented in the \texttt{anchors} Python library,} are reported in Table~\ref{tab:anchors}{, averaged over 100 test instances}.
{For ANCHORS, increasing the required minimum confidence (metric $\textit{MC}_F$) reduces the coverage (metric $c$) of the resulting rules, i.e.,~they are less general. For \reasonx{}, instead, coverage increases, and this occurs at any depth $D$. This can be attributed, however, to a lower number of answers (metric $S/N$). In fact, ANCHORS searches for a rule starting from the instance to be explained and a reference dataset, and it always returns one rule for every instance considered. By contrast, \reasonx{} may fail to return a factual rule when the surrogate DT does not classify the instance with sufficient confidence. Accordingly, $S/N$ for \reasonx{} decreases as the minimum-confidence threshold increases, although this effect can be mitigated by increasing the maximum depth $D$ of the surrogate DT.}
{Rule length increases with the minimum-confidence threshold for ANCHORS, and with the tree depth $D$ for \reasonx{}. ANCHORS produces shorter rules on average. This can be attributed to its instance-specific search procedure, although the resulting rules still admit a global interpretation. Instead, \reasonx{} statically derives rules from the fixed rule set induced by the surrogate DT. On the other hand, the rules produced by \reasonx{} achieve higher precision and coverage, indicating that they are, on average, both more precise and more general.}

\paragraph*{Runtime.}
We report the elapsed runtimes of {\reasonx{}}, 
DiCE, and ANCHORS in Tables~\ref{tab:dice} and \ref{tab:anchors}.
{Compared with DiCE, \reasonx{} is faster in most cases, particularly when immutability constraints are imposed. Since \reasonx{} is natively based on constraint solving, such constraints can be incorporated with minimal additional computational cost. Moreover, \reasonx{} optimizes the distance function directly over the polyhedra of feasible solutions, whereas DiCE searches over a set of candidate instances synthetically generated at runtime. The elapsed time of \reasonx{} increases with the maximum depth $D$, as deeper surrogate trees induce larger symbolic representations, and when optimizing the $L_{\infty}$ distance. We refer the reader to Section~\nameref{sec:appendix_parameter_testing} in the \supplement\ for additional experiments analyzing the effect of varying $D$.}
{Compared with ANCHORS, \reasonx{} is approximately one-two orders of magnitude faster. This can be attributed to the fact that ANCHORS searches for rules dynamically by repeatedly querying the black-box model on synthetically generated instances in order to characterize its decision boundary.}

\subsection{Completeness, Coverage, and Constraint Validity}

The following statements regarding the completeness, coverage and validity of {\reasonx{}} hold, assuming that its base model was learned successfully, i.e.,~that accuracy (case \textbf{(DT-M)}) or fidelity (case \textbf{(DT-GS)} and \textbf{(DT-LS)}) are sufficiently high. This is guaranteeing a certain level of correctness between the base model's prediction and the original class (case \textbf{(DT-M)}) or the predicted class (case \textbf{(DT-GS)} and \textbf{(DT-LS)}):
\textit{first}, under no additional constraints, for every data instance, {\reasonx{}} can find the factual, and one or more contrastive rules. It can also find one or more contrastive data instances. 
%
If no factual rule is returned, this is because the prediction of the base model does not agree with the data label (case \textbf{(DT-M)}) or the prediction of the approximated ML model (case \textbf{(DT-GS)} and \textbf{(DT-LS)}), or because the parameter of the minimum confidence value is too high. However, this is an intentional design choice and {\reasonx{}} can still retrieve the results by a change in the query -- to explain the incorrect prediction of the base model, or to provide an explanation with a lower confidence value.
Additionally, for the returned contrastive rules, constraints, and examples, an explicit validation against the black-box ML model is {offered} in cases \textbf{(DT-GS)} and \textbf{(DT-LS)} to {estimate their fidelity, namely the probability} that they indeed lie on the opposite side of the black-box decision boundary. 
%
\textit{Second}, the average length and the number of the obtained (factual or contrastive) rules depend on the depth and the width of the base model (see also Section~\nameref{sec:reasonx_evaluation}), and the distance between the original and the contrastive instances may not be minimal w.r.t. the original decision boundary, as the base model only approximates the boundary.
\textit{Third}, all data instances in the input space of the base model are covered by factual rules which have no overlap -- only one rule is the factual rule. 
Other XAI approaches based on rules do not have this property, therefore~\citep{DBLP:conf/kdd/LakkarajuBL16,DBLP:journals/frai/ViloneL21} introduce a metric called \enquote{fraction overlap}.
\textit{Fourth}, if constraints are enforced in {\reasonx{}}, the rules -- if they exist -- align with the constraints.
This is not always the case for XAI approaches based on rules, thus,~\citep{DBLP:conf/nips/PawelczykBHRK21} introduce a metric called \enquote{constraint violation} specifically for contrastive explanations.

\section{Limitations and Future Work}
\label{sec:reasonx_limitations}

Let us discuss here the limitations and future work of our approach regarding its evaluation, technical aspects, the user-perspective, and social assumptions.

\paragraph*{Evaluation.}

From the theoretical side, {\reasonx{}} advances other methods in qualitative terms {(cf. Table~\ref{tab:reasonx_vs_other_methods}). 
} 
However, further work is needed to verify this also from a practical side~\citep{DBLP:conf/ijcai/Byrne23}: how the tool is perceived by expert or lay users, how easy it is to encode and reason about domain knowledge as constraints, and how the approach scales in presence of noise and non-separable classes. This validation can be conducted through user studies (see \cite{DBLP:journals/pami/RongLNFQUSKK24} for a survey specifically on XAI) and through real-world case studies. 
Regarding the application setting of this paper, namely the credit domain, some reference literature on credit scoring and ML/XAI and can be found 
at~\cite{DBLP:journals/jors/BuckerSGB22,demajo_explainable_2020,DBLP:journals/nca/ShiTLD022}. We further re-emphasize that the cases presented in this paper are constructed and do not represent real-world conditions.

\paragraph*{Technical.}
We demonstrated the capabilities of {\reasonx{}} on 
tabular data, and for binary decision problems. While an extension to multi-class problems is straightforward, the extension to unstructured data, such as text and images, {is non-trivial and} needs to be strictly formalized.
In the domain of images, a solution can be the integration of concepts, as shown, e.g.,~by~\citet{DBLP:conf/aiia/DonadelloD20}.
Moreover, the extension of the approach to other constraint (logic programming) schemes beyond linear constraints over the reals is a relevant direction, including, for instance, finite domains, non-linear constraints, or probabilistic/fuzzy logic programming. The literature on such extensions is large~\citep{DBLP:conf/ijcai/MorettinMKP21,DBLP:journals/tplp/KornerLBCDHMWDA22}. To adapt our framework to such extensions, the key requirement is to preserve the closedness property of the extended language, ensuring that the optimization and reasoning capabilities remain 
tractable.

\paragraph*{User-centered.}
We aim at extending {\reasonx{}} along the perspective of the Human-XAI interaction \citep{DBLP:conf/interact/ChromikB21}.
First, by improving the Python layer through the integration of a plotting functionality (to produce similar plots as in Section~\nameref{sec:reasonx_synthetic_dataset}), and  additional syntactic sugar functionalities, e.g.,~to express a class label of the form \enquote{not class X} for the generation of contrastive rules.
Second, by adding a natural language layer on top of the Python layer for an easier communication with users (see also next section).
Third, by performing user studies testing the quality of {the interaction between users the system, including potential risks associated with XAI}, such as automation bias~\citep{DBLP:journals/ai/VeredLHMS23}.


\paragraph*{Social.}

Two basic assumptions of {\reasonx{}} must be taken care of in socially-sensitive applications, where contrastive rules might be used as recommendations to change the decision (a.k.a.~for \textit{algorithmic recourse}). 
%
%
First, we assume that the ML model is \textit{stable over time}, which implies that a contrastive rule may change the model's decision in future applications.
%
While this might hold in a toy setting, it likely does not in practice and can create some wrong promises~\citep{DBLP:conf/fat/BarocasSR20,DBLP:conf/iclp/State21}.
Using {\reasonx{}} to retrospectively reason over time is a first step in the correct direction, but we cannot know how a ML model will evolve in the future.
Second, by querying the minimal contrastive rules, there is the implicit assumption that the closer the result to the instance under analysis, the easier the proposed change. This does not necessarily hold in practice, as pointed out already~\citep{DBLP:journals/csur/KarimiBSV23}.
While an extension of {\reasonx{}} to account for a variety of norms in the optimization, i.e.,~beyond the 
$L_1$~norm and the $L_{\infty}$~norm, is possible and part of future work, we acknowledge the importance of the choice of a suitable distance function, depending on the application context.
{Lastly, we point to the inherent tension between transparency and privacy that needs to be considered when using XAI methods in practice~\citep{DBLP:journals/ci/FerryAGHS25}.}

\section{Concluding Remarks}
\label{sec:reasonx_conclusion}

We introduced {\reasonx{}} (\textit{reason to explain}), a declarative explanation tool grounded on constraint logic programming.
{\reasonx{}} explains decision tree classifiers and, via {high-fidelity} surrogate trees, any tabular-data ML classifier at both the local and global level.
The tool generates explanations in a declarative and computationally principled manner, integrating user requirements and symbolic knowledge encoded in the form of linear constraints. Moreover, {\reasonx{}} supports higher-level reasoning capabilities, such as reasoning over time and across model types, and can operate on data profiles rather than individual instances in under-specified information settings.
{\reasonx{}} produces interpretable explanations in the form of decision rules or contrastive examples, built upon a well-defined algebra of operators, and supports interactive exploration to foster enhanced communication with the user. This design makes {\reasonx{}} not only a versatile and practically useful tool but also one that explicitly centers the user and her needs.

Finally, we present a fictional dialog between {\reasonx{}} and a user, designed to illustrate how the system could evolve through the integration of large language models (LLMs) to bridge natural and formal languages -- an emerging direction within the rapidly growing field of ``LLMs for XAI" (e.g.,~\citet{DBLP:journals/corr/abs-2405-06064,DBLP:journals/corr/abs-2401-13110,DBLP:conf/extraamas/XuCDD23}). Although {\reasonx{}} currently lacks natural language processing capabilities, the explanatory content presented in the dialog faithfully reflects the types of computations that the tool can already perform. Determining how to best adapt and fine-tune LLMs to improve natural language explanation quality, however, remains open for future research~\citep{DBLP:journals/corr/abs-2601-02224}.
The dialog is between {\reasonx{}} and a developer as a user of the tool. The questions aim to provide a first understanding of the model, which is again situated in the credit domain.

\begin{mdframed}\small
\texttt{USER}: I have two ML models that perform similarly well on my dataset. For a single data instance, what is its classification and the intersection of the factual rules between these two models?\\
\texttt{REASONX}: Both models assign the same label to the instance (credit denial). {There is a minimal intersection between these rules that contains the instance in focus.} \\
\texttt{USER}: Is there a sensitive attribute appearing in the factual rules? \\
\texttt{REASONX}: No, no sensitive attribute appears in the factual rules. \\
\texttt{USER}: Is there an intersection between the contrastive rules of the two models? \\
\texttt{REASONX}: Yes, two of the constrative rules intersect. The intersection is given by the following new rule: if there was no lease on the car and if the credit requested was between $8,000$~EUR and $10,000$~EUR, the credit would have been granted in both models.\\
\texttt{USER}: Is there also a contrastive instance that is valid in both models? \\
\texttt{REASONX}: Yes, there is one contrastive instance. The changed features compared to the factual data instance are: a credit amount of $8,500$~EUR and no lease of a car.
\end{mdframed}


\begin{acks}
This work has received funding from the European Union’s Horizon 2020 research and innovation program under Marie Sklodowska-Curie Actions (g.a. number 860630) for the project ``NoBIAS - Artificial Intelligence without Bias'',
and by PNRR - M4C2 - Investimento 1.3, Partenariato Esteso PE00000013 - ``FAIR - Future Artificial Intelligence Research'' - Spoke~1 ``Human-centered AI'', funded by the European Commission under the NextGeneration EU programme.
Views and opinions expressed are however those of the authors only and do not necessarily reflect those of the EU. Neither the EU nor the granting authority can be held responsible for them.
\end{acks}



\bibliographystyle{apalike}
\bibliography{ref.bib}

@article{DBLP:journals/corr/abs-2601-02224,
  author       = {Fabian Lukassen and
                  Jan Herrmann and
                  Christoph Weisser and
                  Benjamin S{\"{a}}fken and
                  Thomas Kneib},
  title        = {From {XAI} to Stories: {A} Factorial Study of {LLM}-Generated Explanation
                  Quality},
  journal      = {CoRR},
  volume       = {abs/2601.02224},
  year         = {2026}
}

@article{DBLP:journals/ci/FerryAGHS25,
  author       = {Julien Ferry and
                  Ulrich A{\"{\i}}vodji and
                  S{\'{e}}bastien Gambs and
                  Marie{-}Jos{\'{e}} Huguet and
                  Mohamed Siala},
  title        = {Taming the Triangle: On the Interplays Between Fairness, Interpretability,
                  and Privacy in Machine Learning},
  journal      = {Comput. Intell.},
  volume       = {41},
  number       = {4},
  year         = {2025}
}

@article{DBLP:journals/tplp/TakemuraI24,
  author       = {Akihiro Takemura and
                  Katsumi Inoue},
  title        = {Generating Global and Local Explanations for Tree-Ensemble Learning
                  Methods by Answer Set Programming},
  journal      = {Theory Pract. Log. Program.},
  volume       = {24},
  number       = {5},
  pages        = {973--1010},
  year         = {2024}
}

@article{DBLP:journals/ijon/MershaLWAK24,
  author       = {Melkamu Mersha and
                  Khang Nhut Lam and
                  Joseph Wood and
                  Ali K. AlShami and
                  Jugal Kalita},
  title        = {Explainable artificial intelligence: {A} survey of needs, techniques,
                  applications, and future direction},
  journal      = {Neurocomputing},
  volume       = {599},
  pages        = {128111},
  year         = {2024},
  url          = {https://doi.org/10.1016/j.neucom.2024.128111},
  doi          = {10.1016/J.NEUCOM.2024.128111},
  bibsource    = {dblp computer science bibliography, https://dblp.org}
}

@article{DBLP:journals/nca/BhuyanRTS24,
  author       = {Bikram Pratim Bhuyan and
                  Amar Ramdane{-}Cherif and
                  Ravi Tomar and
                  T. P. Singh},
  title        = {Neuro-symbolic artificial intelligence: a survey},
  journal      = {Neural Comput. Appl.},
  volume       = {36},
  number       = {21},
  pages        = {12809--12844},
  year         = {2024},
  url          = {https://doi.org/10.1007/s00521-024-09960-z},
  doi          = {10.1007/S00521-024-09960-Z},
  bibsource    = {dblp computer science bibliography, https://dblp.org}
}

@article{DBLP:journals/tnn/RamachandranpillaiBH25,
  author       = {Resmi Ramachandranpillai and
                  Ricardo Baeza{-}Yates and
                  Fredrik Heintz},
  title        = {Fair{XAI} - {A} Taxonomy and Framework for Fairness and Explainability
                  Synergy in Machine Learning},
  journal      = {{IEEE} Trans. Neural Networks Learn. Syst.},
  volume       = {36},
  number       = {6},
  pages        = {9819--9836},
  year         = {2025}
}

@inproceedings{DBLP:conf/aaai/IzzaIS024,
  author       = {Yacine Izza and
                  Alexey Ignatiev and
                  Peter J. Stuckey and
                  Jo{\~{a}}o Marques{-}Silva},
  title        = {Delivering Inflated Explanations},
  booktitle    = {{AAAI}},
  pages        = {12744--12753},
  publisher    = {{AAAI} Press},
  year         = {2024}
}

@inproceedings{DBLP:conf/aies/MouganCRS23,
  author       = {Carlos Mougan and
                  Jos{\'{e}} M. {\'{A}}lvarez and
                  Salvatore Ruggieri and
                  Steffen Staab},
  title        = {Fairness Implications of Encoding Protected Categorical Attributes},
  booktitle    = {{AIES}},
  pages        = {454--465},
  publisher    = {{ACM}},
  year         = {2023}
}

@article{DBLP:journals/dss/MalandriMMS24,
  author       = {Lorenzo Malandri and
                  Fabio Mercorio and
                  Mario Mezzanzanica and
                  Andrea Seveso},
  title        = {Model-contrastive explanations through symbolic reasoning},
  journal      = {Decis. Support Syst.},
  volume       = {176},
  pages        = {114040},
  year         = {2024}
}

@article{Sabbatini2025,
author = {Sabbatini, Federico},
title = {Four Decades of Symbolic Knowledge Extraction from Sub-Symbolic Predictors. A Survey},
year = {2025},
volume = 58,
number = 3,
pages = "61",
journal = {ACM Comput. Surv.}
}

@inproceedings{DBLP:conf/ijcai/AudemardLMS24a,
  author       = {Gilles Audemard and
                  Jean{-}Marie Lagniez and
                  Pierre Marquis and
                  Nicolas Szczepanski},
  title        = {Deriving Provably Correct Explanations for Decision Trees: The Impact
                  of Domain Theories},
  booktitle    = {{IJCAI}},
  pages        = {3688--3696},
  publisher    = {ijcai.org},
  year         = {2024}
}

@book{fra-handbook-discrimination,
    title = {Handbook on European non-discrimination law},
    author = {{European Union Agency for Fundamental Rights}},
    publisher = {Publications Office of the European Union},
    year = {2018}
}

@article{DBLP:journals/ml/GoethalsMC24,
  author       = {Sofie Goethals and
                  David Martens and
                  Toon Calders},
  title        = {{PreCoF}: counterfactual explanations for fairness},
  journal      = {Mach. Learn.},
  volume       = {113},
  number       = {5},
  pages        = {3111--3142},
  year         = {2024}
}

@article{DBLP:journals/ai/CooperS23,
  author       = {Martin C. Cooper and
                  Jo{\~{a}}o Marques{-}Silva},
  title        = {Tractability of explaining classifier decisions},
  journal      = {Artif. Intell.},
  volume       = {316},
  pages        = {103841},
  year         = {2023}
}

@inproceedings{DBLP:conf/ijcnn/FragkathoulasP25,
  author       = {Christos Fragkathoulas and
                  Evaggelia Pitoura},
  title        = {{UGCE:} User-Guided Incremental Counterfactual Exploration},
  booktitle    = {{IJCNN}},
  pages        = {1--8},
  publisher    = {{IEEE}},
  year         = {2025}
}

@inproceedings{DBLP:conf/ijcai/JiangL0T24,
  author       = {Junqi Jiang and
                  Francesco Leofante and
                  Antonio Rago and
                  Francesca Toni},
  title        = {Robust Counterfactual Explanations in Machine Learning: {A} Survey},
  booktitle    = {{IJCAI}},
  pages        = {8086--8094},
  publisher    = {ijcai.org},
  year         = {2024}
}

@inproceedings{DBLP:conf/ijcai/AlfanoGL0T25,
  author       = {Gianvincenzo Alfano and
                  Adam Gould and
                  Francesco Leofante and
                  Antonio Rago and
                  Francesca Toni},
  title        = {Counterfactual Explanations Under Model Multiplicity and Their Use
                  in Computational Argumentation},
  booktitle    = {{IJCAI}},
  pages        = {4321--4329},
  publisher    = {ijcai.org},
  year         = {2025}
}

@inproceedings{DBLP:conf/icml/BewleyAMMV24,
  author       = {Tom Bewley and
                  Salim I. Amoukou and
                  Saumitra Mishra and
                  Daniele Magazzeni and
                  Manuela Veloso},
  title        = {Counterfactual Metarules for Local and Global Recourse},
  booktitle    = {{ICML}},
  series       = {Proceedings of Machine Learning Research},
  volume       = {235},
  pages        = {3707--3724},
  publisher    = {{PMLR} / OpenReview.net},
  year         = {2024}
}

@incollection{Cook2009,
    title={Intensional Definition},
author={Roy T. Cook},
	booktitle = {A Dictionary of Philosophical Logic},
publisher = {Edinburgh University Press},
	year = {2009},
   pages = {155}
}

@incollection{Kristina2025,
    title={Unjustified untrue ``beliefs'': {AI} hallucinations and justification logics},
author={Kristina {\v{S}}ekrst},
	editor = {Kordula \'{S}wi\k{e}torzecka and Filip Grgi\'c and Anna Brozek},
	booktitle = {Logic, Knowledge, and Tradition. Essays in Honor of Srecko Kovac},
    series = {Poznań Studies in the Philosophy of the Sciences and the Humanities},
    volume = 127,
    publisher = {Brill Academic Publishers},
	year = {2026}
}

@inproceedings{DBLP:conf/interact/ChromikB21,
  author       = {Michael Chromik and
                  Andreas Butz},
  title        = {Human-XAI Interaction: {A} Review and Design Principles for Explanation
                  User Interfaces},
  booktitle    = {{INTERACT} {(2)}},
  series       = {Lecture Notes in Computer Science},
  volume       = {12933},
  pages        = {619--640},
  publisher    = {Springer},
  year         = {2021}
}

@article{DBLP:journals/pami/RongLNFQUSKK24,
  author       = {Yao Rong and
                  Tobias Leemann and
                  Thai{-}trang Nguyen and
                  Lisa Fiedler and
                  Peizhu Qian and
                  Vaibhav V. Unhelkar and
                  Tina Seidel and
                  Gjergji Kasneci and
                  Enkelejda Kasneci},
  title        = {Towards Human-Centered Explainable {AI:} {A} Survey of User Studies
                  for Model Explanations},
  journal      = {{IEEE} Trans. Pattern Anal. Mach. Intell.},
  volume       = {46},
  number       = {4},
  pages        = {2104--2122},
  year         = {2024}
}

@inproceedings{DBLP:conf/coco/Karp72,
  author       = {Richard M. Karp},
  title        = {Reducibility Among Combinatorial Problems},
  booktitle    = {Complexity of Computer Computations},
  series       = {The {IBM} Research Symposia Series},
  pages        = {85--103},
  publisher    = {Plenum Press, New York},
  year         = {1972}
}

@inproceedings{DBLP:conf/rweb/Silva22,
  author       = {Jo{\~{a}}o Marques{-}Silva},
  title        = {Logic-Based Explainability in Machine Learning},
  booktitle    = {{RW}},
  series       = {Lecture Notes in Computer Science},
  volume       = {13759},
  pages        = {24--104},
  publisher    = {Springer},
  year         = {2022}
}

@inproceedings{DBLP:conf/clar/LuoSD23,
  author       = {Jieting Luo and
                  Thomas Studer and
                  Mehdi Dastani},
  title        = {Providing Personalized Explanations: {A} Conversational Approach},
  booktitle    = {{CLAR}},
  series       = {Lecture Notes in Computer Science},
  volume       = {14156},
  pages        = {121--137},
  publisher    = {Springer},
  year         = {2023}
}

@book{Artemov2019,
  author = "Sergei Artemov and Melvin Fitting",
  title = "Justification Logic: {R}easoning with Reasons",
  publisher = " Cambridge University Press",
  year = "2019"
}

@book{dbscb2008,
  author = "Hector Garcia-Molina and Jeffrey D. Ullman  and Jennifer Widom",
  title = "Database Systems: {T}he Complete Book",
  publisher = "Pearson College",
  edition = {2},
  year = "2008"
}

@article{DBLP:journals/eswa/CarrizosaRM24,
  author       = {Emilio Carrizosa and
                  Jasone Ram{\'{\i}}rez{-}Ayerbe and
                  Dolores Romero Morales},
  title        = {Generating collective counterfactual explanations in score-based classification
                  via mathematical optimization},
  journal      = {Expert Syst. Appl.},
  volume       = {238},
  number       = {Part {E}},
  pages        = {121954},
  year         = {2024}
}

@inproceedings{DBLP:conf/iccbr/WarrenDGK24,
  author       = {Greta Warren and
                  Eoin Delaney and
                  Christophe Gu{\'{e}}ret and
                  Mark T. Keane},
  title        = {Explaining Multiple Instances Counterfactually:User Tests of Group-Counterfactuals
                  for {XAI}},
  booktitle    = {{ICCBR}},
  series       = {Lecture Notes in Computer Science},
  volume       = {14775},
  pages        = {206--222},
  publisher    = {Springer},
  year         = {2024}
}

@article{DBLP:journals/ai/VeredLHMS23,
  author       = {Mor Vered and
                  Tali Livni and
                  Piers Douglas Lionel Howe and
                  Tim Miller and
                  Liz Sonenberg},
  title        = {The effects of explanations on automation bias},
  journal      = {Artif. Intell.},
  volume       = {322},
  pages        = {103952},
  year         = {2023}
}

@inproceedings{DBLP:conf/ijcai/MorettinMKP21,
  author       = {Paolo Morettin and
                  Pedro Zuidberg Dos Martires and
                  Samuel Kolb and
                  Andrea Passerini},
  title        = {Hybrid Probabilistic Inference with Logical and Algebraic Constraints:
                  a Survey},
  booktitle    = {{IJCAI}},
  pages        = {4533--4542},
  publisher    = {ijcai.org},
  year         = {2021}
}

@inproceedings{DBLP:conf/gecco/BarbosaSF24,
  author       = {Pedro Barbosa and
                  Rosina Savisaar and
                  Alcides Fonseca},
  title        = {Semantically Rich Local Dataset Generation for Explainable {AI} in
                  Genomics},
  booktitle    = {{GECCO}},
  publisher    = {{ACM}},
  year         = {2024}
}

@inproceedings{DBLP:conf/ijcai/AmgoudMT23,
  author       = {Leila Amgoud and
                  Philippe Muller and
                  Henri Trenquier},
  title        = {Leveraging Argumentation for Generating Robust Sample-based Explanations},
  booktitle    = {{IJCAI}},
  pages        = {3104--3111},
  publisher    = {ijcai.org},
  year         = {2023}
}

@inproceedings{DBLP:conf/ecai/AudemardLMS23,
  author       = {Gilles Audemard and
                  Jean{-}Marie Lagniez and
                  Pierre Marquis and
                  Nicolas Szczepanski},
  title        = {On Contrastive Explanations for Tree-Based Classifiers},
  booktitle    = {{ECAI}},
  series       = {Frontiers in Artificial Intelligence and Applications},
  volume       = {372},
  pages        = {117--124},
  publisher    = {{IOS} Press},
  year         = {2023}
}

@article{DBLP:journals/dke/AudemardBBKLM22,
  author       = {Gilles Audemard and
                  Steve Bellart and
                  Louenas Bounia and
                  Fr{\'{e}}d{\'{e}}ric Koriche and
                  Jean{-}Marie Lagniez and
                  Pierre Marquis},
  title        = {On the explanatory power of Boolean decision trees},
  journal      = {Data Knowl. Eng.},
  volume       = {142},
  pages        = {102088},
  year         = {2022}
}

@article{DBLP:journals/jair/IzzaIM22,
  author       = {Yacine Izza and
                  Alexey Ignatiev and
                  Jo{\~{a}}o Marques{-}Silva},
  title        = {On Tackling Explanation Redundancy in Decision Trees},
  journal      = {J. Artif. Intell. Res.},
  volume       = {75},
  pages        = {261--321},
  year         = {2022}
}

@inproceedings{DBLP:conf/ifip12/FerreiraCB22,
  author       = {Caique Augusto Ferreira and
                  Adriano Henrique Cant{\~{a}}o and
                  Jos{\'{e}} Augusto Baranauskas},
  title        = {Decision Tree Induction Through Meta-learning},
  booktitle    = {{AIAI} {(2)}},
  series       = {{IFIP} Advances in Information and Communication Technology},
  volume       = {647},
  pages        = {101--111},
  publisher    = {Springer},
  year         = {2022}
}

@article{DBLP:journals/ai/WaaNCN21,
  author       = {Jasper van der Waa and
                  Elisabeth Nieuwburg and
                  Anita H. M. Cremers and
                  Mark A. Neerincx},
  title        = {Evaluating {XAI:} {A} comparison of rule-based and example-based explanations},
  journal      = {Artif. Intell.},
  volume       = {291},
  pages        = {103404},
  year         = {2021}
}

@article{DBLP:journals/pacmhci/PiorkowskiVCDA23,
  author       = {David Piorkowski and
                  Inge Vejsbjerg and
                  Owen Cornec and
                  Elizabeth M. Daly and
                  {\"{O}}znur Alkan},
  title        = {{AIMEE:} An Exploratory Study of How Rules Support {AI} Developers
                  to Explain and Edit Models},
  journal      = {Proc. {ACM} Hum. Comput. Interact.},
  volume       = {7},
  number       = {{CSCW2}},
  pages        = {1--25},
  year         = {2023}
}

@inproceedings{DBLP:conf/iclr/LeeJ20,
  author       = {Guang{-}He Lee and
                  Tommi S. Jaakkola},
  title        = {Oblique Decision Trees from Derivatives of {ReLU} Networks},
  booktitle    = {{ICLR}},
  publisher    = {OpenReview.net},
  year         = {2020}
}

@inproceedings{DBLP:conf/extraamas/XuCDD23,
  author       = {Yifan Xu and
                  Joe Collenette and
                  Louise A. Dennis and
                  Clare Dixon},
  title        = {Dialogue Explanations for Rule-Based {AI} Systems},
  booktitle    = {{EXTRAAMAS}},
  series       = {Lecture Notes in Computer Science},
  volume       = {14127},
  pages        = {59--77},
  publisher    = {Springer},
  year         = {2023}
}

@inproceedings{DBLP:journals/corr/abs-2401-13110,
  author       = {Philip Mavrepis and
                  Georgios Makridis and
                  Georgios Fatouros and
                  Vasileios Koukos and
                  Spyros Theodoropoulos and
                  Maria Margarita Separdani and
                  Dimosthenis Kyriazis},
  title        = {{XAI} for All: {C}an Large Language Models Simplify Explainable {AI}?},
  booktitle    = {{NLAI}},
  series       = {CS \& IT},
  volume       = {14},
  issue = 22,
  pages        = {127--144},
  publisher    = {AIRCC},
  year         = {2024}
}

@article{DBLP:journals/corr/abs-2405-06064,
  author       = {Alexandra Zytek and
                  Sara Pid{\`{o}} and
                  Kalyan Veeramachaneni},
  title        = {{LLM}s for {XAI:} Future Directions for Explaining Explanations},
  journal      = {CoRR},
  volume       = {abs/2405.06064},
  year         = {2024}
}

@article{DBLP:journals/csur/KarimiBSV23,
  author       = {Amir{-}Hossein Karimi and
                  Gilles Barthe and
                  Bernhard Sch{\"{o}}lkopf and
                  Isabel Valera},
  title        = {A Survey of Algorithmic Recourse: Contrastive Explanations and Consequential
                  Recommendations},
  journal      = {{ACM} Comput. Surv.},
  volume       = {55},
  number       = {5},
  pages        = {95:1--95:29},
  year         = {2023}
}

@inproceedings{DBLP:conf/fat/LaugelJLMD23,
  author       = {Thibault Laugel and
                  Adulam Jeyasothy and
                  Marie{-}Jeanne Lesot and
                  Christophe Marsala and
                  Marcin Detyniecki},
  title        = {Achieving Diversity in Counterfactual Explanations: a Review and Discussion},
  booktitle    = {FAccT},
  pages        = {1859--1869},
  publisher    = {{ACM}},
  year         = {2023}
}

@article{LOREsa,
  author       = {Riccardo Guidotti and
                  Anna Monreale and
                  Salvatore Ruggieri and
                  Francesca Naretto and
                  Franco Turini and
                  Dino Pedreschi and
                  Fosca Giannotti},
  title        = {Stable and actionable explanations of black-box models through factual
                  and counterfactual rules},
  journal      = {Data Min. Knowl. Discov.},
  volume       = {38},
  number       = {5},
  pages        = {2825--2862},
  year         = {2024}
}

@article{Rudin2019_StopExplainingML,
  author       = {Cynthia Rudin},
  title        = {Stop explaining black box machine learning models for high stakes
                  decisions and use interpretable models instead},
  journal      = {Nat. Mach. Intell.},
  volume       = {1},
  number       = {5},
  pages        = {206--215},
  year         = {2019}
}

@incollection{DBLP:books/sp/datamining2005/RokachM05,
  author       = {Lior Rokach and
                  Oded Maimon},
  editor       = {Oded Maimon and
                  Lior Rokach},
  title        = {Decision Trees},
  booktitle    = {The {D}ata {M}ining and {K}nowledge {D}iscovery {H}andbook},
  pages        = {165--192},
  publisher    = {Springer},
  year         = {2005}
}

@inproceedings{DBLP:conf/ijcai/DudyrevK21,
  author       = {Egor Dudyrev and
                  Sergei O. Kuznetsov},
  title        = {Summation of Decision Trees},
  booktitle    = {FCA4AI@IJCAI},
  series       = {{CEUR} Workshop Proceedings},
  volume       = {2972},
  pages        = {99--104},
  publisher    = {CEUR-WS.org},
  year         = {2021}
}

@article{DBLP:journals/jbd/WeinbergL19,
  author       = {Abraham Itzhak Weinberg and
                  Mark Last},
  title        = {Selecting a representative decision tree from an ensemble of decision-tree
                  models for fast big data classification},
  journal      = {J. Big Data},
  volume       = {6},
  pages        = {23},
  year         = {2019}
}

@inproceedings{DBLP:conf/dis/BonsignoriGM21,
  author       = {Valerio Bonsignori and
                  Riccardo Guidotti and
                  Anna Monreale},
  title        = {Deriving a Single Interpretable Model by Merging Tree-Based Classifiers},
  booktitle    = {{DS}},
  series       = {Lecture Notes in Computer Science},
  volume       = {12986},
  pages        = {347--357},
  publisher    = {Springer},
  year         = {2021}
}

@inproceedings{DBLP:conf/icml/VidalS20,
  author       = {Thibaut Vidal and
                  Maximilian Schiffer},
  title        = {Born-Again Tree Ensembles},
  booktitle    = {{ICML}},
  series       = {Proceedings of Machine Learning Research},
  volume       = {119},
  pages        = {9743--9753},
  publisher    = {{PMLR}},
  year         = {2020}
}

@article{DBLP:journals/air/CostaP23,
  author       = {Vin{\'{\i}}cius G. Costa and
                  Carlos Eduardo Pedreira},
  title        = {Recent advances in decision trees: an updated survey},
  journal      = {Artif. Intell. Rev.},
  volume       = {56},
  number       = {5},
  pages        = {4765--4800},
  year         = {2023}
}

@inproceedings{DBLP:conf/vmcai/MarechalP17,
  author       = {Alexandre Mar{\'{e}}chal and
                  Micha{\"{e}}l P{\'{e}}rin},
  title        = {Efficient Elimination of Redundancies in Polyhedra by Raytracing},
  booktitle    = {{VMCAI}},
  series       = {Lecture Notes in Computer Science},
  volume       = {10145},
  pages        = {367--385},
  publisher    = {Springer},
  year         = {2017}
}

@inproceedings{DBLP:conf/fat/KarimiSV21,
  author       = {Amir{-}Hossein Karimi and
                  Bernhard Sch{\"{o}}lkopf and
                  Isabel Valera},
  title        = {Algorithmic Recourse: From Counterfactual Explanations to Interventions},
  booktitle    = {FAccT},
  pages        = {353--362},
  publisher    = {{ACM}},
  year         = {2021}
}

@inproceedings{DBLP:conf/fat/UstunSL19,
  author       = {Berk Ustun and
                  Alexander Spangher and
                  Yang Liu},
  title        = {Actionable Recourse in Linear Classification},
  booktitle    = {{FAT}},
  pages        = {10--19},
  publisher    = {{ACM}},
  year         = {2019}
}

@inproceedings{DBLP:conf/fat/MothilalST20,
  author       = {Ramaravind Kommiya Mothilal and
                  Amit Sharma and
                  Chenhao Tan},
  title        = {Explaining machine learning classifiers through diverse counterfactual
                  explanations},
  booktitle    = {FAT*},
  pages        = {607--617},
  publisher    = {{ACM}},
  year         = {2020}
}

@article{DBLP:journals/csur/GuidottiMRTGP19,
  author       = {Riccardo Guidotti and
                  Anna Monreale and
                  Salvatore Ruggieri and
                  Franco Turini and
                  Fosca Giannotti and
                  Dino Pedreschi},
  title        = {A Survey of Methods for Explaining Black Box Models},
  journal      = {{ACM} Comput. Surv.},
  volume       = {51},
  number       = {5},
  pages        = {93:1--93:42},
  year         = {2019}
}

@inproceedings{DBLP:conf/kdd/Ribeiro0G16,
  author       = {Marco T{\'{u}}lio Ribeiro and
                  Sameer Singh and
                  Carlos Guestrin},
  title        = {``{W}hy Should {I} Trust You?'': Explaining the Predictions of Any Classifier},
  booktitle    = {{KDD}},
  pages        = {1135--1144},
  publisher    = {{ACM}},
  year         = {2016}
}

@inproceedings{DBLP:conf/xai/StateRT23,
  author       = {Laura State and
                  Salvatore Ruggieri and
                  Franco Turini},
  title        = {Reason to Explain: Interactive Contrastive Explanations {(REASONX)}},
  booktitle    = {xAI {(1)}},
  series       = {Communications in Computer and Information Science},
  volume       = {1901},
  pages        = {421--437},
  publisher    = {Springer},
  year         = {2023}
}

@article{DBLP:journals/ai/Miller19,
  author       = {Tim Miller},
  title        = {Explanation in artificial intelligence: Insights from the social sciences},
  journal      = {Artif. Intell.},
  volume       = {267},
  pages        = {1--38},
  year         = {2019}
}

@inproceedings{DBLP:conf/aiia/DonadelloD20,
  author       = {Ivan Donadello and
                  Mauro Dragoni},
  title        = {{SeXAI}: Introducing Concepts into Black Boxes for Explainable {A}rtificial
                  {I}ntelligence},
  booktitle    = {{XAI.it@AI*IA}},
  series       = {{CEUR} Workshop Proceedings},
  volume       = {2742},
  pages        = {41--54},
  publisher    = {CEUR-WS.org},
  year         = {2020}
}

@article{DBLP:journals/corr/abs-1711-00399,
  title={Counterfactual Explanations without Opening the Black Box},
  author={Wachter, Sandra and others},
  journal={Harv. JL \& Tech.},
  volume={31},
  pages={841},
  year={2017},
  publisher={HeinOnline}
}

@inproceedings{DBLP:conf/ijcai/Byrne23,
  author       = {Ruth M. J. Byrne},
  title        = {Good Explanations in Explainable Artificial Intelligence {(XAI):}
                  Evidence from Human Explanatory Reasoning},
  booktitle    = {{IJCAI}},
  pages        = {6536--6544},
  publisher    = {ijcai.org},
  year         = {2023}
}

@inproceedings{DBLP:conf/ijcai/KeaneKDS21,
  author       = {Mark T. Keane and
                  Eoin M. Kenny and
                  Eoin Delaney and
                  Barry Smyth},
  title        = {If Only We Had Better Counterfactual Explanations: Five Key Deficits
                  to Rectify in the Evaluation of Counterfactual {XAI} Techniques},
  booktitle    = {{IJCAI}},
  pages        = {4466--4474},
  publisher    = {ijcai.org},
  year         = {2021}
}

@inproceedings{DBLP:conf/aistats/KarimiBBV20,
  author       = {Amir{-}Hossein Karimi and
                  Gilles Barthe and
                  Borja Balle and
                  Isabel Valera},
  title        = {Model-Agnostic Counterfactual Explanations for Consequential Decisions},
  booktitle    = {{AISTATS}},
  series       = {Proceedings of Machine Learning Research},
  volume       = {108},
  pages        = {895--905},
  publisher    = {{PMLR}},
  year         = {2020}
}

@inproceedings{DBLP:conf/ijcai/KanamoriTKA20,
  author    = {Kentaro Kanamori and
               Takuya Takagi and
               Ken Kobayashi and
               Hiroki Arimura},
  title     = {{DACE:} Distribution-Aware Counterfactual Explanation by Mixed-Integer
               Linear Optimization},
  booktitle = {{IJCAI}},
  pages     = {2855--2862},
  publisher = {ijcai.org},
  year      = {2020}
}

@inproceedings{DBLP:conf/fat/BarocasSR20,
  author    = {Solon Barocas and
               Andrew D. Selbst and
               Manish Raghavan},
  title     = {The hidden assumptions behind counterfactual explanations and principal
               reasons},
  booktitle = {FAT*},
  pages     = {80--89},
  publisher = {{ACM}},
  year      = {2020}
}

@inproceedings{DBLP:conf/iclp/State21,
  author    = {Laura State},
  title     = {Logic Programming for {XAI:} {A} Technical Perspective},
  booktitle = {{ICLP} Workshops},
  series    = {{CEUR} Workshop Proceedings},
  volume    = {2970},
  publisher = {CEUR-WS.org},
  year      = {2021}
}

@inproceedings{DBLP:journals/corr/abs-2105-10172,
  author       = {Katharina Beckh and
                  Sebastian M{\"{u}}ller and
                  Matthias Jakobs and
                  Vanessa Toborek and
                  Hanxiao Tan and
                  Raphael Fischer and
                  Pascal Welke and
                  Sebastian Houben and
                  Laura von R{\"{u}}den},
  title        = {Harnessing Prior Knowledge for Explainable Machine Learning: An Overview},
  booktitle    = {SaTML},
  pages        = {450--463},
  publisher    = {{IEEE}},
  year         = {2023}
}

@article{DBLP:journals/access/StepinACP21,
  author       = {Ilia Stepin and
                  Jos{\'{e}} Maria Alonso and
                  Alejandro Catal{\'{a}} and
                  Martin Pereira{-}Fari{\~{n}}a},
  title        = {A Survey of Contrastive and Counterfactual Explanation Generation
                  Methods for Explainable Artificial Intelligence},
  journal      = {{IEEE} Access},
  volume       = {9},
  pages        = {11974--12001},
  year         = {2021}
}

@Article{Guidotti2022,
  author       = {Riccardo Guidotti},
  title        = {Counterfactual explanations and how to find them: literature review
                  and benchmarking},
  journal      = {Data Min. Knowl. Discov.},
  volume       = {38},
  number       = {5},
  pages        = {2770--2824},
  year         = {2024}
}

@inproceedings{DBLP:conf/fat/PaniguttiPP20,
  author       = {Cecilia Panigutti and
                  Alan Perotti and
                  Dino Pedreschi},
  title        = {Doctor {XAI:} an ontology-based approach to black-box sequential data
                  classification explanations},
  booktitle    = {FAT*},
  pages        = {629--639},
  publisher    = {{ACM}},
  year         = {2020}
}

@article{DBLP:journals/tkdd/RuggieriPT10,
  author    = {Salvatore Ruggieri and
               Dino Pedreschi and
               Franco Turini},
  title     = {Data mining for discrimination discovery},
  journal   = {{ACM} Trans. Knowl. Discov. Data},
  volume    = {4},
  number    = {2},
  pages     = {9:1--9:40},
  year      = {2010}
}

@article{DBLP:journals/ki/SokolF20,
  author       = {Kacper Sokol and
                  Peter A. Flach},
  title        = {One Explanation Does Not Fit All},
  journal      = {K{\"{u}}nstliche Intell.},
  volume       = {34},
  number       = {2},
  pages        = {235--250},
  year         = {2020}
}

@inproceedings{DBLP:conf/ijcai/SokolF18a,
  author       = {Kacper Sokol and
                  Peter A. Flach},
  title        = {Glass-Box: Explaining {AI} Decisions With Counterfactual Statements
                  Through Conversation With a Voice-enabled Virtual Assistant},
  booktitle    = {{IJCAI}},
  pages        = {5868--5870},
  publisher    = {ijcai.org},
  year         = {2018}
}

@article{DBLP:journals/corr/abs-2205-08974,
  author       = {Andr{\'{e}} Artelt and
                  Stelios G. Vrachimis and
                  Demetrios G. Eliades and
                  Marios M. Polycarpou and
                  Barbara Hammer},
  title        = {One Explanation to Rule them All - Ensemble Consistent Explanations},
  journal      = {CoRR},
  volume       = {abs/2205.08974},
  year         = {2022}
}

@inproceedings{DBLP:conf/nips/RawalL20,
  author       = {Kaivalya Rawal and
                  Himabindu Lakkaraju},
  title        = {Beyond Individualized Recourse: Interpretable and Interactive Summaries
                  of Actionable Recourses},
  booktitle    = {NeurIPS},
  year         = {2020}
}

@article{DBLP:journals/cacm/WeldB19,
  author       = {Daniel S. Weld and
                  Gagan Bansal},
  title        = {The challenge of crafting intelligible intelligence},
  journal      = {Commun. {ACM}},
  volume       = {62},
  number       = {6},
  pages        = {70--79},
  year         = {2019}
}

@article{DBLP:journals/corr/abs-2202-01875,
  author       = {Himabindu Lakkaraju and
                  Dylan Slack and
                  Yuxin Chen and
                  Chenhao Tan and
                  Sameer Singh},
  title        = {Rethinking Explainability as a Dialogue: A Practitioner's Perspective},
  journal      = {CoRR},
  volume       = {abs/2202.01875},
  year         = {2022}
}

@article{DBLP:journals/ia/CalegariCO20,
  author    = {Roberta Calegari and
               Giovanni Ciatto and
               Andrea Omicini},
  title     = {On the integration of symbolic and sub-symbolic techniques for {XAI:}
               {A} survey},
  journal   = {Intelligenza Artificiale},
  volume    = {14},
  number    = {1},
  pages     = {7--32},
  year      = {2020}
}

@article{DBLP:journals/tplp/WielemakerSTL12,
  author       = {Jan Wielemaker and
                  Tom Schrijvers and
                  Markus Triska and
                  Torbj{\"{o}}rn Lager},
  title        = {{SWI-P}rolog},
  journal      = {Theory Pract. Log. Program.},
  volume       = {12},
  number       = {1-2},
  pages        = {67--96},
  year         = {2012}
}

@article{nnccp,
    author = "R. Bagnara and P. M. Hill and E. Ricci and E. Zaffanella",
    title = "Not necessarily closed convex polyhedra and the double description method",
    journal = {Formal Aspects Comput.},
    volume = 17,
    number = 2,
    pages = {222-257},
    year      = {2005}
}

@article{DBLP:journals/tplp/KornerLBCDHMWDA22,
  author    = {Philipp K{\"{o}}rner and
               Michael Leuschel and
               Jo{\~{a}}o Barbosa and
               V{\'{\i}}tor Santos Costa and
               Ver{\'{o}}nica Dahl and
               Manuel V. Hermenegildo and
               Jos{\'{e}} F. Morales and
               Jan Wielemaker and
               Daniel Diaz and
               Salvador Abreu},
  title     = {Fifty Years of {P}rolog and Beyond},
  journal   = {Theory Pract. Log. Program.},
  volume    = {22},
  number    = {6},
  pages     = {776--858},
  year      = {2022}
}

@inproceedings{DBLP:journals/corr/abs-2308-15893,
  author       = {Theresa Swift and
                  Carl Andersen},
  title        = {The Janus System: Multi-paradigm Programming in {P}rolog and {P}ython},
  booktitle    = {{ICLP}},
  series       = {{EPTCS}},
  volume       = {385},
  pages        = {241--255},
  year         = {2023}
}

@article{DBLP:journals/amai/Greenberg96,
  author    = {Harvey J. Greenberg},
  title     = {Consistency, Redundancy, and Implied Equalities in Linear Systems},
  journal   = {Ann. Math. Artif. Intell.},
  volume    = {17},
  number    = {1-2},
  pages     = {37--83},
  year      = {1996}
}

@article{DBLP:journals/jlp/JaffarM94,
  author    = {Joxan Jaffar and
               Michael J. Maher},
  title     = {Constraint Logic Programming: {A} Survey},
  journal   = {J. Log. Program.},
  volume    = {19/20},
  pages     = {503--581},
  year      = {1994}
}

@article{DBLP:journals/jlp/JaffarMMS98,
  author    = {Joxan Jaffar and
               Michael J. Maher and
               Kim Marriott and
               Peter J. Stuckey},
  title     = {The Semantics of Constraint Logic Programs},
  journal   = {J. Log. Program.},
  volume    = {37},
  number    = {1-3},
  pages     = {1--46},
  year      = {1998}
}

@inproceedings{Tschantz2022_ProxyDisc,
  author    = {Michael Carl Tschantz},
  title     = {What is Proxy Discrimination?},
  booktitle = {FAccT},
  pages     = {1993--2003},
  publisher = {{ACM}},
  year      = {2022}
}

@inproceedings{DBLP:conf/icml/OatesJ97,
  author       = {Tim Oates and
                  David D. Jensen},
  title        = {The Effects of Training Set Size on Decision Tree Complexity},
  booktitle    = {{ICML}},
  pages        = {254--262},
  publisher    = {Morgan Kaufmann},
  year         = {1997}
}

@article{DBLP:journals/corr/abs-2005-01427,
  author       = {Kacper Sokol and
                  Peter A. Flach},
  title        = {{LIMEtree}: {C}onsistent and Faithful Surrogate Explanations of Multiple Classes},
  journal      = {Electronics},
  volume       = {14},
  number = {5},
  pages = {929},
  year         = {2025}
}

@phdthesis{sokol2021intelligible,
title
= {{T}owards Intelligible and Robust Surrogate Explainers:
{A} Decision Tree Perspective},
author
= {Sokol, Kacper},
school
= {School of Computer Science, Electrical and Electronic
Engineering, and Engineering Maths, University of Bristol},
year
= {2021}
}

@article{DBLP:journals/tplp/Bertossi23,
  author       = {Leopoldo E. Bertossi},
  title        = {Declarative Approaches to Counterfactual Explanations for Classification},
  journal      = {Theory Pract. Log. Program.},
  volume       = {23},
  number       = {3},
  pages        = {559--593},
  year         = {2023}
}

@article{DBLP:journals/tplp/BenoyKM05,
  author       = {Florence Benoy and
                  Andy King and
                  Fr{\'{e}}d{\'{e}}ric Mesnard},
  title        = {Computing convex hulls with a linear solver},
  journal      = {Theory Pract. Log. Program.},
  volume       = {5},
  number       = {1-2},
  pages        = {259--271},
  year         = {2005}
}

@article{DBLP:journals/expert/GuidottiMGPRT19,
  author       = {Riccardo Guidotti and
                  Anna Monreale and
                  Fosca Giannotti and
                  Dino Pedreschi and
                  Salvatore Ruggieri and
                  Franco Turini},
  title        = {Factual and Counterfactual Explanations for Black Box Decision Making},
  journal      = {{IEEE} Intell. Syst.},
  volume       = {34},
  number       = {6},
  pages        = {14--23},
  year         = {2019}
}

@article{DBLP:journals/ai/SetzuGMTPG21,
  author       = {Mattia Setzu and
                  Riccardo Guidotti and
                  Anna Monreale and
                  Franco Turini and
                  Dino Pedreschi and
                  Fosca Giannotti},
  title        = {{GLocalX} - From Local to Global Explanations of Black Box {AI} Models},
  journal      = {Artif. Intell.},
  volume       = {294},
  pages        = {103457},
  year         = {2021}
}

@inproceedings{DBLP:conf/aaai/Ribeiro0G18,
  author       = {Marco T{\'{u}}lio Ribeiro and
                  Sameer Singh and
                  Carlos Guestrin},
  title        = {Anchors: High-Precision Model-Agnostic Explanations},
  booktitle    = {{AAAI}},
  pages        = {1527--1535},
  publisher    = {{AAAI} Press},
  year         = {2018}
}

@inproceedings{DBLP:conf/cpaior/BonfiettiLM15,
  author       = {Alessio Bonfietti and
                  Michele Lombardi and
                  Michela Milano},
  title        = {Embedding Decision Trees and Random Forests in Constraint Programming},
  booktitle    = {{CPAIOR}},
  series       = {Lecture Notes in Computer Science},
  volume       = {9075},
  pages        = {74--90},
  publisher    = {Springer},
  year         = {2015}
}

@book{DBLP:books/mk/Quinlan93,
  author       = {J. Ross Quinlan},
  title        = {{C4.5:} Programs for Machine Learning},
  publisher    = {Morgan Kaufmann},
  year         = {1993}
}

@book{DBLP:books/wa/BreimanFOS84,
  author       = {Leo Breiman and
                  J. H. Friedman and
                  R. A. Olshen and
                  C. J. Stone},
  title        = {Classification and Regression Trees},
  publisher    = {Wadsworth},
  year         = {1984}
}

@article{DBLP:journals/ml/BertsimasD17,
  author       = {Dimitris Bertsimas and
                  Jack Dunn},
  title        = {Optimal classification trees},
  journal      = {Mach. Learn.},
  volume       = {106},
  number       = {7},
  pages        = {1039--1082},
  year         = {2017}
}

@article{DBLP:journals/widm/NtoutsiFGINVRTP20,
  author       = {Eirini Ntoutsi and
                  Pavlos Fafalios and
                  Ujwal Gadiraju and
                  Vasileios Iosifidis and
                  Wolfgang Nejdl and
                  Maria{-}Esther Vidal and
                  Salvatore Ruggieri and
                  Franco Turini and
                  Symeon Papadopoulos and
                  Emmanouil Krasanakis and
                  Ioannis Kompatsiaris and
                  Katharina Kinder{-}Kurlanda and
                  Claudia Wagner and
                  Fariba Karimi and
                  Miriam Fern{\'{a}}ndez and
                  Harith Alani and
                  Bettina Berendt and
                  Tina Kruegel and
                  Christian Heinze and
                  Klaus Broelemann and
                  Gjergji Kasneci and
                  Thanassis Tiropanis and
                  Steffen Staab},
  title        = {Bias in data-driven artificial intelligence systems - An introductory
                  survey},
  journal      = {WIREs Data Mining Knowl. Discov.},
  volume       = {10},
  number       = {3},
  year         = {2020}
}

@article{DBLP:journals/jair/MurthyKS94,
  author       = {Sreerama K. Murthy and
                  Simon Kasif and
                  Steven Salzberg},
  title        = {A System for Induction of Oblique Decision Trees},
  journal      = {J. Artif. Intell. Res.},
  volume       = {2},
  pages        = {1--32},
  year         = {1994}
}

@article{DBLP:journals/ml/FrankWIHW98,
  author       = {Eibe Frank and
                  Yong Wang and
                  Stuart Inglis and
                  Geoffrey Holmes and
                  Ian H. Witten},
  title        = {Using Model Trees for Classification},
  journal      = {Mach. Learn.},
  volume       = {32},
  number       = {1},
  pages        = {63--76},
  year         = {1998}
}

@incollection{DBLP:conf/iclp/BrogiMPT91,
  author       = {Antonio Brogi and
                  Paolo Mancarella and
                  Dino Pedreschi and
                  Franco Turini},
  title        = {Theory Construction in Computational Logic},
  editor       = {Jean{-}Marie Jacquet},
  booktitle        = {Constructing Logic Programs},
  pages        = {241--250},
  publisher    = {Wiley},
  year         = {1993}
}

@article{DBLP:journals/air/MinhWLN22,
  author       = {Dang Minh and
                  H. Xiang Wang and
                  Y. Fen Li and
                  Tan N. Nguyen},
  title        = {Explainable artificial intelligence: A comprehensive review},
  journal      = {Artif. Intell. Rev.},
  volume       = {55},
  number       = {5},
  pages        = {3503--3568},
  year         = {2022}
}

@book{DBLP:series/cogtech/FurnkranzGL12,
  author       = {Johannes F{\"{u}}rnkranz and
                  Dragan Gamberger and
                  Nada Lavrac},
  title        = {Foundations of Rule Learning},
  series       = {Cognitive Technologies},
  publisher    = {Springer},
  year         = {2012}
}

@article{DBLP:journals/bioinformatics/PaulJSWN24,
  author       = {Richard D. Paul and
                  Johann F. Jadebeck and
                  Anton Stratmann and
                  Wolfgang Wiechert and
                  Katharina N{\"{o}}h},
  title        = {hopsy - a methods marketplace for convex polytope sampling in {P}ython},
  journal      = {Bioinform.},
  volume       = {40},
  number       = {6},
  year         = {2024}
}

@article{DBLP:journals/tcs/JerrumVV86,
  author       = {Mark Jerrum and
                  Leslie G. Valiant and
                  Vijay V. Vazirani},
  title        = {Random Generation of Combinatorial Structures from a Uniform Distribution},
  journal      = {Theor. Comput. Sci.},
  volume       = {43},
  pages        = {169--188},
  year         = {1986}
}

@article{DBLP:journals/jossw/SunC25,
  author       = {Benny Sun and
                  Yuansi Chen},
  title        = {PolytopeWalk: Sparse {MCMC} Sampling over Polytopes},
  journal      = {J. Open Source Softw.},
  volume       = {10},
  number       = {112},
  pages        = {7957},
  year         = {2025}
}

@inproceedings{DBLP:conf/issac/NalbachAK21,
  author       = {Jasper Nalbach and
                  Erika {\'{A}}brah{\'{a}}m and
                  Gereon Kremer},
  title        = {Extending the Fundamental Theorem of Linear Programming for Strict
                  Inequalities},
  booktitle    = {{ISSAC}},
  pages        = {313--320},
  publisher    = {{ACM}},
  year         = {2021}
}

@Book{Sch:theory,
  author =       "A. Schrijver",
  title =        "Theory of Linear and Integer Programming",
  year =         "1987",
  address =      "New York, NY",
  publisher =    "John Wiley and Sons"
}

@article{DBLP:journals/tvcg/MingQB19,
  author       = {Yao Ming and
                  Huamin Qu and
                  Enrico Bertini},
  title        = {RuleMatrix: {V}isualizing and Understanding Classifiers with Rules},
  journal      = {{IEEE} Trans. Vis. Comput. Graph.},
  volume       = {25},
  number       = {1},
  pages        = {342--352},
  year         = {2019}
}

@inproceedings{mahajan2019preserving,
author = {Mahajan, Divyat and Tan, Chenhao and Sharma, Amit},
title = {Preserving Causal Constraints in Counterfactual Explanations for Machine Learning Classifiers},
booktitle = {CausalML: Machine Learning and Causal Inference for Improved Decision Making Workshop, NeurIPS 2019},
year = {2019},
month = dec,
}

@book{molnar2019,  
  title={Interpretable machine learning},
  author={Molnar, Christoph},
  publisher={Lulu. com},
  year={2019}
}

@inproceedings{DBLP:conf/nips/CravenS95,
  author       = {Mark W. Craven and
                  Jude W. Shavlik},
  title        = {Extracting Tree-Structured Representations of Trained Networks},
  booktitle    = {{NIPS}},
  pages        = {24--30},
  publisher    = {{MIT} Press},
  year         = {1995}
}

@inproceedings{DBLP:conf/fat/Russell19,
  author    = {Chris Russell},
  title     = {Efficient Search for Diverse Coherent Explanations},
  booktitle = {{FAT}},
  pages     = {20--28},
  publisher = {{ACM}},
  year      = {2019}
}

@inproceedings{DBLP:conf/kdd/CuiCHC15,
  author    = {Zhicheng Cui and
               Wenlin Chen and
               Yujie He and
               Yixin Chen},
  title     = {Optimal Action Extraction for Random Forests and Boosted Trees},
  booktitle = {{KDD}},
  pages     = {179--188},
  publisher = {{ACM}},
  year      = {2015}
}

@article{DBLP:journals/frai/ViloneL21,
  author       = {Giulia Vilone and
                  Luca Longo},
  title        = {A Quantitative Evaluation of Global, Rule-Based Explanations of Post-Hoc,
                  Model Agnostic Methods},
  journal      = {Frontiers Artif. Intell.},
  volume       = {4},
  pages        = {717899},
  year         = {2021}
}

@inproceedings{DBLP:conf/nips/PawelczykBHRK21,
  author       = {Martin Pawelczyk and
                  Sascha Bielawski and
                  Johannes van den Heuvel and
                  Tobias Richter and
                  Gjergji Kasneci},
  title        = {{CARLA:} {A} Python Library to Benchmark Algorithmic Recourse and
                  Counterfactual Explanation Algorithms},
  booktitle    = {NeurIPS Datasets and Benchmarks},
  year         = {2021}
}

@inproceedings{DBLP:conf/jelia/StateRT23,
  author       = {Laura State and
                  Salvatore Ruggieri and
                  Franco Turini},
  title        = {Declarative Reasoning on Explanations Using Constraint Logic Programming},
  booktitle    = {{JELIA}},
  series       = {Lecture Notes in Computer Science},
  volume       = {14281},
  pages        = {132--141},
  publisher    = {Springer},
  year         = {2023}
}

@article{DBLP:journals/ml/LampridisSGR23,
  author       = {Orestis Lampridis and
                  Laura State and
                  Riccardo Guidotti and
                  Salvatore Ruggieri},
  title        = {Explaining short text classification with diverse synthetic exemplars
                  and counter-exemplars},
  journal      = {Mach. Learn.},
  volume       = {112},
  number       = {11},
  pages        = {4289--4322},
  year         = {2023}
}

@article{DBLP:journals/jors/BuckerSGB22,
  author       = {Michael B{\"{u}}cker and
                  Gero Szepannek and
                  Alicja Gosiewska and
                  Przemyslaw Biecek},
  title        = {Transparency, auditability, and explainability of machine learning
                  models in credit scoring},
  journal      = {J. Oper. Res. Soc.},
  volume       = {73},
  number       = {1},
  pages        = {70--90},
  year         = {2022}
}

@article{DBLP:journals/access/FerrarioL22,
  author       = {Andrea Ferrario and
                  Michele Loi},
  title        = {The Robustness of Counterfactual Explanations Over Time},
  journal      = {{IEEE} Access},
  volume       = {10},
  pages        = {82736--82750},
  year         = {2022}
}

@inproceedings{DBLP:conf/uai/PawelczykBK20,
  author       = {Martin Pawelczyk and
                  Klaus Broelemann and
                  Gjergji Kasneci},
  title        = {On Counterfactual Explanations under Predictive Multiplicity},
  booktitle    = {{UAI}},
  series       = {Proceedings of Machine Learning Research},
  volume       = {124},
  pages        = {809--818},
  publisher    = {{AUAI} Press},
  year         = {2020}
}

@inproceedings{DBLP:conf/kdd/LakkarajuBL16,
  author       = {Himabindu Lakkaraju and
                  Stephen H. Bach and
                  Jure Leskovec},
  title        = {Interpretable Decision Sets: {A} Joint Framework for Description and
                  Prediction},
  booktitle    = {{KDD}},
  pages        = {1675--1684},
  publisher    = {{ACM}},
  year         = {2016}
}

@article{DBLP:journals/nca/ShiTLD022,
  author       = {Si Shi and
                  Rita Tse and
                  Wuman Luo and
                  Stefano D'Addona and
                  Giovanni Pau},
  title        = {Machine learning-driven credit risk: A systemic review},
  journal      = {Neural Comput. Appl.},
  volume       = {34},
  number       = {17},
  pages        = {14327--14339},
  year         = {2022}
}

@inproceedings{demajo_explainable_2020,
	title = {Explainable {AI} for {Interpretable} {Credit} {Scoring}},
	doi = {10.5121/csit.2020.101516},
	urldate = {2022-10-05},
	booktitle = {Computer {Science} \& {Information} {Technology} ({CS} \& {IT})},
	publisher = {AIRCC Publishing Corporation},
	author = {Demajo, Lara Marie and Vella, Vince and Dingli, Alexiei},
	month = nov,
	year = {2020},
	pages = {185--203},
}

@misc{charter2007,
    author = {{The European Union}},
    title = {Charter of Fundamental Rights of the European Union},
    note = {{OJ C C326/391. 2012/C 326/02}},
    year = {2012},
}

@misc{Groemping2019,
  author  = {Ulrike Gr\"omping},
  title   = {South German Credit Data: Correcting a Widely Used Data Set},
  year    = {2019},
  note = {Reports in Mathematics, Physics and Chemistry, Department II, Beuth University of Applied Sciences Berlin. \url{http://www1.beuth-hochschule.de/FB\_II/reports/Report-2019-004.pdf}}
}

@MISC{dccc_1,
    title = {{Default of Credit Card Clients}},
    author = {{UCI Machine Learning Repository}},
    year=2025,
    note = {Accessed 2025-12, \url{https://archive.ics.uci.edu/dataset/350/default+of+credit+card+clients}}
}

@MISC{aca,
    title = {{Australian Credit Approval}},
    author = {{UCI Machine Learning Repository}},
    year=2025,
    note ={Accessed 2025-12, \url{http://archive.ics.uci.edu/dataset/143/statlog+australian+credit+approval}}
}

@MISC{gmsc,
    title = {{Give Me Some Credit}},
    author = {{Kaggle}},
    year=2025,
    note = {Accessed 2025-12, \url{https://www.kaggle.com/datasets/brycecf/give-me-some-credit-dataset?select=cs-test.csv}}
}

@MISC{sgc,
    title = {{South German Credit}},
    author = {{UCI Machine Learning Repository}},
    year=2025,
    note ={Accessed 2025-12, \url{https://archive.ics.uci.edu/dataset/522/south+german+credit}}
}

@MISC{adult,
    title ={{Adult Dataset}},
    author ={{UCI Machine Learning Repository}},
    year=2025,
    note = {Accessed 2025-12, \url{https://archive.ics.uci.edu/ml/datasets/Adult}}
}


\newpage
\begin{sm}

\section{Contrasting to Sufficient Reasons}\label{app:suff}

We {integrate the related work description by contrasting} here our notion of factual constraints (see e.g., Table \ref{tab:my_label}) with the one of \textit{sufficient reasons} from~\citet{DBLP:journals/jair/IzzaIM22}.
We argue that while sufficient reasons are meaningful in the context of fully specified instances, they are not equipped to handle under-specified instances.

\paragraph*{Fully-specified instances}

Sufficient reasons are defined for fully-specified instances, and they consists of a \textit{minimal} subset of feature values of the instance to explain, such that the prediction of the DT is the same as for the instance whatever the values of the remaining features are. The general assumption here is that succinctness of explanations is a desired property, for which the decision path in a decision tree cannot lead to interpretability. However, the minimality requirement introduces notable challenges. In particular, multiple minimal subsets may exist for a single instance, leading to ambiguity in the explanation~\citep{DBLP:journals/dke/AudemardBBKLM22}. Furthermore, computing sufficient reasons is generally intractable~\citep{DBLP:journals/ai/CooperS23}, limiting their practical applicability to sub-classes of decision trees. {Finally, enforcing subset-minimality would compromise the compositionality of the algebra, since minimality is not generally preserved under the subsequent composition of explanations with additional constraints or operators.}
Our definition of a factual constraint reflects the intuitive decision path followed by the instance. It consists of the conjunction of split conditions (linear constraints) in the path from the root node to a leaf determined by the values of the instance. While these constraints are outputed in non-redundant form (e.g., split conditions $x \geq 0$ and $x \geq 5$ lead to factual constraint $x \geq 5$), an answer constraint is not necessarily minimal in the sense of sufficient reasons. This is because it is derived solely from the observed decision path, without considering alternative paths in the decision tree that might yield the same prediction.

\paragraph*{Under-specified instances}

The notion of sufficient reasons is not defined for under-specified instances, as there is no guarantee that all possible groundings of such instances will lead to the same prediction. 
Consider for example the decision tree with a single split: the predicted class is $l=1$ if $x_2 \geq 1$, and $l=0$ otherwise. Now take the under-specified instance $I_1$ such that only $I_1.x_1 = 2$ is known. If we further specify $I_1.x_1 = 2, I_1.x_2 = 0$, the decision tree outputs $l=0$. Instead, by considering $I_1.x_1 = 2, I_1.x_2 = 1$, the output is $l=1$. This shows that multiple predictions are possible depending on how the unspecified features are instantiated. Consequently, the concept of sufficient reasons cannot be straightforwardly extended to under-specified instances, as the decision tree may yield different outcomes for different completions of the same partial input. A preliminary study on how to generalize abductive explanations is due to \citet{DBLP:conf/aaai/IzzaIS024}, who introduce the notion of inflated explanations. Here, however, the input is still a fully specified instance.
{\reasonx{}} is instead able to deal with the example above by returning two answers. The first one has answer constraint $I_1.x_1 = 2, I_1.x_2 \geq 1$ and factual rule $I_1.x_2 \geq 1 \rightarrow l=1$. The second one has answer constraint $I_1.x_1 = 2, I_1.x_2 < 1$ and factual rule $I_1.x_2 < 1 \rightarrow l=0$.

\section{Constraint Logic Programming over the Reals}
\label{sec:reasonx_setting_the_stage}

{To keep the paper self-contained, we summarize here the key notions of CLP over the reals.} The CLP scheme defines a family of languages that is parametric in a constraint
domain \citep{DBLP:journals/jlp/JaffarM94,DBLP:journals/tplp/KornerLBCDHMWDA22}. We consider the constraint domain over the reals CLP($\mathbb{R}$), and adhere to the SWI Prolog implementation~\citep{DBLP:journals/tplp/WielemakerSTL12}. We assume that the underlying constraint solver is ideal, that is, without rounding errors. 

\paragraph*{Syntax}

%
A term is an expression over a finite set of function symbols (lower case initial) and variables (upper case initial) and constants ($0$-ary function symbols). A list \texttt{[}$t_1, \ldots, t_n$\texttt{]} is a term built by appending to the empty list \texttt{[]} the terms $t_i$'s using the list constructor 
\texttt{[}$h$\texttt{|}$t$\texttt{]}
where the element $h$ is prefixed to the list $t$.
An atom $p(t_1, \ldots, t_k)$ consists of a predicate $p$ of arity $k$ and terms $t_1, \ldots, t_k$, with $k \geq 0$. 
A CLP($\mathbb{R})$ program is a finite set of clauses of the form:
\begin{center}
    $A$ \texttt{:-} \texttt{\{}$c$\texttt{\}}$, B_1, \ldots, B_n$\texttt{.} 
\end{center}
with $n \geq 0$, where $A$ (the head) is an atom, $c$ a (possibly empty) linear
constraint, and $B_1, \ldots, B_n$ (the body) a sequence of atoms. If $n=0$, the clause is called a fact, and if $c$ is empty, it is written as $A$\texttt{.} (without the \texttt{:-} operator). A query \texttt{\{}$c$\texttt{\}}$, B_1, \ldots, B_n$ consists of a
linear constraint and a sequence of atoms.






\paragraph*{Semantics}

The operational semantics of CLP($\mathbb{R})$ is an extension of the SLD-resolution of logic programming. We assume the (Prolog) leftmost selection rule~\citep{DBLP:journals/jlp/JaffarMMS98}.
%
A state $\langle Q, cs \rangle$ consists of a query $Q$ and a set of linear constraints $cs$, called the constraint store. In initial states, the constraint store is empty. A derivation is a sequence of states starting from initial ones through the following transitions:
\begin{itemize}
    \item For constraints: $\langle (\texttt{\{}$c$\texttt{\}}, C_1, \ldots, C_m), cs \rangle \rightarrow \langle (C_1, \ldots, C_m), cs \cup \{c\} \rangle$ if adding the constraint $c$ to $cs$ yields a satisfiable constraint store, namely $\models \exists (cs \wedge  c)$. 
    If satisfiability does not occur, the derivation  fails.
    \item For atoms: $\langle (C_1, \ldots, C_m), cs \rangle \rightarrow \langle ($\texttt{\{}$c$\texttt{\}}$, B_1, \ldots, B_n, C_{2}, \ldots, C_m)\theta, cs\theta \rangle$ if the atom $C_1$ unifies with the head of a (renamed apart) clause $A$ \texttt{:-} \texttt{\{}$c$\texttt{\}}$, B_1,$ $\ldots,$  $B_n.$ with the most general unifier $\theta$. If no unification is possible, the derivation  fails.
\end{itemize}

In the rewriting of atoms, more than one clause may unify, making the derivation non-deterministic: 
the transition diagram is a tree, which is explored depth-first, trying clauses in the order of how they appear in the program. 
Backtracking occurs if a failed transition is reached or if more answers are required.

A state for which no further transition is possible is of the form $\langle \emptyset, cs \rangle$, and it is called a success state. The \textit{answer constraint} returned is the projection of $cs$ over the variables $\mathbf{x}$ of the initial query, i.e.,~$\exists_{-\mathbf{x}} cs$ where $-\mathbf{x}$ denotes the variables of $cs$ that are not in $\mathbf{x}$. Constraint stores and answer constraints are kept in a minimal representation, e.g.,~by removing redundant inequalities~\citep{DBLP:journals/amai/Greenberg96,nnccp,DBLP:conf/vmcai/MarechalP17}.

\paragraph*{MILP optimization}

On top of the constraint store, CLP($\mathbb{R})$ implementations offer mixed integer linear programming (MILP) optimization functionalities. 
Common built-in predicates include the calculation of the supremum and the infimum of an expression w.r.t.~the solutions of the constraint store. 
An example is the query \texttt{\{X >= 1.5\}, inf(X, R)}, which returns the infimum of the expression \texttt{X} in \texttt{R}, therefore \texttt{R=1.5}. 
We will use the predicate \texttt{bb\_inf} which is similar to \texttt{inf}, but additionally accepts a list of variables in the domain of integers. For example, the query \texttt{\{X >= 1.5\}, bb\_inf([X], X, R)}, i.e.,~the query previously discussed updated with \texttt{bb\_inf}, now returns \texttt{R=2}. 

\section{Distance Functions}
\label{sec:appendix_norm}

{We provide here further details on the implementations of the two distance functions offered by {\reasonx{}}.} The first distance function is a combination of the $L_1$~norm for the ordinal and continuous features and a simple matching distance for the nominal features: 
\begin{eqnarray*}
L_1(F, CE) = \min \hspace{-0.2cm}  \sum_{i\ \textrm{\tiny nominal}} \mathds{1} (CE.x_{i} \ne F.x_{i}) + \sum_{i\ \textrm{\tiny ord., cont.}}  | CE.x_{i} - F.x_{i} |
\end{eqnarray*}
The second one is the $L_{\infty}$ norm, where nominal features are compared by simple matching: 
\begin{eqnarray*}
L_{\infty}(F, CE) = \max\{\ \ \max_{i\ \textrm{\tiny ord., cont.}}  | CE.x_{i} - F.x_{i} |, \max_{i\ \textrm{\tiny nominal}} \mathds{1} (CE.x_{i} \ne F.x_{i}) \ \ \}
\end{eqnarray*}
%
%
For both distance functions, continuous and ordinal features are min-max normalized. Thus each feature in isolation has a maximum distance of $1$, and then it contributes the same to the norm. 
%
%
The $L_{\infty}$ norm penalizes maximum changes over all features while the $L_1$ norm penalizes the average change over all features. See also~\citet{DBLP:conf/aistats/KarimiBBV20,DBLP:journals/corr/abs-1711-00399} for a discussion of how the different norms influence the outcome.

In order to solve the MILP problem of minimizing distance, we need to linearize each norm. This leads to the introduction of additional constraints and slack variables.
The $L_1$ norm can be minimized by introducing the slack variables $t_i$'s as follows:
\begin{eqnarray*}
\min \sum_i t_i \quad
s.t. \quad \forall i.\ (CE.x_{i} - F.x_{i} \leq t_i \wedge F.x_{i} - CE.x_{i} \leq t_i)
\end{eqnarray*}
Similarly, for the minimization of $L_{\infty}$ norm, we introduce the slack variable $s$ and define:
\begin{eqnarray*}
\min  s \quad
s.t. \quad \forall i.\ ( CE.x_{i} - F.x_{i} \leq s \wedge F.x_{i} - CE.x_{i} \leq s)
\end{eqnarray*}
Listing~\ref{lst:l1_linf_con} report the Prolog code for the $L_1$ and the $L_{\infty}$ norm respectively. 
Both predicates define a relation between the two instances (\texttt{Inst1}, \texttt{Inst2}) for which the distance is calculated, the list of the constraints that must be enforced, following the linerization above (\texttt{Cs}) and the expression to be minimized (\texttt{Norm}).
The predicate \texttt{norm\_weights} refers to the feature-specific weights, computed at the Python layer, such that the maximum normalized value of a features is always one. For continuous and ordinal features, this is the range of values. For nominal features (which are one-hot encoded), it is $0.5$ for $L_1$ (because if the nominal values differ, then \textit{two} one-hot encoded values differ) and $1$ for  $L_{\infty}$.
%
%

\begin{lstlisting}[basicstyle=\ttfamily\scriptsize, caption = {Definition of \texttt{l1\_con} and of \texttt{linf\_con}.}, label = {lst:l1_linf_con}]
l1_con(Inst1, Inst2, Cs, Norm) :-
    norm_weights(W),
    l1_con(W, Inst1, Inst2, Cs, Norm).	
l1_con([], [], [], [], 0).    
l1_con([W|Ws], [X|Xs], [Y|Ys], [S >= X-Y, S >= Y-X|Cs], W*S+Sum) :-
    l1_con(Ws, Xs, Ys, Cs, Sum).

linf_con(Inst1, Inst2, Cs, Norm) :-
    norm_weights(W),
    linf_con(W, Inst1, Inst2, Cs, Norm).	
linf_con([], [], [], [], _).
linf_con([W|Ws], [X|Xs], [Y|Ys], [S >= W*(X-Y), S >= W*(Y-X)|Cs], S) :-
    linf_con(Ws, Xs, Ys, Cs, S).
\end{lstlisting}

\section{Experimental Settings}\label{app:expsettings}

{We present here the datasets, metrics, parameters and experimental settings used in the Sections \nameref{sec:reasonx_demonstrations} and \nameref{sec:reasonx_evaluation} of the main text.}

\subsection{Dataset Information}
\label{sec:appendix_datasets}

\begin{table}[t]
    \centering
    \scriptsize
    \begin{tabular}{rccccc}
    \toprule
    Name & $N$ & Class ratio & $n_{\mathit{nominal}}$ & $n_{\mathit{ordinal}}$ & $n_{\mathit{continuous}}$ \\
    \midrule
    \texttt{ADULT} & $48,842$ & $37,155:11,687$, $(\leq50K:> 50K)$ & 3 & 1 & 4 \\
    \texttt{SGC} & $1,000$ & $700:300$, $(1:0)$ & 0 & 17 & 3 \\
    \texttt{GMSC} & $150,000$ & $139,974:10,024$, $(0:1)$ & 0 & 0 & 10 \\
    \texttt{DCCC} & $30,000$ & $23,364:6,636$, $(0:1)$ & {3} & {0} & {8} \\
    \texttt{ACA} & $690$ & $383:307$, $(0,1)$ & $8$ & 0 & 6 \\
    \bottomrule
    \end{tabular}
    \caption{Dataset summary statistics{: number $N$ of instances, class distribution, and number of nominal, ordinal and continuous features}.}
    \label{tab:dataset_statistics}
\end{table}

We experiment with the Adult Income (\texttt{ADULT}) dataset~\citep{adult}, the South German Credit (\texttt{SGC}) dataset~\citep{sgc,Groemping2019}, the Give Me Some Credit (\texttt{GMSC}) dataset~\citep{gmsc}, the Default of Credit Cards Clients (\texttt{DCCC}) dataset~\citep{dccc_1}, 
and the Australian Credit Approval~(\texttt{ACA}) dataset~\citep{aca}.
Table~\ref{tab:dataset_statistics} reports  dataset summary statistics.
The preprocessing of the datasets contains several steps in accordance with the capacities of {\reasonx{}}. 
All steps are reported in the Jupyter notebooks in the released repository.

\subsection{Metrics and Parameters}
\label{sec:appendix_metrics}

\begin{table}[t]
    \centering
    \scriptsize
    \textbf{Evaluation metrics}\\
    \begin{tabular}{lll}
    \toprule
    \textbf{Applies to} & \textbf{Metric} & \multicolumn{1}{c}{\textbf{Notation}} \\
    \midrule
    \textit{Classification models} & accuracy & $acc$ \\
    \textit{Surrogate models} & fidelity ($^{\star}$) & $f$ \\  
    \textit{Set of queries} & fraction of queries with at least one answer  & $S/N$ \\
    \textit{Set of explanation rules} & mean number of contrastive rules per query & $N_{\mathit{CE}}$\\
    &  mean factual rule length & $l_{\mathit{F}}$\\
    & mean contrastive rule length & $l_{\mathit{CE}}$\\
    \textit{Set of CE or MCE rules} & mean $L_1$ distance  & $d_{\mathit{CE}, L_1}$, $d_{\mathit{MCE}, L_1}$\\
     & mean $L_{\infty}$ distance  & $d_{\mathit{CE}, L_{\infty}}$, $d_{\mathit{MCE}, L_{\infty}}$\\
    \bottomrule
    \end{tabular}
    \mbox{}\\[2ex]
    \textbf{{\reasonx{}} parameters}\\    \begin{tabular}{ll}
    \toprule
    \textbf{Parameter} & \textbf{Notation}\\
    \midrule
    maximum tree depth & $D$ \\
    minimum confidence factual & $\mathit{MC}_F$ \\
    minimum confidence contrastive & $\mathit{MC}_{CE}$ \\
    $\mathit{relax}$ parameter & $\epsilon$\\
    \bottomrule
    \end{tabular}
    \caption{Notation for evaluation metrics and parameters of {\reasonx{}}. ($^{\star}$) The fidelity metric applies only to surrogate models, i.e.,~in cases \textbf{(DT-GS)} and \textbf{(DT-LS)}.}
    \label{tab:evaluation}
\end{table}

To the best of our knowledge, no evaluation framework specifically tailored to explanation tools based on CLP has been proposed so far. We therefore adopt a set of standard XAI evaluation metrics~\citep{DBLP:journals/frai/ViloneL21,DBLP:conf/nips/PawelczykBHRK21,DBLP:journals/expert/GuidottiMGPRT19}, adapting them where necessary to the characteristics of \reasonx{}. The resulting metrics are summarized in Table~\ref{tab:evaluation}.
%
{For a model, accuracy} refers to the ratio of correctly predicted class labels of a classification model w.r.t.~the ground-truth labels, while fidelity refers to the ratio of correctly predicted class labels of the surrogate model w.r.t.~the class predicted by the black-box model (applicable only in cases \textbf{(DT-GS)} and \textbf{(DT-LS)}). {For a collection of $N$ queries, $S/N$ refers to fraction of queries with at least one answer constraint, where $S$ is the number of them with at least one answer constraint. A query can have no answer if there is no factual rule or contrastive rule satisfying the required class labels or minimum confidence threshold. For the typical queries looking for (minimal) contrastive explanations of a given factual instance, $N_{\mathit{CE}}$ is then the mean number of (minimal) contrastive rules found. }
%
%
{For an (factual/constrative) explanation rule, the length of the rule is the number of primitive linear constraints in its left-hand side. $l_F$ and $l_{CE}$ denote the mean rule length for a set of factual and constrative rules respectively. For (minimal) contrastive explanation rules, we also consider the mean $L_1$ or $L_{\infty}$ distance values minimized by those rules.} 
{Table~\ref{tab:evaluation} also summarizes the {\reasonx{}} parameters that can be set in experiments.}

%
%
%

\subsection{Experimental Settings for the Demonstrators}
\label{sec:demonstrator_settings}

\begin{table}[t!]
    \centering
    \scriptsize
    \begin{tabular}{lccccccc}
    \toprule
    \textbf{Demonstration} & \textbf{Dataset} & \textbf{Splits} & \textbf{Instances} & \textbf{Case} & \textbf{Distance} & $\epsilon$  & $D$ \\
    \midrule
    Synthetic dataset case & synthetic & train/test & test & (DT-M) & $L_1$ & $\epsilon_{\text{mach}}$ & 3\\
    Adult income case & \texttt{ADULT} & train/test & test & all & $L_1$ & $0.01$ & 3\\
    Reason over time & \texttt{ADULT} & train/test$^{(1)}$ & test$^{(2)}$ & (DT-M) & $L_1$ & $0.01$  & 3\\
    Reason over models & \texttt{ADULT} & train/test/expl & expl & (DT-M), (DT-GS) & $L_1$ & $0.01$  & 3\\
    Diversity optimization & \texttt{ADULT} & train/test/expl & test/expl & (DT-M), (DT-GS) & $L_1$ & $0.01$ &  3\\
    Detecting bias & \texttt{ADULT} & train/test & test & (DT-M) & n.a. & n.a. & 5 \\
    \bottomrule
    \end{tabular}
    \caption{Overview of experimental settings for the demonstrations.
    {The first column reports the demonstration, the second the underlying dataset. The third column indicates how the dataset was split. \enquote{train/test} indicates that there was a single split into training and test set with ratio 7:3, \enquote{train/test/expl} indicates that the dataset was split into training and test set with ratio 7:3, with a subsequently split of the test set into an explanation training and test set again with ratio 7:3, $^{(1)}$ indicates that the dataset was split twice into training and test (to simulate changes in the data over time). The next column describes from which set the instance to explain was taken. $^{(2)}$ indicates that instances shared by the two test sets were considered. Moreover, the column lists two cases if {\reasonx{}} was used in two different cases. These cases are indicated in the next column. Finally, the table indicates the distance function used in the optimization,}
    the value of the parameter $\epsilon$, {where $\epsilon_{\text{mach}}$ is the machine epsilon (default)}, and the maximum depth of the base model $D$. 
    n.a. $=$ not applicable, because minimal contrastive rules were not used.
    }
    \label{tab:reasonx_demonstrations_settings}
\end{table}

{Table~\ref{tab:reasonx_demonstrations_settings} reports the experimental settings used in the examples of the Section \nameref{sec:reasonx_demonstrations} of the main text. The $\mathit{MC}_F$ and $\mathit{MC}_{CE}$ parameters are specified in queries of each demonstration.}

\subsection{Experimental Settings for the Quantitative Evaluation}
\label{sec:appendix_experiments}

Here, we provide the {experimental settings for the Section \nameref{sec:reasonx_evaluation} of the main text}. For all experiments, we set $\epsilon=0.01$. {Results are averaged over 100 test instances.}

\paragraph*{{{\reasonx{}}} only.}

We compute all evaluation metrics for case \textbf{(DT-M)}, \textbf{(DT-GS)} and \textbf{(DT-LS)} and on five datasets in the credit domain, 
and under no user constraints. 
The datasets are split into training and test set with a ratio of $7:3$, the test set further divided into an explanation training set {(for training the surrogate DT)} and an explanation test set with a ratio of $7:3$ for case \textbf{(DT-GS)}. In case \textbf{(DT-LS)}, we apply a split with the same conditions on a synthetically generated neighborhood {(of 10,000 instances)} of the instance to explain.
Results are averaged over $100$ data instances from the test set, for $D = 3$, and for the XGB black-box model in case \textbf{(DT-GS)} and \textbf{(DT-LS)}. {We use a simple neighborhood-generation procedure that keeps nominal and ordinal features fixed, bounds continuous features between one standard deviation from the feature value of the factual instance, and relies on MCMC-based uniform sampling from polytopes. This is close the procedure adopted for fidelity estimation in Section~\nameref{sec:fidelity}.}

\paragraph*{Contrastive examples.}
We evaluate {\reasonx{}} against DiCE~\citep{DBLP:conf/fat/MothilalST20}.
Besides the case with no user constraints, we test two cases with different immutability constraints. Since the number of contrastive examples $N_{CE}$ is a parameter in DiCE and not a metric to be measured, we run DiCE under different settings, $N_{CE} \in \{2,3,4,5\}$.
Further, DiCE allows to specify parameters that determine the weight of the proximity and the diversity term in the optimization. To ensure a fair comparison, we set the weight of the diversity term to zero. 
Other parameters of DiCE are left to their default values.
We transform results of DiCE using the preprocessing of {\reasonx{}} and compute both distances as in {\reasonx{}}.
Evaluations are done on the \texttt{ADULT} dataset, case \textbf{DT-GS} with an XGB black-box model, and with $D = 3$ for {\reasonx{}}. 
%

\paragraph*{Factual rules.}
Here, we evaluate {\reasonx{}} against ANCHORS~\citep{DBLP:conf/aaai/Ribeiro0G18}.
ANCHORS has a parameter for the precision{, which is essentially the minimum confidence $\mathit{MC}_F$. We vary this parameter, $\mathit{MC}_F \in \{0.80, 0.90, 0.95\}$.} We obtain the actual precision $p$ and coverage of the produced rules
directly from the ANCHORS package. Other parameters for ANCHORS are left to their default values.
To ensure a fair comparison to {\reasonx{}}, we run our tool for different tree depths $D$ (implying different rule lengths) and for the same values of the minimum confidence $MC_F$.
Results are computed under no user constraints, on the \texttt{ADULT} dataset, case \textbf{DT-GS} with an XGB black-box model, and with $D = 3$ for {\reasonx{}}. 

\paragraph*{Runtime.}
%
Elapsed runtimes per query are computed as the difference between the complete explanation pipeline minus the steps before explainers are initialized (preprocessing/model learning).
%
%
{
The experimental machine was an Intel Xeon w5-2465X CPU at 3.10~GHz and 32~GiB of RAM, running Windows~11 and Python~3.14, using single-threaded, single-core execution.}

\section{More on Demonstrations}\label{app:moredemo}

{In the following, we report additional qualitative examples on the \texttt{ADULT} dataset, and adds details on the \nameref{sec:reasonx_demonstration_diversity} and on the \nameref{sec:reasonx_detecting_bias} subsections of the main text.}

\subsection{Adult Income Dataset}
\label{sec:reasonx_demonstration_adult}

This example focuses on explanations for a hypothetical automated decision-making (ADM) system that assesses credit applications, for example, provided by a bank.
The main purpose of the explanation is for the decision maker to make sense of the decision and for the applicant to receive information on possible opportunities for action.
Secondary purposes are legal compliance of the ADM, as well as increasing the trust of the bank's stakeholders in internal decisions. {The experimental settings are included in Table~\ref{tab:reasonx_demonstrations_settings}.}


We run the example on a processed version of the Adult Income dataset \citep{adult},
which has a binary response variable -- whether the income of a person is $\leq 50$K USD, or $> 50$K USD. 
It contains twelve features, and we restrict ourselves to a subset: three nominal features (\texttt{workclass}, \texttt{race}, \texttt{sex}), one ordinal (\texttt{education}) and four continuous (\texttt{age}, \texttt{capitalgain}, \texttt{capitalloss}, \texttt{hoursperweek}).
Nominal features are transparently one-hot encoded by the {\reasonx{}} utility functions. 
%

We split the dataset into 70\% training and 30\% test set. A DT is built on the training set. It has an accuracy over the test set of 80.2\%. In the following, an instance is chosen from the test set with predicted income $\leq 50$K USD. 

After initialization {setting the predictive and class attributes, and passing the DT classifier under analysis}, we turn towards a first question of the user of {\reasonx{}}, in this scenario the applicant (who corresponds to the factual instance): \textit{Why was my application rejected?} 
It can be answered by computing the rule behind the decision.
This requires naming an instance (\texttt{F} in the code below, for \enquote{factual}) and passing the feature values (\texttt{InstFeatures}) and the decision of the base model (\texttt{InstDecision}). 

\begin{lstlisting}[basicstyle=\ttfamily\scriptsize]
USER:    r = reasonx.ReasonX(pred_atts, target, df_code)
         r.model(clf)
         r.instance('F', features=InstFeatures, label=InstDecision)
         r.solveopt()
\end{lstlisting}
\begin{lstlisting}[basicstyle=\ttfamily\scriptsize, backgroundcolor=\color{answertoquery}, caption = {Inititalisation and factual instance definition.}]
REASONX: Answer constraint: 
           F.race=AsianPacIslander,F.sex=Male,F.workclass=Private,F.education=10.0,
           F.age=19.0,F.capitalgain=0.0,F.capitalloss=0.0,F.hoursperweek=40.0
         Rule satisfied by F: 
           IF F.capitalgain<=5119.0,F.education<=12.5,F.age<=30.5 
           THEN <=50K [0.9653]
\end{lstlisting}

{The answer constraint states the values of the features set for the instance.} The factual rule explains the classification of the instance. It refers to a specific region in the data input space as characterized by the base decision tree.
A second answer can be given by comparing the factual instance against a contrastive rule, using the differences as an explanation. This requires naming a second instance (\texttt{CE} in the code below, for \enquote{contrastive example}), and possibly a minimum confidence of the rule leading to the contrastive decision.
We obtain two rules by running the following query:

\begin{lstlisting}[basicstyle=\ttfamily\scriptsize]
USER:    r.instance('CE', label=1-InstDecision, minconf=0.8)
         r.solveopt(project=['CE'])
\end{lstlisting}
\begin{lstlisting}[basicstyle=\ttfamily\scriptsize, backgroundcolor=\color{answertoquery}, caption = {Contrastive instance definition.}]
REASONX: Contrastive rule satisfied by CE: 
           IF CE.capitalgain>5119.0,CE.capitalgain<=5316.5 
           THEN >50K [1.000]
         Contrastive rule satisfied by CE: 
           IF CE.capitalgain>7055.5,CE.age>20.0 
           THEN >50K [1.000]
\end{lstlisting}

By comparing this output with the answer to the previous question, the user can understand the factual decision of the ADM better.
This is especially relevant to answer a second question: \textit{What are my options to change the outcome of the ADM, and receive the credit?} 
For example, comparing the first contrastive with the factual rule, an increase of the feature \texttt{capitalgain} will lead to a change in the predicted outcome of the ADM. 
Similarly, an increase in \texttt{capitalgain}, and an increase in  \texttt{age} ({the instance \texttt{F} has \texttt{age=19})} will alter it.

In the next step, we add some background knowledge to the explanations. 
We apply an immutability constraint on the feature \texttt{age}:

\begin{lstlisting}[basicstyle=\ttfamily\scriptsize]
USER:    r.constraint('CE.age = F.age')
         r.solveopt(project=['CE'])
\end{lstlisting}
\begin{lstlisting}[basicstyle=\ttfamily\scriptsize, backgroundcolor=\color{answertoquery}, caption = {Setting user constraints.}]
REASONX: Contrastive rule satisfied by CE: 
           IF CE.age=19.0,CE.capitalgain>5119.0,CE.capitalgain<=5316.5 
           THEN >50K [1.000]
\end{lstlisting}

As expected, the solution has changed: by adding the above-stated constraint, the admissible region for contrastive explanations becomes smaller, and only one contrastive rule remains.
Last, we ask for the contrastive explanation that is \textit{closest} to the factual instance, {and use the $\epsilon$ parameter to cope with minimization in presence of strict inequalities.}

\begin{lstlisting}[basicstyle=\ttfamily\scriptsize]
USER:    r.solveopt(minimize='l1norm(F, CE)', project=['CE'], eps=0.01)
\end{lstlisting}
\begin{lstlisting}[basicstyle=\ttfamily\scriptsize, backgroundcolor=\color{answertoquery}, caption = {Minimal contrastive explanation rule.}]
REASONX: Contrastive rule satisfied by CE: 
           IF CE.race=AsianPacIslander, CE.sex=Male,CE.workclass=Private,
             CE.education=10.0,CE.age=19.0,CE.capitalgain=5119.01,
             CE.capitalloss=0.0,CE.hoursperweek=40.0   
           THEN >50K [1.0000]
\end{lstlisting}

{\reasonx{}} returned the \textit{minimal} contrastive constraint. This corresponds to the most common notion of a contrastive example in the literature. However, {\reasonx{}} also took care of the specified background knowledge - the contrastive example is returned under the above specified constraint on the feature \texttt{age}.
We extend the example to account for under-specified information. e.g.,~it can be interesting to ask for the minimal contrastive constraint in the case the feature \texttt{age} is not fixed, but restricted to $19$ years or lower. 

\begin{lstlisting}[basicstyle=\ttfamily\scriptsize]
USER:    r.retract('F.age=19.0')
         r.constraint('F.age<=19.0')
         r.solveopt(minimize='l1norm(F, CE)', project=["CE", "F.age"], eps=0.01)
\end{lstlisting}
\begin{lstlisting}[basicstyle=\ttfamily\scriptsize, backgroundcolor=\color{answertoquery}, caption = {Reasoning over an under-specified instance.}]
REASONX: Answer constraint:
           CE.race=AsianPacIslander,CE.sex=Male,CE.workclass=Private,
           CE.education=10.0, CE.capitalgain=5119.01,CE.capitalloss=0.0,
           CE.hoursperweek=40.0,CE.age=F.age, CF.age<=19.0
\end{lstlisting}

The returned solution is similar to the previous one, but we observe that also in the answer constraint, the feature \texttt{age} is not fixed to a single value but is provided as an equality region. 

Since queries work iteratively, the flow of the above corresponds exactly to how an \textit{interaction} with {\reasonx{}} could look like, forming what we call an \textit{explanation dialogue}. 
Repeated specification of background knowledge and querying are a central part of this dialog between the user and {\reasonx{}}, allowing to build individual, personalized explanations.
A variant of this example for the cases \textbf{(DT-GS)} and \textbf{(DT-LS)} 
is discussed in the next section.

\subsubsection{Global and Local Surrogate Models}
\label{sec:reasonx_global_local}


We extend the example from the previous section 
to cases \textbf{(DT-GS)} and \textbf{(DT-LS)}. We focus on the lines of code that are relevant to change and use the example of a neural network (NN) as a ML model.
In cases \textbf{(DT-GS)} and \textbf{(DT-LS)} {\reasonx{}} produces explanations via a global or local surrogate model, respectively. By this, any type of ML model on tabular data can be explained.

\paragraph*{Global surrogate}
An extension of case \textbf{(DT-M)} to this is straightforward: what is needed for case~\textbf{(DT-GS)} is to learn the base model not using the original class labels but the ML model's predicted class labels. After this, {\reasonx{}} can be initialized and queried.
Thus, the crucial lines are only in the Python code. They read as follows:

\begin{lstlisting}[basicstyle=\ttfamily\scriptsize, caption = {Initialization of {\reasonx{}} in case \textbf{(DT-GS)}.}]
USER:    mlp = MLPClassifier(random_state=0)           
         mlp.fit(X1, y1)
         mlp_label = mlp.predict(X1)                     
         clf1 = DecisionTreeClassifier(max_depth=3)      
         clf1.fit(X1, mlp_label)
         r = reasonx.ReasonX(pred_atts, target, df_code) 
         r.model(clf1, bb=mlp)
\end{lstlisting}

In line 1-2 the NN is initialized and trained, in line 3 class labels are predicted. In line 4-5 the base model is initialized and trained with these labels, and in line 6-7 {\reasonx{}} is initialized. {Specifially, line 7 binds \texttt{mlp} as the black-box model of the (surrogate) decision tree \texttt{clf1}. The black-box is used when computing the fidelity of (minimal) constrative rules.}

\paragraph*{Local surrogate}
In case \textbf{(DT-LS)} such an extension means that \textit{after} the instance to explain is chosen, a local neighborhood has to be created. 
Here, we rely on a random neighborhood generation.
After all instances of that neighborhood are classified by the ML model, the base model can be learned using these neighborhood instances and their labels as predicted by the ML model. {\reasonx{}} has to be instantiated with this base model, and the produced explanations are only valid for the specific instance.
The important code lines are the following:

\begin{lstlisting}[basicstyle=\ttfamily\scriptsize, caption = {Initialization of {\reasonx{}} in case \textbf{(DT-LS)}.}]
USER:    neigh = neighborhood_generation(...)
         mlp = MLPClassifier(random_state=0)
         mlp.fit(X1, y1)
         neigh_mlp_label = mlp.predict(neigh)
         clf1 = DecisionTreeClassifier(max_depth=3)
         clf1.fit(neigh, neigh_mlp_label)
         r = reasonx.ReasonX(pred_atts, target, df_code)
         r.model(clf1, bb=mlp)
\end{lstlisting}

The local neighborhood is generated in line 1, around the specified data instance and according to some neighborhood generation algorithm\footnote{While we have already integrated a simple neighborhood generation, relying on a more stable implementation such as the method used in LORE~\citep{DBLP:journals/expert/GuidottiMGPRT19} is a natural extension of {\reasonx{}}.}.
In line 2-3, the NN is initialized and trained, in line 4 class labels of the neighborhood are predicted. In line 5-6 the base model is initialized and trained with the neighborhood and its previously predicted labels, and in line 7-8 {\reasonx{}} is initialized.

\subsection{Diversity Optimization}
\label{appendix:diversity}

In Figure~\ref{fig:diversity_optimization}, we compare the result of the diversity optimization (optimizing proximity \textit{and} diversity) with the classical optimization (optimizing only proximity). The figure is commented in the main text in Section~\nameref{sec:reasonx_demonstration_diversity}.
%

\begin{figure}[t]
    \centering
    \includegraphics[width = 0.3\linewidth]{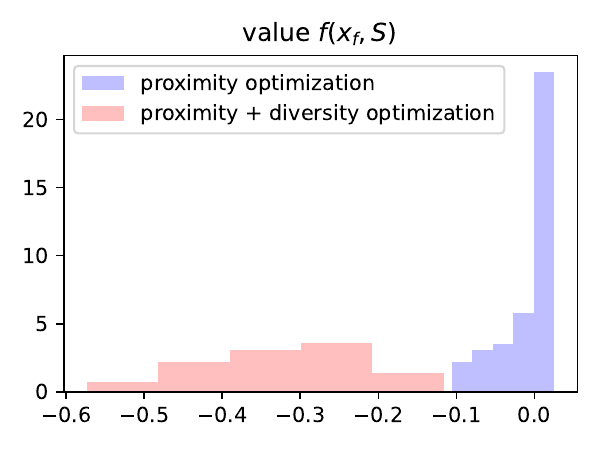}
    \includegraphics[width = 0.3\linewidth]{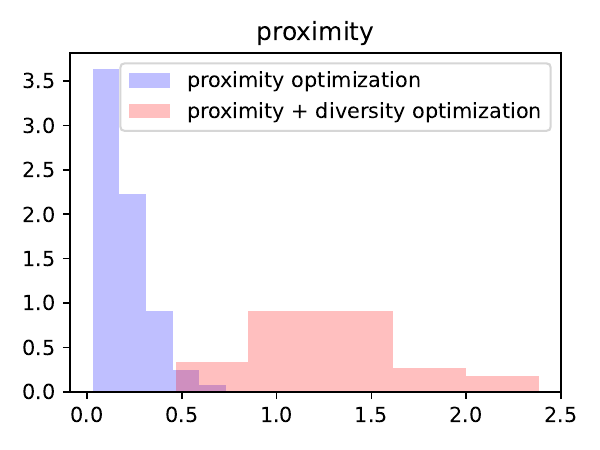}
    \includegraphics[width = 0.3\linewidth]{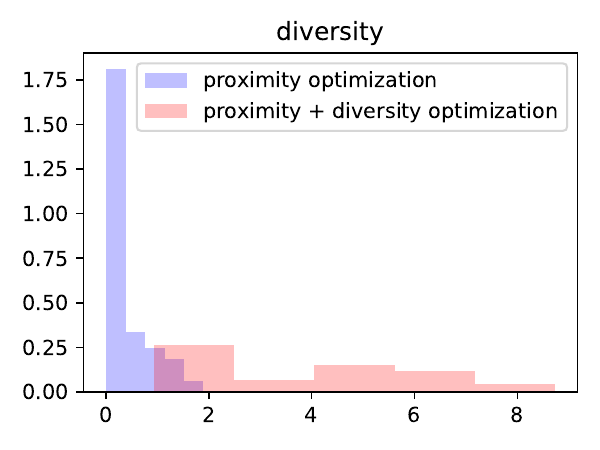}
    \includegraphics[width = 0.3\linewidth]{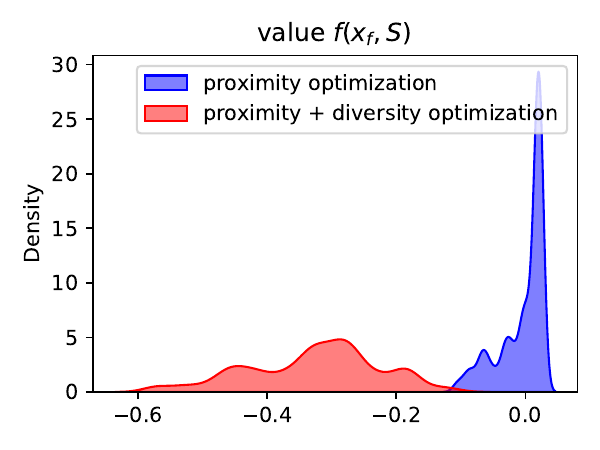}
    \includegraphics[width = 0.3\linewidth]{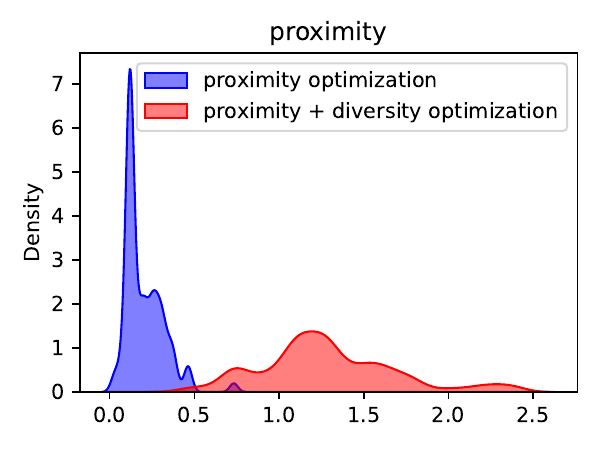}
    \includegraphics[width = 0.3\linewidth]{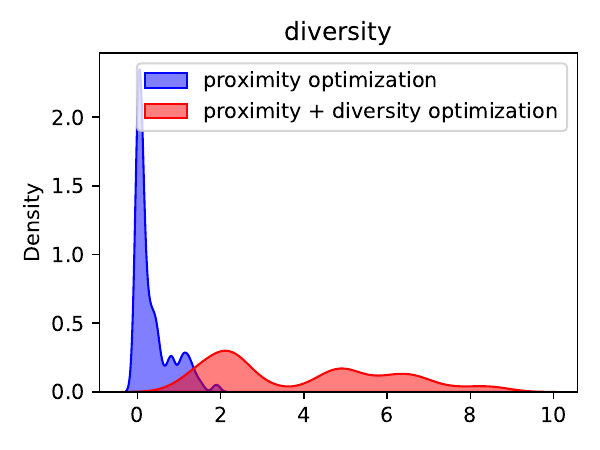}
    \caption{Diversity optimization: top row shows histograms, bottom row the density functions. The left column shows the value of the optimization function $f(x_f, S)$, the middle column the value of only the proximity term and the right column the values of only the diversity term. Color codes indicate the type of optimization.}
    \label{fig:diversity_optimization}
\end{figure}

\subsection{Detecting Biases}
\label{appendix:detecting_biases}
Explanations help users understand the reasoning behind a machine learning model’s outcome. Here, we discuss how such explanations can support the detection of biases -- for an in-depth discussion, cf.~\citet{DBLP:journals/tnn/RamachandranpillaiBH25}.
An important question towards this aim is: \textit{Is a sensitive feature (e.g.,~gender, race, age~\citep{charter2007}) determining the decision outcome?}\footnote{Alternative names for sensitive features are: protected variable/feature or identity category. Moreover, in this work, we use the original variable names from the Adult Income dataset, i.e.,~\texttt{sex} instead of \texttt{gender}.}
Computing contrastive explanations of a data instance and noting any change in the sensitive feature is one way to answer the question.
Below, we describe the experimental setup to obtain such an answer via {\reasonx{}}. Our approach is close to the definition of \enquote{explicit bias} in~\citet{DBLP:journals/ml/GoethalsMC24}: the authors define bias based on changes of the sensitive attribute between the data instance and the contrastive examples. However, they focus on negative decisions and distinguish between groups when aggregating results which is different to us.

Furthermore, our approach is closely related to the legal definition of \textit{direct discrimination} as well as to the concept of \textit{prima facie} discrimination. While the first refers to differential treatment on the basis of a sensitive feature~\citep{fra-handbook-discrimination}, the second builds on the information asymmetry between those deploying the AI system and affected users: in cases of algorithmic discrimination, the party accused typically holds the majority of relevant information. A claim can therefore be established via a presumption of discrimination. The alleged defendant then has to prove that no discrimination took place or that it was justified~\citep{fra-handbook-discrimination}.

Our approach is not taking into account \textit{proxy variables} \citep{Tschantz2022_ProxyDisc}, nor does it refer to \textit{indirect discrimination}~\citep{fra-handbook-discrimination}.

\paragraph*{Experimental setup.}
Our experiment starts
from the following simplistic assumption: if the decision outcome can be altered by \textit{only} changing a sensitive feature, or a set of such features, we take this as a \textit{prima facie} evidence of discriminatory behavior of the ML model that must be further investigated.
Here, we focus on those features that are protected by the EU law~\citep{charter2007}.
This assumption translates into the following experiment, testing the previous statement for one or two sensitive attributes.
We first initialize {\reasonx{}} with the ML model that we want to test and a data instance. 
In a second step, we test whether a query for a contrastive decision rule succeeds, and under the following constraints settings: asserting a set of equality constraints to keep all features between the factual and contrastive instance fixed and excluding one or two sensitive features. 
This means, we assert the constraint $F.x_i = CE.x_i$ for all $x_i$'s excluding the features \texttt{age}, \texttt{sex} or \texttt{race}, or pairwise combinations of these three features.
%
%
We repeat the experiment for a total of $N$ data instances, and compute the ratio $S/N$ of the number $S$ of queries with at least one answer constraint over the total $N$.
Further, after running the experiments for a DT, i.e.,~case \textbf{(DT-M)} and for {$N=1,000$}, we repeat it on {the surrogate DT of an} XGB model, i.e.,~case \textbf{(DT-GS)}. 

\paragraph*{Results.}
Results of this experiment are reported in Table~\ref{tab:reasonx_fairness} {for different maximum depths of the DT model (case \textbf{DT-M}) and of the surrogate tree of the XGB black-box (case \textbf{DT-GS}). The sensitive feature \texttt{sex} has the largest impact, followed by \texttt{age}, whereas changes to \texttt{race} alone never lead to a change in prediction. For the DT, the percentages generally increase with the maximum depth; for XGB, instead, larger surrogate depths tend to yield lower percentages. Notice that, because only sensitive attributes are allowed to change, the percentages need not vary monotonically with tree depth: even if a deeper decision tree refines a shallower one structurally, a contrastive rule available at depth~3 need not remain valid at depth~4. Overall, the DT exhibits a substantial dependence on sensitive attributes, which frequently appear among the features whose modification reverses the prediction. By contrast, the (surrogate of the) XGB model appears less affected by this form of bias.}
%
%
%

\begin{table}[t]
    \centering
    \small
    \scriptsize
    \begin{tabular}{rcccccccc}
        \toprule
        \textbf{Case} & $D$ & $f$ & \textbf{\texttt{age}} & \textbf{\texttt{race}} & \textbf{\texttt{sex}} & \textbf{\texttt{age, race}} & \textbf{\texttt{age, sex}} & \textbf{\texttt{race,sex}}\\
        \midrule
DT-M & 3 &  & 4.3\% & 0.0\% & 0.0\% & 4.3\% & 4.3\% & 0.0\%\\
DT-M & 4 &  & 8.1\% & 0.0\% & 11.5\% & 8.1\% & 16.5\% & 11.5\%\\
DT-M & 5 &  & 9.5\% & 0.0\% & 12.0\% & 9.5\% & 17.5\% & 12.0\%\\
\midrule
DT-GS & 3 & 0.921 & 5.1\% & 0.0\% & 6.9\% & 5.1\% & 13.2\% & 6.9\%\\
DT-GS & 4 & 0.940 & 3.9\% & 0.0\% & 5.2\% & 3.9\% & 10.3\% & 5.2\%\\
DT-GS & 5 & 0.948 & 4.2\% & 0.0\% & 4.8\% & 4.2\% & 10.1\% & 4.8\%\\
        \bottomrule
    \end{tabular}
    \caption{{Percentage of cases over $N=1,000$ instances with a contrastive rule that changes sensitive features only. $D$ indicates the depth of the (surrogate) DT, $f$ its fidelity (case \textbf{DT-GS}). For case \textbf{DT-GS}, the black-box model is XGB.}}
    \label{tab:reasonx_fairness}
\end{table}

\section{{More on Quantitative Evaluation}}
\label{sec:appendix_parameter_testing}


\begin{figure}[t]
    \centering
    \includegraphics[width = 0.95\linewidth]{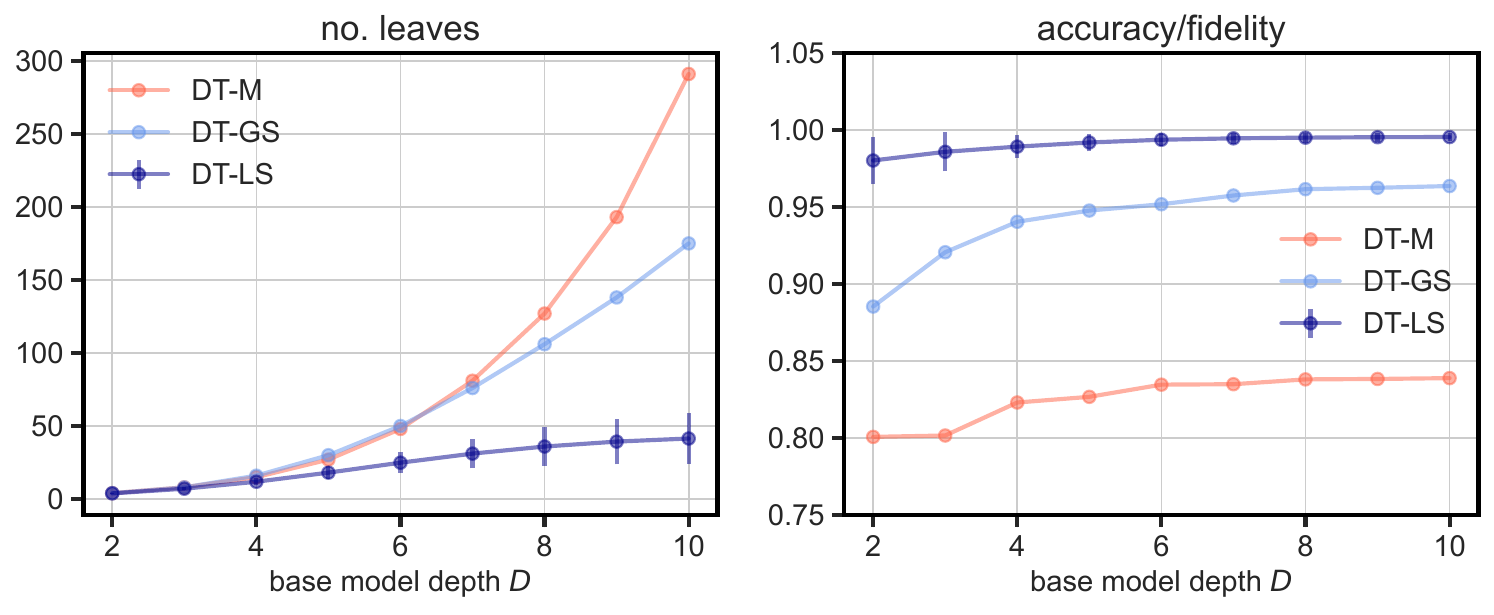}
    \caption{{Dataset \texttt{ADULT}. Number of leaves of DTs and their accuracy (for \textbf{(DT-M)}) / fidelity (for \textbf{(DT-GS)} and \textbf{(DT-LS)}). In case \textbf{(DT-LS)} metrics are averaged over 100 instances.}}
    \label{fig:further1}
\end{figure}

{We present additional experiments to evaluate the behavior of \reasonx{} as the maximum depth $D$ of the base models varies. The experiments use the \texttt{ADULT} dataset, with the same settings reported in Section~\nameref{sec:appendix_experiments}, except that $D$ now varies.
%
%
}

{First, we discuss the size and predictive performance of the base DT models. Figure~\ref{fig:further1} reports the number of leaves and the accuracy (for \textbf{(DT-M)}) or fidelity (for \textbf{(DT-GS)} and \textbf{(DT-LS)}) as the maximum depth of the base DT varies. For global models, the number of leaves grows exponentially with depth. For local models, instead, the base DTs remain considerably smaller, since the black-box decision boundary is generally easier to approximate within a neighborhood of the instance to be explained. This is a well-known advantage of local surrogate approaches. 
Regarding predictive performance, the accuracy of \textbf{(DT-M)} increases with tree complexity until reaching a plateau, a behavior well known in the literature~\citep{DBLP:conf/icml/OatesJ97}. A similar pattern is observed for the fidelity of global surrogate DTs in \textbf{(DT-GS)}. By contrast, the fidelity of the local surrogate models in \textbf{(DT-LS)} is already high for shallow trees and rapidly reaches a plateau. In summary, DTs used as local surrogate models tend to provide high fidelity while remaining relatively small, which supports their use in local explanation methods. For global explanations, instead, surrogate fidelity must be balanced against the increasing complexity of the DT, which also increase the complexity of explanations as we will discuss next.
}

\begin{figure}[t]
    \centering
    \includegraphics[width = 0.95\linewidth]{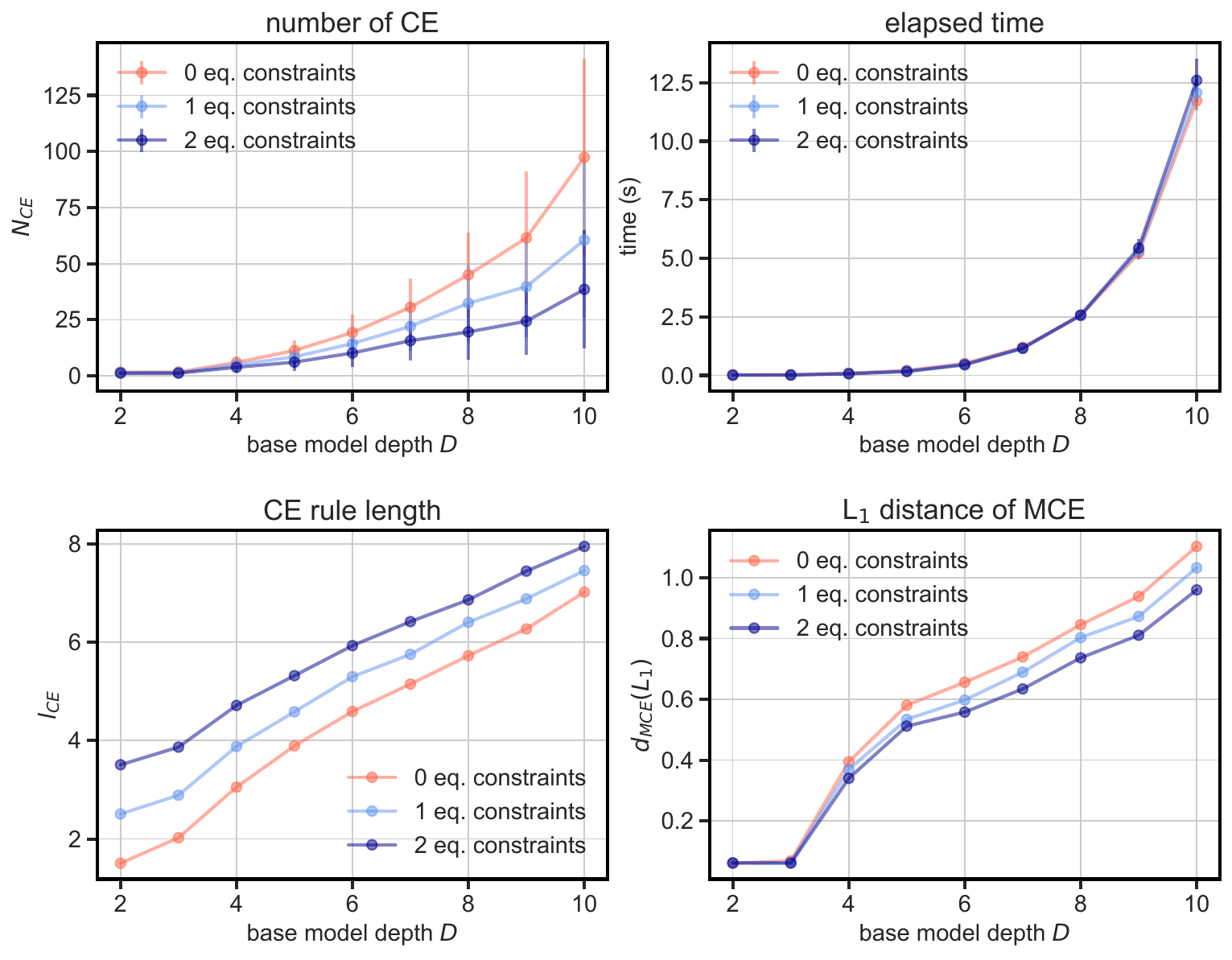}
    \caption{{Dataset \texttt{ADULT} and case \textbf{(DT-M)}. Number of answers, elapsed time, size of contrastive explanation rules, and $L_1$ distance in minimal contrastive explanation rules. Metrics are averaged over 100 instances.}}
    \label{fig:further2}
\end{figure}

{Let us now focus on the \textbf{(DT-M)} case. Figure~\ref{fig:further2} reports four analyses as the maximum depth of the model varies, also considering the effect of user constraints. Specifically, when searching for a contrastive explanation for a given factual instance, we impose zero, one, or two equality constraints expressing the immutability of randomly selected features.  The second constraint is added to the first, so changes are incremental. The experiments are conducted over $100$ factual instances, and the plots report the mean values together with $\pm$ one standard deviation.}

{The first analysis (top left) reports the number $N_{\textit{CE}}$ of contrastive explanation rules per factual instance. As expected, imposing additional user constraints reduces the number of admissible answers, which otherwise grows with the size of the DT. Allowing users to specify an arbitrary number of constraints therefore provides a direct mechanism for narrowing the search space and focusing the contrastive explanation problem on solutions that satisfy their requirements.}

{The second analysis (top right) reports the elapsed time per factual instance. The cost of searching for contrastive instances grows with the size of the base DT model, while the presence of user constraints does not substantially reduce the elapsed time. This again favors the use of smaller base DTs. The apparent exponential growth with maximum depth is also related to the specific DT embedding currently adopted by \reasonx{}, which represents each leaf as a separate Prolog fact; since the number of leaves may itself grow exponentially with depth, so can the running time. An alternative embedding could represent the DT directly through Prolog rules, one per decision node, and perform consistency checks while traversing the tree. This could prune unsatisfiable branches earlier, at the cost of performing a larger number of satisfiability tests.}

{The third analysis (bottom left) reports the size $l_{\mathit{CE}}$ of the contrastive explanation rules, measured as the number of constraints in their left-hand side. As expected, deeper DTs tend to produce larger rules. The presence of user constraints also increases the rule size proportionally. From a practical perspective, larger contrastive rules may be more difficult for users to parse and understand, further motivating the use of compact base models.}

{The last analysis (bottom right) focuses on a \textit{curse of dimensionality} effect in minimal contrastive rules. The plot reports the minimized $L_1$ distance between the factual and contrastive instances. As the depth of the base model increases, more features tend to appear in decision splits and, consequently, in the contrastive rules. This leaves fewer features \enquote{free} to remain unchanged between the factual and contrastive instances, which can increase the minimum $L_1$ distance. This effect is less pronounced in the presence of equality constraints, since such constraints explicitly force selected features to remain unchanged.}

\end{sm}

\end{document}